\documentclass[fleqn,usenatbib]{mnras}

\usepackage{newtxtext,newtxmath}
\usepackage{xcolor}%
\usepackage{subcaption}

\usepackage[T1]{fontenc}

\DeclareRobustCommand{\VAN}[3]{#2}
\let\VANthebibliography\thebibliography
\def\thebibliography{\DeclareRobustCommand{\VAN}[3]{##3}\VANthebibliography}

\usepackage{graphicx}	
\usepackage{amsmath}	
\usepackage{import}
\usepackage[nobiblatex]{xurl}
\usepackage{makecell}
\usepackage{tabularx}

\title[Lunar topography impact for 21-cm arrays]{The impact of lunar topography on the 21-cm power spectrum for grid-based arrays : Insights for the Dark-ages EXplorer (DEX)}

\author[Ghosh et al.]{S. Ghosh,$^{1}$\thanks{E-mail: soniaghosh@astro.rug.nl (SG)}
L. V. E. Koopmans,$^{1}$
C. Brinkerink,$^{3}$
A. R. Offringa,$^{1,2}$
A. J. Boonstra$^{2}$\thanks{Deceased},
S. A. Brackenhoff,$^{1,4}$
\newauthor E. Ceccotti,$^{1,5}$
J. K. Chege,$^{1}$
L. Y. Gao,$^{1,12}$
B. K. Gehlot,$^{1}$
L. I. Gurvits,$^{9,10}$
C. Höfer,$^{1}$
F. G. Mertens,$^{1,6}$
\newauthor M. Mevius,$^{2}$
S. Munshi,$^{1}$
A. Saxena,$^{8}$
J. A. Tauber,$^{11}$
H. Vedantham,$^{1,2}$
S. Yatawatta,$^{2}$
S. Zaroubi$^{1,7}$
\\
$^{1}$ Kapteyn Astronomical Institute, University of Groningen, PO Box 800, NL-9700 AV Groningen, The Netherlands\\
$^{2}$ Netherlands Institute for Radio Astronomy (ASTRON), PO Box 2, NL-7990 AA Dwingeloo, The Netherlands\\
$^{3}$ Radboud Radio Lab, Radboud University, Heyendaalseweg 135, 6525 AJ Nijmegen, The Netherlands\\
$^{4}$ Faculty of Electrical Engineering, Mathematics and Computer Science, Delft University of Technology, Mekelweg 4, 2628 CD Delft, The Netherlands\\
$^{5}$ INAF - Istituto di Radioastronomia, Via P. Gobetti 101, 40129 Bologna, Italy\\
$^{6}$  LUX, Observatoire de Paris, PSL Research University, CNRS, Sorbonne Université, 75014 Paris, France\\
$^{7}$ ARCO (Astrophysics Research Center), Department of Natural Sciences, The Open University of Israel, 1 University Road, PO Box 808, Ra’anana 4353701, Israel\\
$^{8}$ Kavli Institute for Cosmology, Madingley Road, CB3 0HA Cambridge, UK\\
$^{9}$ Faculty of Aerospace Engineering, Delft University of Technology, Kluyverweg 1, 2629 HS Delft, The Netherlands\\
$^{10}$ Joint Institute for VLBI ERIC, Oude Hoogeveensedijk 4, 7991 PD Dwingeloo, The Netherlands\\
$^{11}$ Leiden Observatory, Leiden University, P.O. Box 9513, 2300 RA Leiden, The Netherlands\\
$^{12}$ Liaoning Key Laboratory of Cosmology and Astrophysics, College of Sciences, Northeastern University, Shenyang 110819, China
}

\date{Accepted XXX. Received YYY; in original form ZZZ}

\pubyear{\the\year{}}

\begin{document}
\label{firstpage}
\pagerange{\pageref{firstpage}--\pageref{lastpage}}
\maketitle

\begin{abstract}
The Dark Ages (DA) provides a crucial window into the physics of the infant Universe, with the 21-cm signal offering the only direct probe for mapping out the three-dimensional distribution of matter at this epoch. To measure this cosmological signal, the Dark-ages EXplorer (DEX) has been proposed as a compact, grid-based radio array on the lunar farside. The minimal design consists of a 32 $\times$ 32 array of 3-m dipole antennas, operating in the 7 -- 50 MHz band. 
A practical challenge on the lunar surface is that the antennas may get displaced from their intended positions due to deployment imprecisions and non-coplanarity arising from local surface undulations. We present, for the first time, an end-to-end simulation pipeline, called SPADE-21cm, that integrates a sky model with a DA 21-cm signal model simulated in the lunar frame and incorporating lunar topography data. We study the effects of both lateral (xy) and vertical (z) offsets on the two-dimensional power spectra across the 7 -- 12 MHz and 30 -- 35 MHz spectral windows, with tolerance thresholds derived only for the latter. 
Our results show that positional offsets bias the power spectrum by 10 -- 30 per cent relative to the expected 21-cm power spectrum during DA. Lateral offsets within $\sigma_{xy}/\lambda \lesssim 0.027$ (at 32.5 MHz) keep the fraction of Fourier modes with strong contamination (> 50 per cent of the signal) to less than 1 per cent, whereas vertical height offsets affect a larger fraction. This conclusion holds for the 21-cm window with $k_\parallel > 0.5$ $h$ cMpc$^{-1}$ over the range of $k_\perp = 0.003 - 0.009$ $h$ cMpc$^{-1}$.




\end{abstract}




\section{Introduction}
 The launch of the James Webb Space Telescope (JWST; \citealt{gardner2006james}) has enabled us to push the redshift limit of direct observations. With reports of high redshift candidates at \textit{z} $\sim$ 20 \citep{yan2022first} to spectroscopically confirmed galaxies at redshift \textit{z} $\sim$ 13 \citep{curtis2023spectroscopic,robertson2023identification}, JWST has exceeded the observational capabilities of its predecessor, the Hubble Space Telescope (HST). Despite these observations, the infant Universe remains largely unexplored, including the crucial periods of the Dark Ages (DA; $200 \gtrsim z \gtrsim30$) and the Cosmic Dawn (CD; $30 \gtrsim z \gtrsim15$). The DA defines the period of cosmic history of the Universe before the first stars "turned on" \citep{varshalovich1977distortion,rees2000first}. During this time, the underlying physics of the Universe was believed to be relatively simple and described by the physics of the standard $\Lambda$ Cold Dark Matter ($\Lambda$CDM) model. This era was followed by the CD, a period marked by the emergence of the first sources of radiation invoking structural complexity of the Universe \citep{pritchard2010constraining}.


Even with the capabilities of JWST, observing sources beyond the early CD is not feasible. The very first luminous objects, although intrinsically massive and extremely bright, either appear faint owing to their large cosmological distances or they are obscured by the intergalactic medium (IGM). Moreover, probing the DA remains entirely unattainable due to lack of any known luminous sources.

Currently, the spin-flip, hyperfine transition of neutral hydrogen (HI), with a wavelength of 21 cm in the rest frame, is the only known direct probe for studying the properties of the DA. It also provides one of the most comprehensive probes of the subsequent CD and the Epoch of Reionization (EoR) eras \citep{barkana2005detecting,pritchard200721}. The spatial distribution of HI is measured indirectly by radio interferometers through its differential brightness temperature ($\delta T_{b}$) against the Cosmic Microwave Background (CMB) radiation. This signal not only depends on the density and velocity fields of HI, but also on the spin temperature and ionization fraction. By mapping $\delta T_{b}$ across the sky and over frequency (a proxy of redshift and co-moving distance), we can reconstruct a three-dimensional (3D) map of HI.

During the DA, baryonic matter consists mostly of HI and helium, providing a relatively clean measurement window of cosmology compared to complex astrophysical probes of the present-day Universe. The Universe is also believed to be nearly homogeneous and isotropic during this period \citep{koopmans2021peering}. Hence, the small spatial fluctuations of the 21-cm signal from the DA are predominantly influenced by the dark-matter field power spectrum and remain well within the linear regime, even at very small scales (see \citealt{liu2020data}). This implies that modelling and interpreting them does not require addressing the complex non-linearities and physical processes encountered in the more recent Universe. The 21-cm signal observations from the DA can thus provide a unique way to test the standard $\Lambda$CDM model. Any deviations from this well-established model could provide valuable insights into the physics of structure formation, by providing clues about the nature of dark matter \citep{slatyer2013energy, slatyer2016indirect2, slatyer2016indirect, tashiro2014effects, short2020enlightening, hiroshima2021impacts, mondal2024constraining}, early dark energy \citep{hill2018can}, or any exotic physics \citep{clark201821, theriault2021global, yang2022impact}. The large number of linear modes measured by the 21-cm signal from the DA could also enable us to test the primordial non-Gaussianity present in the initial density field \citep{cooray200621,munoz2015primordial,meerburg2017prospects,floss2022dark}, and even reveal the signatures of primordial gravitational waves \citep{book2012lensing,schmidt2014large, masui2017two,ansari2018inflation}.


Given the enormous potential of 21-cm observations to inform us about the physical processes in the infant Universe, constrain or estimate cosmological parameters, and distinguish between various theoretical reionization models, there have been worldwide efforts to design either low-frequency radio arrays measuring spatial fluctuations of the 21-cm signal, or single radio antenna experiments measuring the spatially-averaged 21-cm signal. Most of these current ground-based radio interferometers primarily focus on EoR, placing increasingly tighter constraints on the 21-cm signal power spectrum. However, a confident detection remains elusive. Although the EDGES experiment \citep{bowman2018absorption} claimed a detection, this has been refuted by SARAS2 experiment with 95 per cent confidence \citep{singh2022detection}. 
While the first generation of radio instruments are making substantial progress, the next-generation ground-based radio interferometers such as SKA\footnote{\url{http://www.skatelescope.org}} (Square Kilometre Array, \citealt{dewdney2009square}) are projected to achieve significantly higher sensitivity at these redshifts (CD and EoR), particularly through its SKA-Low component, thus offering better constraints on astrophysical models and making tomographic imaging possible \citep{koopmans2015cosmic}. 

One of the primary challenges to an accurate measurement of 21-cm fluctuations, irrespective of it being observed with ground-based or space-based experiments, is the synchrotron and free-free emission from galactic and extragalactic sources. This foreground emission exceeds the 21-cm signal from the EoR by 4 -- 5 orders of magnitude in intensity \citep{furlanetto2006cosmology}. The difficulty of foreground emission is further increased during the CD and DA, as the foreground emission surpasses the 21-cm signal by 6 -- 7 orders of magnitude. This challenge is further compounded by the chromaticity of the instrument and the higher thermal noise at low frequencies. 

In addition to these, ground-based experiments must also overcome other difficulties such as the Earth's ionosphere, and terrestrial radio frequency interference (RFI). The frequency-dependent, direction-dependent, and time-varying effects of the Earth's ionosphere, which by means of refraction, diffraction, absorption, and through its own thermal emission, can potentially introduce chromatic leakage of the bright foregrounds into the 21-cm signal \citep{koopmans2010ionospheric, sokolowski2015impact, vedantham2016scintillation, datta2016effectsionospheregroundbaseddetection, mevius2016probing, shen2021quantifying}. Also, below the plasma frequency ($\sim$10 MHz) of the ionosphere's F-layer, which generally depends on the electron density, the ionosphere is effectively opaque to radio waves. As a result, measuring the highly redshifted 21-cm signal from DA at a few tens of MHz and below is not feasible from the ground. 

Furthermore, human-generated RFI poses an increasingly larger challenge for ground-based 21-cm experiments. RFI can originate from a range of sources such as FM towers, digital TV, audio broadcasts, satellite and aircraft communications. Even infrastructures such as wind and solar farms, transformers add to the interference \citep{offringa2013brightness, offringa2015low, offringa2019impact, whitler2019effects, gehlot2024transient, munshi2025near}. Of growing concern is also the continuing deployment of satellite constellations such as Starlink, which have been found to emit unintended radiation in the frequency band of interest \citep{di2023unintended, grigg2023detection, bassa2024bright, zhang2025broadband}. This issue is becoming increasingly severe and is expected to worsen in the coming years. Without effective mitigation strategies, such emissions could lead to substantial data loss or even the loss of entire frequency ranges.

This compounded set of challenges motivates us to investigate space-based or lunar-based 21-cm DA driven experiments that would eliminate the detrimental impact of the ionosphere, and offer a more pristine radio-environment. The idea of deploying radio telescopes on the Moon \citep{douglas1988very}, or in lunar orbit \citep{schzcerch1968large} has been under consideration for many decades now. In recent years, however, there has been renewed interest in returning to the Moon, driven by both scientific and technological motivations, as well as economic and strategic reasons. This has led to a wide range of proposals for lunar missions that aim to measure radio emission sky (and sometimes sub-surface), including experiments specifically focused on 21-cm cosmology. Among them are DARE (Dark Ages Reionization Explorer, \citealt{burns2012probing}), LuSEE-Night\footnote{\url{https://www.colorado.edu/ness/projects/lunar-surface-electromagnetics-experiment-night-lusee-night}} (Lunar Surface Electromagnetics Experiment, \citealt{bale2023lusee}), DAPPER (Dark Ages Polarimeter PathfindER, \citealt{burns2019dark}), LCRT\footnote{\url{https://www.nasa.gov/general/lunar-crater-radio-telescope-lcrt-on-the-far-side-of-the-moon}} (Lunar Crater Radio Telescope, \citealt{bandyopadhyay2021conceptual}), FARSIDE (Farside Array for Radio Science Investigations of the Dark ages and Exoplanets, \citealt{burns2019farside}), FarView \citep{polidan2024farview}, DSL (Discovering the Sky at Longest wavelength mission, \citealt{xuelei2023discovering}), PRATUSH\footnote{\url{https://wwws.rri.res.in/DISTORTION/pratush.html}}(Probing ReionizATion of the Universe using Signal from Hydrogen, \citealt{sathyanarayana2023pratush}), LARAF (Large-scale Array for Radio Astronomy on the Farside, \citealt{chen2024large}), and TSUKUYOMI \citep{matsumoto2024low}. Another notable space-based low-frequency mission concept is SEAMS (Space Electric and Magnetic Sensor, \citealt{Tanti2023SEAMSAS}), aiming to explore the radio sky at ultra-long wavelengths. At the moment of writing, NCLE (Netherlands-China Low-Frequency Explorer, \citealt{chen2020netherlands}), LFRS (Low Frequency Radio Spectrometer, \citealt{zhu2021ground}), and ROLSES\footnote{\url{https://https://www.colorado.edu/ness/projects/radiowave-observations-lunar-surface-photoelectron-sheath-rolses}}(Radio wave Observations at the Lunar Surface of the photoElectron Sheath, \citealt{burns2021low, hibbard2025results}), are the only missions to have conducted low-frequency radio observations from the Earth-Moon L2 point, and from the farside and near side of the Moon, respectively. These missions have operated in the frequency ranges of 80 kHz -- 80 MHz (NCLE), 100 kHz -- 40 MHz (LFRS) and 5 kHz -- 30 MHz (ROLSES).

To explore the concept of a lunar-based low-frequency radio telescope, the European Space Agency (ESA) formed the Topical Team for an Astronomical Lunar Observatory (ALO) \citep{esa2021astrophysical, klein2024astronomical}, as part of ESA's European Large Logistics Lander (EL3) programme. The topical team proposed the deployment of the Dark-ages EXplorer (DEX), an early-stage concept to build a compact, highly-scalable, grid-based radio array on the lunar farside to study the early CD and DA using the 21-cm signal over a frequency range of 7 -- 50 MHz \citep{brinkerink2025dark}. 


While compelling due to its shielded and almost ionosphere-free environment, a 21-cm experiment on the lunar farside (either on the surface or in orbit) comes with its own set of challenges. It is important to note that for any experiment, whether on Earth or (around) the Moon, the received signal is often compromised by additional spectral features that can leak power into the modes that would otherwise be dominated by 21-cm signal. These artefacts can arise from several sources, including impedance mismatches in the analog signal chain \citep{beardsley2016first, ewall2016hydrogen, o2024understanding}, mutual coupling between antennas \citep{kern2019mitigating, kern2020mitigating, josaitis2022array, rath2024investigating, o2025uncovering}, digital processing artefacts \citep{prabu2015digital, barry2019fhd}, incomplete sky model and beam modeling inaccuracies \citep{barry2016calibration, ewall2017impact, byrne2019fundamental, barry2022role, gan2022statistical, munshi2024first,
brackenhoff2025robust}, and polarization leakage \citep{moore2013effects, asad2016polarization, asad2018polarization, kohn2016constraining}. In addition to this, the `coupling' of antenna properties to the medium over which it resides, especially if no ground plane is present, adds to the complication, regardless of the location of observation \citep{bradley2019ground, mahesh2021validation, spinelli2022antenna, monsalve2024simulating, hendricksen2025estimating}.

The development of radio interferometers on the lunar farside will be implemented in stages, starting with a few-metre scale single element system to a few-kilometre scale array. The site selection for any array of radio antennas, especially those focused on 21-cm signal observations, is a crucial first step. For example, ensuring coplanarity (antenna elements in the same geometric plane) significantly reduces data processing costs by eliminating the need to account for $w$-term effects during imaging. To maintain this, the array requires a relatively flat surface, preferably having perpendicular deviations from the plane well below the observed wavelength. Therefore, a potential site depends on a number of factors, including metre-to-decametre scale roughness, and gradients of the surface.

In addition, most of the rovers used for current Mars and Moon missions by NASA (National Aeronautics and Space Administration) have a maximum safety incline limit of 30$^\circ$ with engineering constraints on the wheel size and speed limit \citep{le2023lunar}. These considerations are critical to ensure the optimal deployment and operation of the radio array. Since DEX is designed as a regular array on the lunar farside, its deployment could result in deviations in antenna positions and orientation from a regular coplanar grid, either due to local undulations, gradient of the surface or deviations in the rover trajectory. This poses a potential issue for calibration and imaging if the relative locations of the antennas are not determined with sufficient precision (i.e., if each antenna's position is not known within a small fraction of a wavelength or if their orientations are uncertain). Furthermore, the resulting irregular array may prohibit the use of FFT-based techniques for correlation and imaging, which are essential for reducing both computational costs and the energy usage of the array on the Moon.

For example, lessons from ground-based radio instruments have shown that antenna feeds can deviate from their intended positions due to misalignments, rotations, or positional displacements \citep{joseph2018bias, orosz2019mitigating}. Additionally, slight variations in electronic gains, mechanical deformations, or even the surrounding environment can introduce nonuniform primary beams \citep{ansah2018simulations, choudhuri2021patterns, kim2022impact, kim2023impact}. By inducing unwanted spectral structure, these perturbations contaminate the Fourier modes used for 21-cm signal measurements, thereby impacting the extraction of the desired signal.




In this paper, we focus on antenna position offsets, both lateral and vertical height, that may arise in lunar-based array during the practical construction and deployment of radio antennas on the surface of the Moon. By systematically introducing controlled perturbations, we assess how these impact the power spectrum estimation. We present, for the first time, a science-driven forward simulation for DEX (and any lunar surface-based array) that incorporates a sky model, a cosmological 21-cm signal model from the DA, as well as a framework for coordinate transformation to a lunar topocentric reference frame, and  lunar topography data from the Lunar Reconnaissance Orbiter (LRO). This work presents an initial step towards understanding this particular systematic effect among the many challenges associated with constructing a large-scale radio array on the farside of the Moon. The findings will inform and guide decisions regarding the array configuration, deployment strategies, and selection of sites on the lunar surface.

This paper is organized as follows. In Section \ref{sec:lunar_topo}, we present an analysis of the lunar topography data obtained from LRO, relevant for a radio array. Section \ref{sec:dex} presents an overview of DEX and its design consideration. The different cases of simulated antenna position errors are described in detail in Section \ref{sec:ant_pos_pert}. In Section \ref{sec:end_to_end}, we describe our forward simulation pipeline. We focus on the impact of antenna position errors on the power spectrum for DEX across two spectral windows, 7 -- 12 MHz ($z$ = 148.7, Z148 hereafter) and 30 -- 35 MHz ($z$ = 42.5, Z42 hereafter), with results primarily focusing on Z42 in Section \ref{sec:results}. Finally, Section \ref{sec:conclusions} presents our summary and main conclusions derived from this study. Throughout this study, we adopt cosmological parameters from the \citealt{planck2016planck}: \(\Omega_{\rm m} = 0.315\), \(\Omega_{\rm b} = 0.049\), and \(H_0 = 67.7\,\mathrm{km\,s^{-1}\,Mpc^{-1}}\).

\section{Lunar Topography}
\label{sec:lunar_topo}

With a diameter of $\sim$ 3474 km, the Moon is the only natural satellite of Earth orbiting at an average distance of $\sim$ 384\,400 km, tidally locked so that the same face always points toward us. Unlike Earth, it is enveloped only by an extremely tenuous ionosphere with electron densities $n_{e}\sim 0.1 - 1$ cm$^{-3}$ \citep{halekas2018tenuous, shen2023limits}, giving a plasma cut-off frequency  $\leq 30$ kHz. Therefore, it is transparent to the long wavelengths needed to probe the 21-cm signal from the DA. The farside of the Moon also provides the best possible natural radio-quiet environment in the inner solar system. This was observed decades ago at low radio frequencies by Radio Astronomy Explorer 2 (RAE-2; \citealt{alexander1975scientific}), the WAVES instrument \citep{bougeret1995waves} and very recently by the Low-Frequency Interferometer and Spectrometer (LFIS) aboard
the satellites Longjiang-1 and Longjiang-2 \citep{yan2023ultra}. Electrodynamic simulations have also demonstrated that terrestrial RFI is attenuated by at least 80 dB \citep{bassett2020characterizing}, required to suppress RFI from Earth below the 21-cm signal, depending on the frequency and altitude of observation on the farside of the Moon.

To deploy low-frequency radio arrays such as DEX on the Moon, it is important to assess the physical characteristics of the lunar surface for rover traversability, safe landings, and stable antenna deployment. Beyond these engineering considerations, the primary motivation for characterising surface roughness is its potential to introduce systematic effects, thereby impacting the accuracy of the measurement of the DA 21-cm signal. A critical aspect of this is the scale-dependent surface roughness, which affects both engineering design and the scientific return of the instrument. The surface of the Moon is broadly divided into two distinct types. Maria, which are basaltic plains formed by volcanic activity, and highlands, which are older, elevated, and more heavily cratered region. Investigation of the topographic roughness of the lunar surface at scales from metres to tens of kilometres using the LRO Lunar Orbiter Laser Altimeter (LOLA; \citealt{smith2010lunar}), and Kaguya Laser Altimeter \citep{noda2009kaguya} digital elevation models (DEMs) has shown that the lunar highlands are generally much rougher than the maria, and that young, large impact craters exhibit the steepest local slopes \citep{rosenburg2011global, cao2015fractal, kreslavsky2016steepest, cai2020meter}. However, the small-scale roughness characteristics remain poorly understood, highlighting the need for further high-resolution studies to support the design of lunar surface infrastructure, and identify deployment sites suitable for a DA 21-cm signal detection experiment.

\subsection{Analysis of the Surface Topography}\label{lunar site}
To quantify the surface topography of the Moon, we have mainly used the data taken by LRO, a NASA mission which has been collecting data since 2009. LRO carries a range of scientific instruments, and has been instrumental in supporting lunar landing site selection, including for the ongoing Artemis program, by providing detailed measurements of lunar topography, surface morphology, illumination conditions, thermal environment, and surface roughness. Notably, the Lunar Reconnaissance Orbiter Camera (LROC) enables production of high-resolution (about 0.5 -- 2 m/pixel) digital terrain models (DTMs) using two Narrow Angle Cameras (NACs) that produce stereo observations \citep{robinson2010lunar}. This allows for detailed characterisation of surface roughness on metre-scales, which is below the scale of the planned antenna elements for DEX. 
In this study, we select Mare Ingenii, an impact basin in the northwestern part of the South Pole-Aitken basin (SPA), as a representative site for the deployment of DEX. This mare comprises a largely flat surface with slopes less than 5$^\circ$, bounded by steep crater walls exceeding 25$^\circ$ in inclination. Flat passageways between these walls offer navigable routes for rover traversability and, as concluded by \citet{le2023lunar}, could feasibly host a 200 km radio array. However, the final site selection will depend on future analyses using higher-resolution topography data and a broader range of engineering and scientific criteria. A detailed discussion of the selection criteria is presented in Section \ref{sec:location}.

\begin{figure*}
    \centering
    \includegraphics[width=\linewidth]{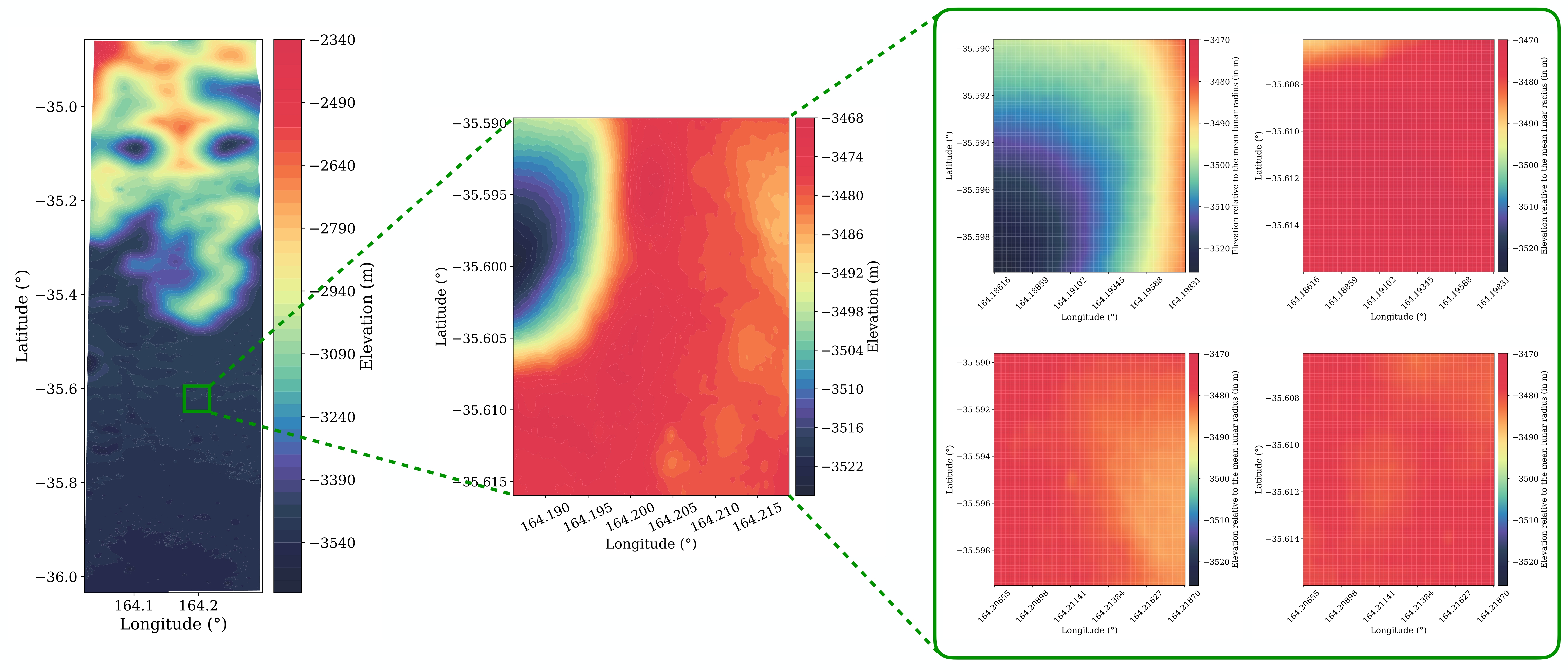}
    \caption{Left Panel: LROC NAC DTM of the Mare Ingenii, with a spatial resolution of 2 m. The colorbar indicates elevation in metres measured relative to the mean lunar radius ($\sim$ 1737.4 km). Middle Panel: The region selected from the full DTM for this work. Right Panel: The selected region in the middle panel is further divided into four sub-regions (Surface 1: top left, Surface 2: top right, Surface 3: bottom left, Surface 4: bottom right), each spanning an area $\sim$ 0.09 km$^2$.   }
    \label{fig:lunar_surface}
\end{figure*}

\subsubsection{Topography Dataset}\label{topography data}
We use the LROC NAC DTM products of Mare Ingenii for first-order characterization of surface roughness. The generated DTMs are archived by Arizona State University (ASU)\footnote{Available at: \url{https://data.lroc.im-ldi.com/lroc/rdr_product_select}} within NASA's Planetary Data System (PDS), where they are provided in the Gridded Data Record (GDR) format and projected using an equirectangular coordinate system. The spatial resolution of the data is 2 m, covering $\sim$ 0.2$^\circ$ in the longitude direction, and $\sim$ 1.2$^\circ$ in the latitude direction with total surface area covered by this dataset being 242.4 km$^2$. However, for the purpose of this study, we have visually selected a smooth region spanning 0.64 km$^2$ from this dataset. This region was further subdivided into four sub-regions as shown in Fig. \ref{fig:lunar_surface}, each with an area $\sim$ 0.09 km$^2$, for a comparative analysis of their surface roughness. We note that the process of selecting "smooth" sites can be automated and extended to higher resolution datasets, but that is beyond the scope of this work.

\begin{table}
 \centering
 \caption{Fractal roughness characterization of the four surfaces, showing the Hurst exponent H, and RMS deviation at reference scale of 4 m ($\sigma_{\rm 4\,m}$), and 175 m ($\sigma_{\rm 175\,m}$)}
 \label{tab:hurst_exponent}
 \def\arraystretch{1}
 \begin{tabularx}{\linewidth}{X|X|X|X}
  \text{Surface} & {\text{H}} & {$\sigma_{\rm 4\,m}$} [m] & {$\sigma_{\rm 175\,m}$} [m]\\
  \hline
  1 & 0.936 & 0.30 & 7.02\\
  2 & 0.917 & 0.26 & 4.12\\
  3 & 0.893 & 0.15 & 1.91\\
  4 & 0.898 & 0.14 & 1.81\\
 \end{tabularx}
\end{table}

\subsubsection{Topography Roughness Parameter}\label{roughness parameter}
\begin{figure*}
    \centering
    \includegraphics[width=\linewidth]{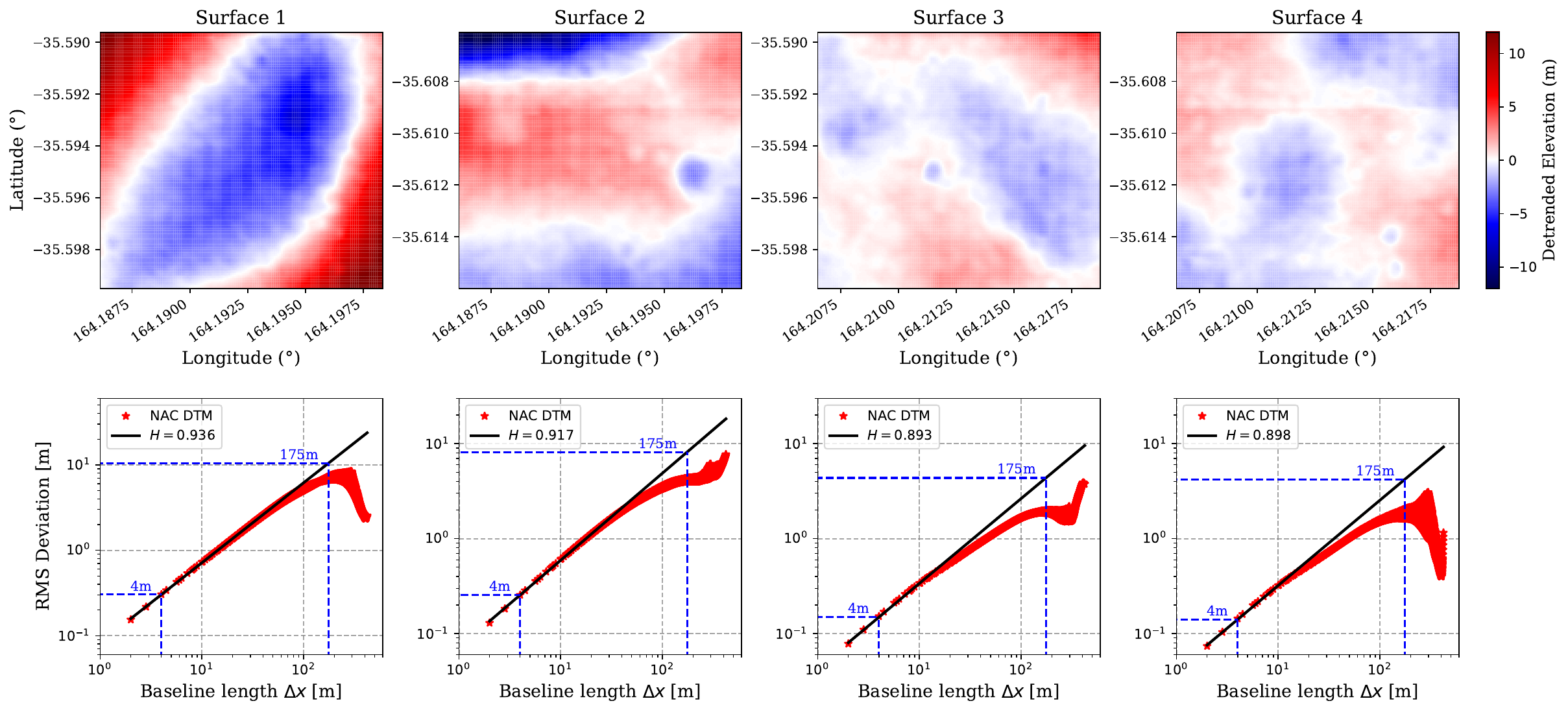}
    \caption{Top Row: The four sub-surfaces that have been detrended by fitting a 2D plane such that the small-scale topographic features are highlighted. Bottom Row: The observed structure functions (or deviograms) of the detrended surfaces. The derived Hurst exponent, $H$ range from 0.898 - 0.936, indicating persistent, and self-affine behaviour. It is observed that at 4 m scale, the RMS deviation of the elevation lie between 0.14 to 0.30 m, while at $\sim$ 175 scale, they reach up to 7 m. Here, the NAC DTM data points are shown in red, and the fitted power-law model in black.}
    \label{fig:surface_analysis}
\end{figure*}

Prior to the analysis, the topographic data are detrended by fitting and subtracting a plane from each of the sub-regions, such that they have a mean value of zero. This step is essential to isolate local surface undulations from broader regional trends, thereby allowing a more accurate characterization of small-scale roughness. 
The top row of Fig. \ref{fig:surface_analysis} shows the detrended elevation maps of the four surfaces.

Natural surfaces are often modeled as stationary Gaussian random fields. The root-mean-square (RMS) difference in elevation (height) between two points separated by distance $r$ can often be well-described by a power law $r$$^H$, where $H$ $\in$ [0,1] is the Hurst exponent \citep{shepard2001roughness}. We calculate the RMS height difference (also known as RMS deviation) as a function of separation $r$ as follows, 

\begin{equation}
    \sigma_{\rm H} (r) = \sigma_{\rm s}\left(\frac{r}{r_{\rm s}} \right)^H,
\label{eq:hurst_scaling}
\end{equation}

\noindent
where \(\sigma_{\rm s}\) is the RMS deviation at a reference distance \(r_{\rm s}\) between two points on the surface. We then analyse the resulting deviogram or structure function to characterise the scale-dependent surface roughness. To limit sample variance and edge effects, it is generally recommended that the maximum baseline (spatial separation between two points on the surface) in the structure function be restricted to less than one-tenth of the total profile length \citep{shepard2001roughness}. The Hurst exponent is then estimated as the slope of a linear fit to the log-log plot (see the bottom row of Fig. \ref{fig:surface_analysis}). A larger Hurst exponent, for a fixed $r{_s}$, implies higher degree of spatial correlations across scales, i.e., nearby points remain similar in elevation over longer distances. \citet{shepard2001roughness, rosenburg2011global} note that the scaling behavior often deviates from linearity at larger scales, therefore the linear fit is applied from the sampling interval (i.e., the resolution of the data) up to a outer scale, beyond which the scaling behavior no longer follows a power-law but in general flattens off. This outer scale maybe interpreted as the characteristic spatial scale at which dominant processes that form or modify the surface undergo a transition \citep{shepard2001roughness}. To quantify the goodness-of-fit, we use the Pearson correlation coefficient with a threshold value of 0.999 to identify the valid linear regime following \citealt{cai2020meter}.

\subsubsection{Topography Roughness Results}\label{roughness result}
We compute the value of $H$ for the four detrended surfaces, and find no discernible difference among them, with values ranging from 0.89 to 0.93. These values are consistent with those reported in the bottom row of Fig. 3 of \citealt{cai2020meter}, and indicate that the surfaces have a similar and strong spatial correlation. Over small geographic separations, as in the present study, this is expected due to the shared geological and morphological context. However, it is well established that different (widely separated) regions on the Moon can exhibit significantly different scaling behaviours of $H$ \citep{rosenburg2011global, cai2020meter, barker2025large}. We observe that Surfaces 1 and 2, show slightly higher values of $H$, suggesting a marginally greater degree of correlation in their scaling behavior compared to Surfaces 3 and 4. In the RMS deviation plots (or structure function) shown in Fig. \ref{fig:surface_analysis}, we observe that towards larger scale, the fitted power-law model (in black) begins to diverge from the NAC DTM data points (in red). This divergence indicates the outer scale beyond which the surface no longer exhibits self-affine fractal behavior, and may instead flatten, saturate, or exhibit more complex topographic behavior, indicating a change in the nature of surface variation. A self-affine surface is one whose statistical properties remain invariant under anisotropic scaling. The outer scale varies from approximately 100 m to 80 m for Surfaces 1 and 2, and 20 m for Surfaces 3 and 4, respectively. If the structure function flattens at the outer scale, then baselines longer than this scale sample approximately baseline-independent RMS height differences. However, if the surface shows long-range correlations or a different scaling law beyond the outer scale, the RMS height difference may continue to grow with baseline length, reflecting more complex topographic structure. 

Next, we analyze \(\sigma_{\rm s}\) at the scale of the smallest baseline $\sim$ 4 m and at the scale of the array's longest baseline $\sim$ 175 m (see Table \ref{tab:hurst_exponent}). Within the self-affine regime, \(\sigma_{\rm s}\) varies from 0.14 m to 0.30 m at a 4 m length scale. However, at larger scales ($\sim$175 m, longest baseline), the measured \(\sigma_{\rm s}\) falls below the power-law extrapolation. We observe that across our scale of interest, \(\sigma_{\rm s}\) of Surfaces 3 and 4 remains consistently lower than that of Surfaces 1 and 2. This implies that, despite their marginally lower $H$,  surfaces 3 and 4 are comparatively flatter in terms of absolute variations of elevation, and are thus more favorable for deployment. The values of $H$ and \(\sigma_{\rm s}\) in our analysis are first-order diagnostics that capture the distinct aspects of surface roughness, sufficient for the purposes of this study.

An optimal deployment site would, therefore, ideally combine both characteristics, namely, \(\sigma_{\rm s}\) (setting absolute height variations) at a reference distance \(r_{\rm s}\) and $H$ (defining spatial scaling). The influence of $H$ depends on the baseline distribution of the array. For experiments with sensitivity concentrated at small baselines, such as those targeting 21-cm fluctuations like DEX, higher values of $H$ are favourable as they suppress variance on the smaller baselines and reduce difference in height between neighboring antennas. Lower $H$ corresponds to larger fluctuations at short length scales, which can increase biases on the small baselines. For long-baseline arrays designed for high-resolution imaging, relatively lower $H$ is favorable because the power spectrum of the surface tends towards baseline-independence, so the error contribution is nearly uniform across all scales. At all times, \(\sigma_{\rm s}\), must be low to ensure that the induced bias remains within tolerance across the baseline length of interest. This emphasises that both the amplitude and scale-dependence of roughness must be considered when evaluating the feasibility of a surface for hosting low-frequency lunar arrays such as DEX. We do not discuss the geological processes responsible for affecting metre-scale surface roughness, as such analysis lies beyond the scope of this work. Note that in applying a fractal structure function description, we assume the absence of large discrete features such as boulders, which may violate the underlying scaling behaviour. 


\begin{figure*}
    \centering
    \includegraphics[width=0.30\textwidth]{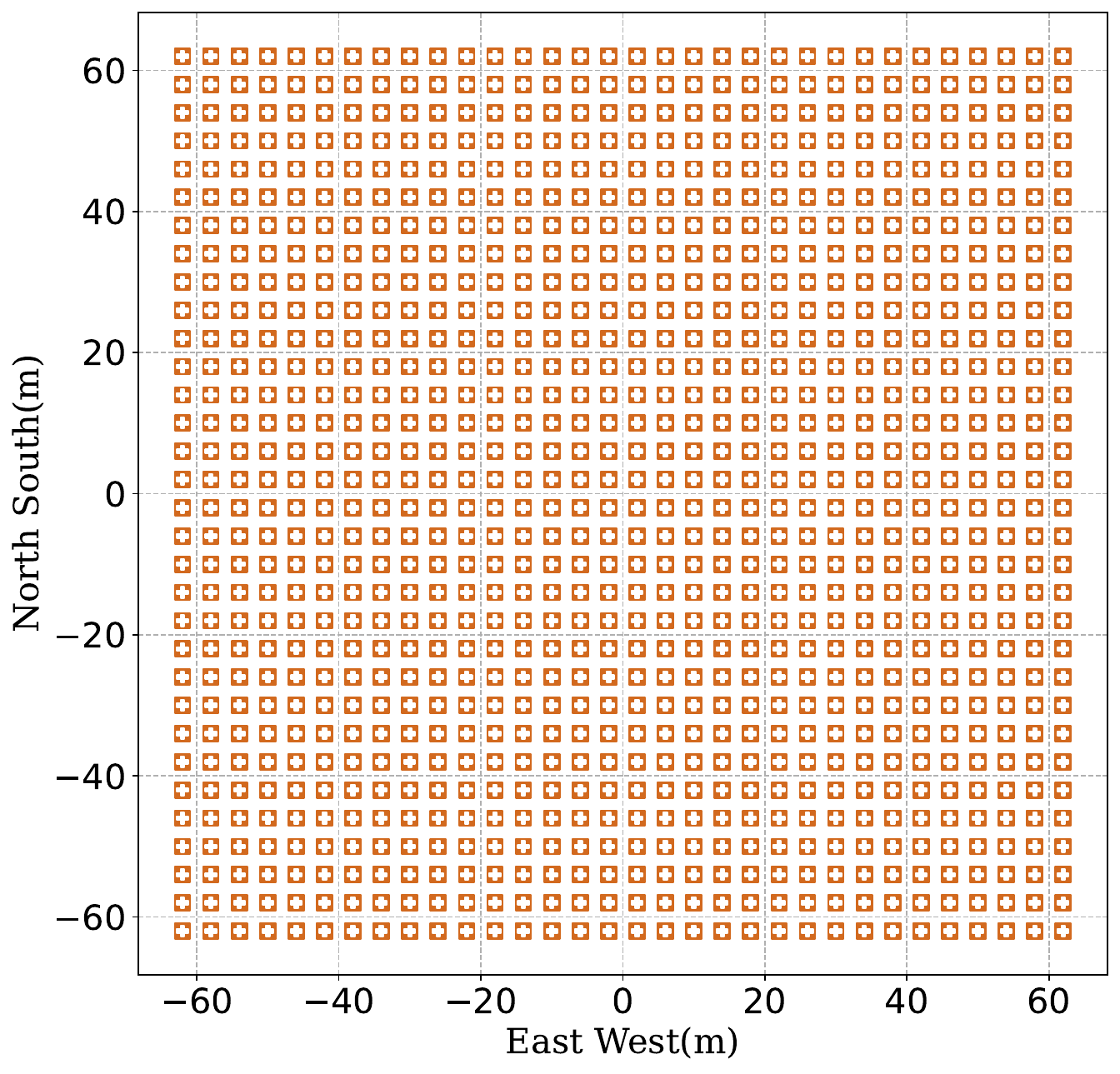}
    \includegraphics[width=0.35\textwidth]{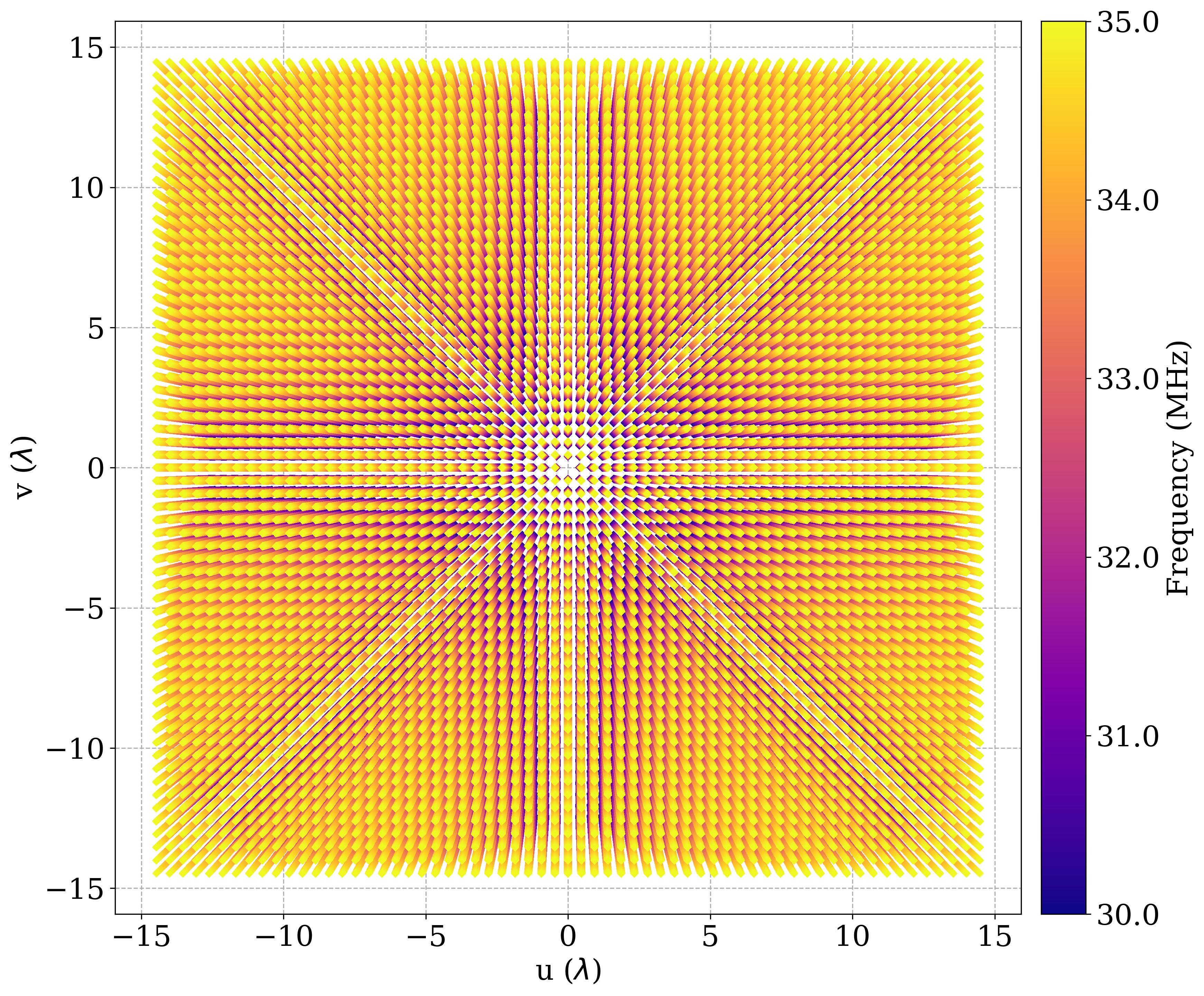}
    \includegraphics[width=0.30\textwidth]{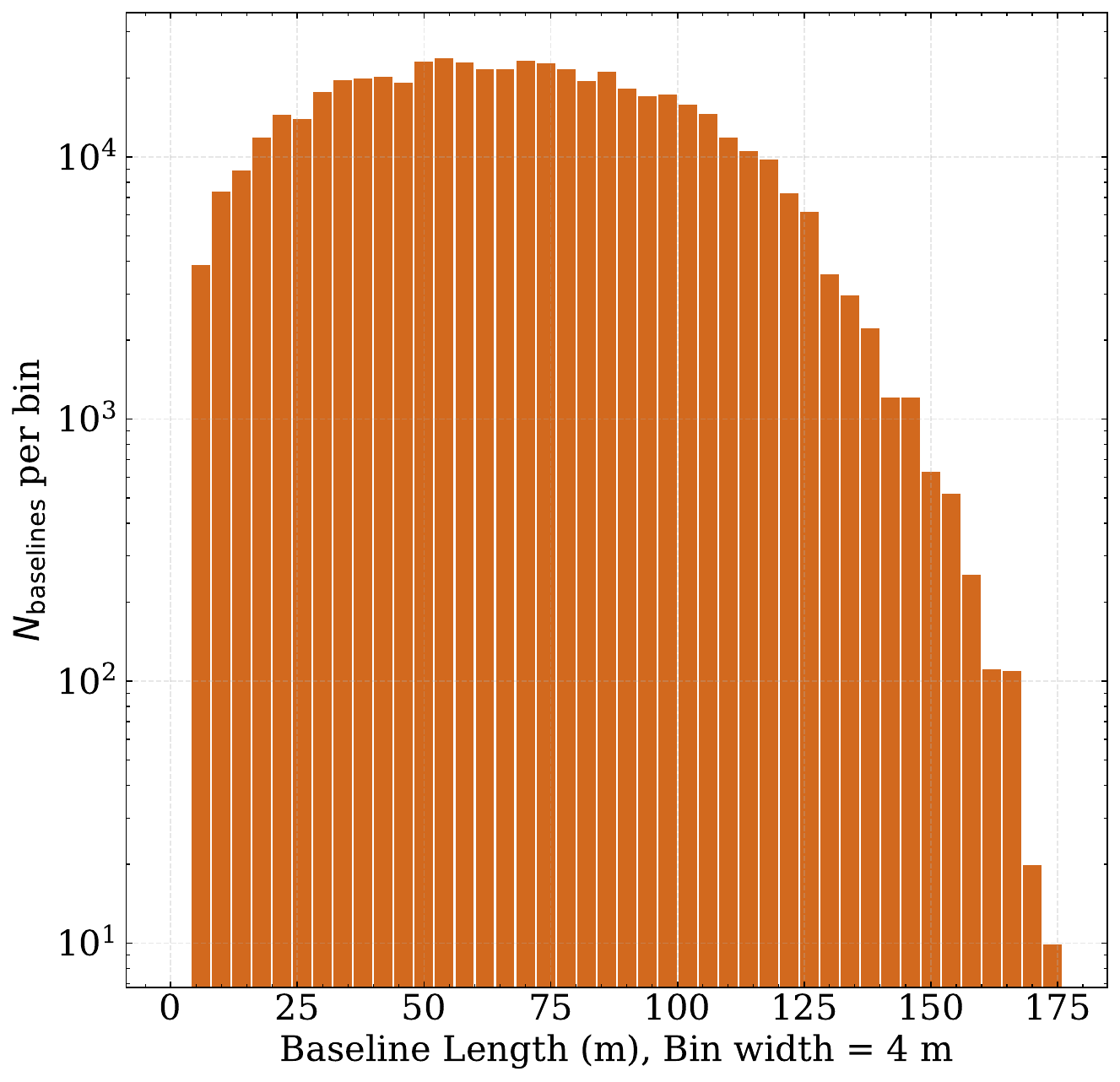}
 \caption{Left Panel: Array layout of the DEX configuration used for the simulation. DEX is planned to be a compact array with 1024 antenna elements, arranged on a regular grid, on the farside of the Moon. Middle Panel: The corresponding instantaneous $u\varv$ coverage of the array for a simulated snapshot of 5 minute within the Z42, demonstrating significant redundancy of the array in $u\varv$ sampling. A color gradient is used to indicate the distribution of independent $u\varv$ samples across the full spectral window. The observation was simulated from Mare Ingenii at RA = 23.052060$^\circ$, DEC = -25.696129$^\circ$, corresponding to an observation time of 16:22:30 UTC on July 30, 2040. Right Panel: A histogram of the baseline lengths ($|\mathbf{u}| = \sqrt{u^{2}+v^{2}}$) distribution (in m) with a bin width of 4 m.}
\label{fig:threefigs}
\end{figure*}


\section{The Dark Ages Explorer}
\label{sec:dex}
The 21-cm signal from the DA is expected to be extremely faint, with brightness temperature fluctuations of the order of only a few mK$^{2}$ at \( k \sim 0.1~\mathrm{Mpc}^{-1} \). To reach the desired sensitivity level in the presence of very bright polarized foregrounds, it is estimated that a minimum of $128\times128$, or more dipoles are necessary for any radio array targeting for detection via power spectrum measurement, as analyzed by \citet{koopmans2021peering}. 
Currently in the early conceptual phase, DEX is envisioned as a compact, grid array of zenith-pointing antennas with a near-unity filling factor, shown in Fig. \ref{fig:threefigs}, deployed on the farside of the Moon \citep{esa2021astrophysical}. This array is primarily designed to observe the Universe in the ultra-long wavelength domain using the highly redshifted 21-cm signal. In its basic and minimal configuration, DEX currently consists of $32\times32$ cross-dipole antennas of length 3 m operating in the radio frequency band of 7 -- 50 MHz. In future iterations of the array design, there should be scope for expansion in both the number of antennas and operational bandwidth (to 100 MHz), which would enable improved sensitivity and enhanced angular resolution. Consequently, the specific configuration parameters used in this work are subject to change as the project continues to mature.

\subsection{Correlation Architecture} \label{FFT correlator}

The data rates increase quadratically with increasing number of antennas when computing the cross-correlated voltages or the visibilities using a standard FX (Fourier Transform followed by cross-multiplication) correlator. This requires considerable computing resources, output data bandwidth, and power consumption. Addressing these challenges requires careful consideration, particularly for any lunar surface-based or space-based missions, where available resources are inherently limited. Therefore, it requires the development of novel and efficient architectures for correlation, data processing, transfer, storage, and minimal power usage, while ensuring that important performance factors such as resolution and sensitivity are not compromised \citep{price2024reduced}. \citet{thyagarajan2025comparison} provides a comparative analysis of the different imaging architectures based on their computational cost efficiency for various planned and proposed radio arrays such as SKA-low, SKA-low-core, LAMBDA-I, CASPA, and FarView-core. One of the key findings is that imaging architectures based on the Fast Fourier Transform (FFT; \citealt{cooley1965algorithm}) offer much greater computational efficiency for large, densely packed arrays across a wide range of observational cadences. This class of architectures, broadly referred to as \textit{Direct Imaging}, avoids the computationally expensive step of correlating all antenna pairs individually \citep{daishido1991direct,tegmark2009fast, tegmark2010omniscopes}. As a result, it eliminates the need to transport and process large volumes of correlation data for subsequent imaging, although access to individual visibility data is lost.

The classification of \textit{Direct Imaging} architecture naturally leads to the exploration of FFT correlators. Unlike the computational cost of the traditional FX and XF (cross-multiplication followed by Fourier Transform) architectures, which scale as $\sim N^2$, the FFT correlator reduces this to $\sim N\log N$, where $N$ is the number of antennas. This improvement is achieved by using the relationship between the electric field measured on the ground and its corresponding image in the sky through a spatial Fourier transform. In its simplest implementation, this approach requires the antennas to be placed on a regular grid. A recent advancement in the direction of this alternative imaging and correlator scheme is the development of E-field Parallel Imaging Correlator (EPIC; \citealt{thyagarajan2017generic}). EPIC uses the Modular Optimal Frequency Fourier (MOFF) algorithm \citep{morales2011enabling}, eliminating restrictions on the antenna being physically placed on a nearly-perfect grid. Instead, the voltages are electronically gridded based on the aperture illumination pattern of the individual antennas, similar to visibility gridding in traditional correlation and imaging steps. We note that this method performs optimally when the array has a near-unity filling factor and closely approximates a grid layout.


In recent years, EPIC has been tested and commissioned at the LWA-SV station \citep{kent2019real,krishnan2023optimization}. Although EPIC is computationally efficient when applied to real data, a full science-driven forward simulation can be computationally demanding for DEX-like arrays. For example, generating stochastic electric field realizations at each antenna by coherently summing contributions from all sky directions across a wide frequency band can be computationally intensive, particularly for large-$N$ arrays.

As a practical workaround for spatially uncorrelated stationary electric fields (and voltages), we adopted a visibility-based approach (as for FX correlator) for this work. We can do this, without loss of generality, because the images produced by a FFT correlator (as in EPIC) are mathematically identical, to a first order, to those produced from a standard FX correlator via the convolution theorem of Fourier Transform. This approach also ensures compatibility with established data formats and facilitates the use of well-tested software tools within the radio astronomy community. 


\subsection{Design Consideration}

The main factors influencing the observational capabilities of a radio interferometer are the FoV of each antenna element (also called the primary beam), total effective collecting area and filling factor, observing frequency range and spectral resolution, the number and arrangement of the antenna elements forming baselines \citep{mellema2013reionization, koopmans2015cosmic}. For DEX, these technical considerations are closely linked to the need for extremely high sensitivity to enable making a statistically significant detection of the extremely faint 21-cm signal from the DA at millikelvin (mK) levels.

\subsubsection{Array Configuration and spatial resolution}
Compact, centrally concentrated configurations improve sensitivity by increasing sampling density and redundancy, as they provide a large number of visibility measurements per $u\varv$-cell within a given integration time. This principle has driven the design of interferometers aimed at detecting the 21-cm signal \citep{bowman2006sensitivity, parsons2012sensitivity, koopmans2015cosmic}. A high-filling-factor array, consisting of 1024 antenna elements, is considered the minimally useful configuration for DEX for 21-cm observations up to the early CD. Although this configuration defines the lower bound for 21-cm cosmology, smaller arrays may still support a range of complementary science cases. 
We also limit ourselves to this array size because simulating cross-correlated visibilities for larger arrays, such as 64 $\times$ 64, 128 $\times$ 128, 512 $\times$ 512, is beyond our current computational capabilities. 

\citet{koopmans2021peering, mondal2023prospects, polidan2024farview} predicted that an array spanning over 100 km$^{2}$, with $\sim$ 10$^6$ antenna elements, will be necessary to probe angular scales down to k$\sim0.1$ Mpc$^{-1}$ at frequencies corresponding to the redshifts ($z \sim$ 50) of the DA. Such a large array on the Moon can only be realized through multiple stages of development and deployment. It is believed that although the current configuration of DEX may not provide access to the DA redshifts assuming standard 21-cm signal models, it can still partially probe the early CD, allowing testing of various early star-formation and cosmological models that predict the power spectrum for this epoch. Also, such a pathfinder array helps investigating various aspects such as the impact of systematics on the science case, antenna deployment techniques, and the implementation of the FFT correlator.

Based on electromagnetic simulations of several cross-dipole antenna models for DEX (Arts, private communication), we find that within the frequency range of the DA, a relatively short dipole of 3 m behaves spectrally better compared to a longer 5 m dipole, assumed earlier, which shows sharper features in its bandpass. Accordingly, in this work, we assume a simple 3 m cross-dipole (measured tip-to-tip). 
We set a minimum baseline of 4 m when measured from antenna center to center (1 m edge-to-edge). To isolate the effects of positional perturbations, we adopt a first-order approximation of primary beam. We use a frequency-independent $\cos^2 \theta$ beam model, where $\theta$ is the angular distance from the zenith, and multiply with the simulated sky map. The chromatic effects of primary beam and mutual coupling between antenna elements will be addressed in details in a subsequent paper. A more complete treatment that incorporates beam patterns, may require revisiting the minimum baseline, as closely spaced elements can alter the effective beam response and impact the scientific performance of the array \citep{virone2018strong, kern2019mitigating, mckinley2020all, bolli2022impact, anstey2024mitigating}. The maximum baseline for DEX is given by the diagonal which is $\sim$ 175 m.

The angular resolutions, characterized by the full width at half maximum (FWHM) of the synthesized beam,  are $\sim$14$^\circ$ at 7 MHz and $\sim$1.9$^\circ$ at 50 MHz. This relatively low resolution is a natural drawback of a compact array at such low frequencies, along with the presence of strong grating lobes in the point-spread function (PSF) due to the regularity of the array. These shortcomings can be mitigated by adding "outrigger" antennas, which improve angular resolution and enable instantaneous wide-field imaging by improving $u\varv$ coverage \citep{dillon2016redundant}. However, these constraints do not fundamentally limit 21-cm power spectrum measurements since the power spectrum is estimated in Fourier space. Furthermore, compact arrays offer enhanced sensitivity and support redundant calibration strategies, which may help to achieve robust statistical measurements. 

\subsubsection{Bandwidth and Spectral Resolution}\label{bandwidth}
DEX is currently planned to operate over a broad frequency range spanning 7 -- 50 MHz, thereby covering the eras of the DA (z $\sim$ 200) and the early CD (z $\sim$ 27). As discussed earlier, this frequency range may be extended to higher frequencies to overlap with current ground-based experiments, allowing for complementary measurements across different redshift regimes and improved cross-validation.

The frequency axis of 21-cm observation captures the spatial fluctuations of the 21-cm field along the line-of-sight (LOS). In a cylindrical 2D power spectrum, which estimates the variance of the field, the spatial modes along the LOS are represented by the Fourier wavenumber, $k_{\parallel}$. The maximum measurable Fourier wavenumber, $k_{\parallel,\rm max}$ is limited by the spectral resolution, whereas the total bandwidth establishes the $k_{\parallel,\rm min}$, which determines the largest LOS scale of the 3D volume being probed. 

The redshift evolution of the cosmological signal is strongly dependent on frequency. When calculating power spectra from observational data, the full bandwidth of the instrument is typically divided into narrower sub-bands. Therefore, the appropriate sub-band width must be chosen carefully, to ensure that the redshift evolution remains negligible within each sub-band. The evolution of the 21-cm signal over a fixed bandwidth during the DA is significantly more pronounced than during EoR, as the fluctuations trace the matter power spectrum \citep{furlanetto2006cosmology, smith2025detecting}. The 21-cm power spectrum during the DA is fully specified by the $\Lambda$CDM cosmological model. For a compact array like DEX, achieving high $k$ values can only be reached by accessing higher $k_{\parallel}$ values because the short baselines limit sensitivity to only the smallest $k_{\perp}$, the Fourier modes perpendicular to the LOS. As a result, sensitivity at large $k$ values is dominated by LOS modes, which also provide a cleaner 21-cm window and enable improved calibration by allowing systematics to be distinguished from foregrounds. Based on the above arguments, we choose a spectral resolution of 50\,kHz for this study, which balances sensitivity and computational cost.

We choose 5 MHz spectral windows with 100 channels, following \citealt{mondal2023prospects}, who showed that excluding the light-cone effect in the analysis causes a small increase in the error of the power spectrum for a 5 MHz bandwidth during DA. The two spectral windows considered, Z148 and Z42, are 7 -- 12 MHz ($z$ $\sim$  201 -- 117), and 30 -- 35 MHz ($z$ $\sim$ 45 -- 39), respectively. Although Z148 corresponds to a very early epoch of the Universe and is of great scientific interest, it presents severe observational challenges. These include stronger foreground signals, increased thermal noise, Galactic self-absorption, and the requirement for a large array footprint to achieve the required signal-to-noise ratio (SNR) at such longer wavelengths (or higher redshifts such as z > 50) \citep{koopmans2021peering}. On the other hand, Z42 lies within a range that may be more observationally accessible. 

\subsubsection{Location}\label{sec:location}
At mid-latitudes, row orientation of an array has minimal impact on the resulting $uv$ coverage as the sky drifts at an angle relative to the local horizon. This angular drift ensures both east-west and north-south baselines sample a range of projected baselines over time. In contrast, near the equator, east-west rows enhance sensitivity via rotational synthesis but north-south baselines experience minimal change in projected length. This produces slower fringe rates, which might be prone to systematics. We therefore prefer sites at mid-latitude. This choice is preferable for several other reasons: i) locations near the poles are more prone to terrestrial RFI that can diffract around the lunar limb and reach high-latitude sites \citep{bassett2020characterizing} ii) flat-mounted solar panels are less efficient near the lunar poles due to the low solar elevation which reduces direct solar incidence on horizontal panels and iii) at near-polar latitudes, reduced sky coverage limits the number of independent modes, making power spectrum measurements increasingly cosmic-variance limited. Note that the rotation period of the Moon is about 27.3 days, the apparent motion of the sky and hence the timescale for rotation synthesis is roughly 28 times slower than on Earth.

Site selection must also consider additional factors such as relatively flat terrain, minimal physical hazards like boulders, craters, and ridges as discussed in Section \ref{sec:lunar_topo}. The site selected for this study, Mare Ingenii, is located 33$^\circ$ south of the lunar equator and 163.5$^\circ$ east of the prime meridian (the meridian that points toward Earth in the Moon Mean Earth/Polar Axis, ME frame). The Moon ME frame has the Sub-Earth point, i.e. the location on the lunar surface directly facing Earth, at Longitude 0$^\circ$ and Latitude 0$^\circ$ (see Fig. \ref{fig:mcmf}). We emphasise that the selected site is chosen solely for demonstrative purposes, as this particular mare has been surveyed at high spatial resolution by the LRO in a localized region. A site located in the lunar southern hemisphere is exposed to the bright Galactic Center during a significant portion of the lunar day. For operational deployment, a site in the lunar northern hemisphere would, therefore, be preferable, assuming that a sufficiently flat site can be identified that meets the engineering and scientific requirements of the array.

\section{Antenna position error}
\label{sec:ant_pos_pert}
In practice, even with high-resolution surface mapping of a given deployment site, there is always a risk that the antennas will deviate from their intended positions due to unresolved surface features. Positioning inaccuracies arising from mechanical tolerances and effects such as slippage during deployment by rovers further contribute to the deviation \citep{li2008characterization, gonzalez2018slippage}.
These positional offsets might perturb the projected baselines of the array, thereby changing the geometric phase as a function of baseline and angular direction in the sky. To model the errors, we categorize the offsets into (i) the horizontal or xy direction (primarily arising from inaccuracies along rover trajectory and potential small obstacles); (ii) the vertical or z direction (primarily arising from roughness of local lunar surface). These perturbations are treated independently to isolate their individual impact on the performance of a grid-based array such as DEX.


\noindent
In this section, we derive an analytical expression for the visibility error introduced by antenna position offsets, expressed in terms of baseline geometry and source direction. This derivation, when used in conjunction, supports the trends observed in our simulations in Section \ref{sec:end_to_end}, and forms the basis for understanding how such errors propagate into the cylindrically averaged 21-cm power spectrum.

\subsection{Analytical Treatment}
Considering a 3D coordinate system, with the baseline coordinates given by $\mathbf{b}=(u,\varv,w)$ in wavelength units, and the unit vector pointing towards the source is given by $\mathbf{s}=(l,m,n)$ with $n=\sqrt{1-l^{2}-m^{2}}$, the complex visibility function is given by



\begin{align}
\label{eq:visibility}
V(\nu, \mathbf{b}) &=\iint A(\mathbf{s}, \nu) I(\mathbf{s},\nu)\exp [-2\pi i \varphi] \frac{dl\,dm}{n},\\
\varphi(\nu,\mathbf{b,s}) &= ul + \varv m + w\left(n - 1 \right)
\end{align}

where $I(\mathbf{s},\nu)$ is the sky brightness distribution as a function of direction cosines $(l,m)$ and frequency $\nu$, and $A(\mathbf{s},\nu)$ is the primary beam response of the antenna. This expression assumes that the data have been phased to the phase center $\hat{\mathbf{p}} = (0,0,1)$, such that the phase contribution from $\hat{\mathbf{p}}$ is zero for all baselines. Equation \ref{eq:visibility} forms the basis of the Radio Interferometric Measurement Equation (RIME) \citep{hamaker1996understanding, smirnov2011revisiting}, which provides a general framework for modelling the response of an interferometer to the sky brightness distribution. In this study, visibilities are simulated using a perturbed array, while the analysis assumes a regular array configuration. The resulting inconsistency in the array configuration captures the effect of antenna position offsets, introducing perturbations in the baseline coordinates given by $\mathbf{b'}=(\Delta u,\Delta \varv,\Delta w)$ that may not be accounted for during calibration or imaging. The perturbed visibility function is then given by

\begin{align}
\label{eq:visibility_pert}
V'(\nu, \mathbf{b}) &=\iint A( \mathbf{s}, \nu) I(\mathbf{s}, \nu)\exp [-2\pi i \varphi'] \frac{dl\,dm}{n},\\
\varphi'(\nu,\mathbf{b,s}) &= (u + \Delta u)l + (\varv + \Delta \varv)m + (w + \Delta w)\left(n - 1 \right) = \varphi + \Delta \varphi ,
\end{align}

where $\Delta \varphi(\nu,\mathbf{b,s}) \equiv$ \(\Delta u\)$l$ + \(\Delta \varv \)$m$ + \(\Delta w\)($n$ - 1) represents the geometric phase error due to positional offsets. For observations when the phase center is at zenith and under the assumption of a coplanar array such that a single delay correction can be applied, the $w$ term and its associated geometric phase error reduce to zero for sources at zenith. In this case, the antenna position errors do not project along the line of sight and therefore do not contribute to visibility phase errors.

In our simulation, first, we define the antenna positions in the local East-North-Up (ENU) coordinate system, a topocentric 3D coordinate system centered at the array location. In this frame, the \(E\) axis points east, the \(N\) axis points north, and the \(U\) axis points upward towards the local zenith. Next, we transform this offset to the Moon-Centered, Moon-Fixed (MCMF) (analogous to the International Terrestrial Reference System, ITRS) coordinate system with its origin at the Moon's center. The $Z$ axis is aligned with the mean rotation axis of the Moon towards the geographic North Pole. The \(X\) axis lies in the equatorial plane along the prime meridian (intersection of the lunar equator and the mean direction to the center of the Earth), and the \(Y\) axis points towards 90\textdegree{} east longitude along the equator such that XYZ is a right-handed system. For reference, Fig. \ref{fig:mcmf}  in Appendix \ref{sec:appendix1} illustrates both the coordinate systems.

The final transformation projects the antenna offset into the $u\varv w$ coordinate system, which is defined relative to the phase center of the observation with right ascension \(\alpha_0\) and declination \(\delta_0\). The $u\varv w$ frame is a right-handed orthonormal basis where the \({w}\) axis points toward the phase center, the \({u}\) axis points along decreasing hour angle (i.e., east on the sky), and the \({\varv}\) axis points northward (increasing declination). A schematic of this coordinate system is presented in Fig. \ref{fig:uvw} in Appendix \ref{sec:appendix1}.

The antenna position offset in the local ENU coordinate system is given by 
\begin{equation} 
\Delta \mathbf{r}_{\mathrm{enu}} = 
(\Delta E, \Delta N, \Delta U)^T, 
\end{equation} 
expressed in metres. 

The offset in $u\varv w$ coordinate system is then:
\begin{equation}
\begin{pmatrix} 
{\Delta u} \\
{\Delta \varv} \\
{\Delta w} 
\end{pmatrix}=
\frac{1}{\lambda}[\mathbf{R}_{\mathrm{u\varv w}} \cdot
\mathbf{R}_{\mathrm{MCMF}} \cdot
\Delta \mathbf{r}_{\mathrm{enu}}],
\label{eq:uvw_offset}
\end{equation}

where \(\mathbf{R}_{\mathrm{u\varv w}}\), \(\mathbf{R}_{\mathrm{MCMF}}\) are the rotation matrices mapping the MCMF frame to $u\varv w$ coordinate system, and the local ENU coordinate system to MCMF frame respectively. For readability, we defer the detailed derivation to Appendix \ref{sec:appendix1}, and present only the key equations. Expanding the matrix multiplication in Equation \ref{eq:uvw_offset} gives

\begin{equation}
\begin{pmatrix}
\Delta u \\
\Delta \varv \\
\Delta w
\end{pmatrix}
=
\frac{1}{\lambda}
\mathbf{A}
\begin{pmatrix}
\Delta E \\
\Delta N \\
\Delta U
\end{pmatrix},
\qquad
\mathbf{A} =
\begin{pmatrix}
a_E & a_N & a_U \\
b_E & b_N & b_U \\
c_E & c_N & c_U
\end{pmatrix}.
\label{eq:uvw_transform}
\end{equation}
where
\begin{equation}
\begin{aligned}
[a_E,a_N,a_U] &= \big[ \,
     \cos H_0, - \sin H_0 \sin\phi, \sin H_0 \cos\phi
\big], \\[1ex]
[b_E,b_N,b_U] &= \big[ \,
     \sin\delta_0 \sin H_0, \\
    &\quad
    \left(
         \cos\delta_0 \cos\phi
        + \sin\delta_0 \cos H_0 \sin\phi
    \right),\\
    &\quad 
    \left(
         \cos\delta_0 \sin\phi
       - \sin\delta_0 \cos H_0 \cos\phi     
    \right)
\big], \\[1ex]
[c_E,c_N,c_U] &= \big[ \,- \cos\delta_0 \sin H_0, \\
    &\quad
    \left(
         \sin\delta_0 \cos\phi
        - \cos\delta_0 \cos H_0 \sin\phi      
    \right), \\
    &\quad 
    \left(
       \sin\delta_0 \sin\phi
       + \cos\delta_0 \cos H_0 \cos\phi      
    \right)
\big].
\end{aligned}
\label{eq:uvw_offsets_explicit}
\end{equation}
Here, \(\lambda\) is the observing wavelength in metres, $H_{0}$ is the local hour angle, and $\phi$ is the latitude of the site. We now define a covariance matrix in the $u\varv w$ coordinate system to capture the induced correlated uncertainties after coordinate transformation as

\begin{equation}
\mathbf{\Sigma}_{\Delta} =
\begin{pmatrix}
\sigma_{uu} & \sigma_{uv} & \sigma_{uw} \\
\sigma_{uv} & \sigma_{vv} & \sigma_{vw} \\
\sigma_{uw} & \sigma_{vw} & \sigma_{ww} 
\end{pmatrix}= \mathrm{Cov}(\Delta u, \Delta v, \Delta w),
\label{eq:covariance_matrix}
\end{equation}
where $\sigma_{uu}, \sigma_{vv}$ and $\sigma_{ww}$ are the variances of the baseline errors along the $u,v$ and $w$ axes, respectively and $\sigma_{uv}, \sigma_{uw}$ and $\sigma_{vw}$ are the covariances between the corresponding pairs of axes. The variance in the phase error can then be written in quadratic form as

\begin{equation}
\mathrm{Var}(\Delta \varphi) = \mathrm{x}^\top \mathbf{\Sigma}_{\Delta} \mathrm{x},
\label{eq:variance_phi}
\end{equation}
where $\mathrm{x} \equiv (l,\, m,\, n - 1)^\top$, since the phase error is a linear combination of the baseline coordinate offsets introduced by errors in the antenna position. From Equation \ref{eq:uvw_offsets_explicit}, we note that $\Delta \varphi$ is now $\Delta \varphi$ = $\Delta \varphi(\nu,\mathbf{b,s}; H_{0},\delta_{0}, \phi)$. These equations are equally valid for an interferometer on the Earth. Note that these equations hold for a fixed array site, using longitude \(\Lambda\) = 0 in a local coordinate frame. However, for analysis involving multiple sites or spatially extended arrays that span a significant range in longitude, it is necessary to retain the general expression (see Equations \ref{eq:rot_mcmf} and \ref{eq:rot_uvw}  in Section \ref{sec:appendix1}) to correctly account for the longitudinal dependence. Note that the ENU and MCMF coordinate systems are fixed to the array location and remain identical for both tracking and drift-scan arrays. In contrast, the $u\varv w$ frame follows the moving phase center in a tracking array, while in a drift-scan array it is fixed, causing sources to drift through the FoV. In the special case when zenith is chosen as the phase center, the transformation further simplifies since the ENU and $u\varv w$ axes align as $H_{0}$ = 0 and $\delta_{0}$ = $\phi$. We now examine how the different types of positional offsets, i.e., along the xy direction or z direction appear in the visibilities.

\subsubsection{Case 1: Perturbation along $xy$-directions }\label{sec:ana_xy_perturb}

To evaluate the antenna position deviations on the xy plane, the positional offsets are modeled using a Gaussian distribution with mean \( \mu = 0 \), and standard deviation \( \sigma_{\text{xy}} \), which parameterizes the magnitude of lateral perturbation. The direction of each displacement is drawn uniformly over azimuth, meaning that the lateral offsets have no preferred direction. We vary \( \sigma_{\text{xy}} \) = [0.05, 0.15, 0.25, 0.55] metres. 
Recent lunar rover navigation studies suggest that while challenging, absolute positional accuracies of a few metres are feasible \citep{cortinovis2024assessment, sabatini2025satellite}. In addition, relative trajectory accuracy on the scales of centimetres is achievable within locally mapped areas \citep{ding2024lunar}. We therefore span perturbation levels from optimistic to conservative bounds to cover both best-case and realistically attainable scenarios.

 For perturbations along xy direction, $\Delta E, \Delta N \sim \mathcal{N}(0,\, \sigma_{\text{xy}}^2)$, whereas $\Delta U$ = 0. The covariance matrix for the xy offsets becomes
\begin{equation}
\mathbf{\Sigma}^{xy}_{\Delta} = \frac{2\sigma_{xy}^2}{\lambda^2}A_{\mathrm{xy}} A_{\mathrm{xy}} ^\top,
\qquad
A_{\mathrm{xy}}  = 
\begin{pmatrix}
a_E & a_N\\
b_E & b_N\\
c_E & c_N
\end{pmatrix}
\label{eq:covar_xy}.
\end{equation}

Here, $A_{\mathrm{xy}} A_{\mathrm{xy}} ^\top$ is a Gram matrix and therefore symmetric and positive semi-definite. The factor of 2 appears because we introduce per-antenna horizontal offsets, and the error in a baseline is therefore the difference of the offsets of its two antennas. By the Gram-quadratic-form identity, Equation \ref{eq:variance_phi} can be expressed as 
\begin{equation}
\mathrm{Var}(\Delta \varphi_{\rm xy}) = \mathrm{x}^\top \mathbf{\Sigma}^{xy}_{\Delta} \mathrm{x}= \frac{2\sigma_{xy}^2}{\lambda^2} \mathrm{x}^\top A_{\mathrm{xy}} A_{\mathrm{xy}}^\top \mathrm{x} =\frac{2\sigma_{xy}^2}{\lambda^2} ||A_{\mathrm{xy}} ^\top \mathrm{x}{||}^2
\label{eq:var_xy},
\end{equation}
where $||\cdot||$ denotes the Euclidean norm.





Following the characteristic function of a zero-mean Gaussian random variable, the ensemble-averaged perturbed visibility for a single baseline takes the form

\begin{equation}
\label{eq:ens_avg_vis_xy}
\langle V'(\nu, \mathbf{b}) \rangle =
\iint K(\mathbf{s},\nu) \, 
\exp[-2\pi i \varphi(\nu,\mathbf{b,s})] \,
P_{xy}\,
\frac{dl \, dm}{n}
\end{equation}

with
\begin{align}
K(\mathbf{s},\nu) &\equiv A(\mathbf{s}, \nu) \, I(\mathbf{s}, \nu),\\
P_{xy}(\mathbf{s},\nu, H_{0},\delta_{0}, \phi) &\equiv 
\exp\left[-2\pi^2 \operatorname{Var}\big(\Delta \varphi_{\rm xy}\big)\right] \in(0,1].
\label{eq:pxy_kernel}
\end{align}

Here, $P_{xy}$ is the additional phase coherence factor introduced by ensemble averaging over random antenna position errors along xy direction. The equation shows that this term acts like a direction-dependent (DD), frequency-dependent, multiplicative attenuation kernel in the image domain. Due to the assumption of independent and identically distributed horizontal offsets, the phase factor factor is baseline-independent but time-dependent (via $H_0$), and its amplitude scales with the $\sigma_{xy}$ of the offsets in the lateral position. In addition to these, $P_{xy}$ also depends on the latitude of the array site, and the phase center. Generally, we average all the visibilities that contribute to an $uv$ cell or average all $uv$ cells within annuli of constant $|\mathbf{u}|$$ (=\sqrt{u^{2}+v^{2}})$. Let $\{b_i\}_{i=1}^{N_{\mathrm{samp}}}$ index all measured visibility samples that fall in a given $uv$ cell, $G$ be the gridding kernel, and $w_{b_i}$ be the weights. The gridded, ensemble-averaged visibility, under the small perturbation approximation, can then be written as
\begin{equation}
\label{eq:avg_vis_xy}
\langle V_g'(\nu, \mathbf{b})\rangle =
\frac{\sum_{i=1}^{N_{\mathrm{samp}}} 
 G(u-u_{b_i},\,v-v_{b_i})\,w_{b_i}\, \langle V'_{b_i}(\nu)\rangle}
{\sum_{i=1}^{N_{\mathrm{samp}}}
G(u-u_{b_i},\,v-v_{b_i})\,w_{b_i}},
\end{equation}

\noindent where $\langle V'_{b_i}(\nu)\rangle$ is the ensemble-averaged, perturbed visibility for baseline $b_i$. Now we see two forms of decorrelation arising from antenna position offsets. The first is the random (stochastic) loss of phase coherence by the phase factor $P_{xy}$ for a single baseline. The second is a decorrelation introduced when visibilities from multiple baselines with slightly different geometric phases or attenuation factors are averaged within a given $uv$ cell or across annuli. This averaging leads to an additional smearing in the $uv$  plane whose strength grows with baseline length, frequency, and the lunar rotation synthesis. Therefore, a closed-form analytical expression after gridding or annular averaging is not straightforward.


\begin{figure*}
    \centering
    \includegraphics[width=\textwidth]{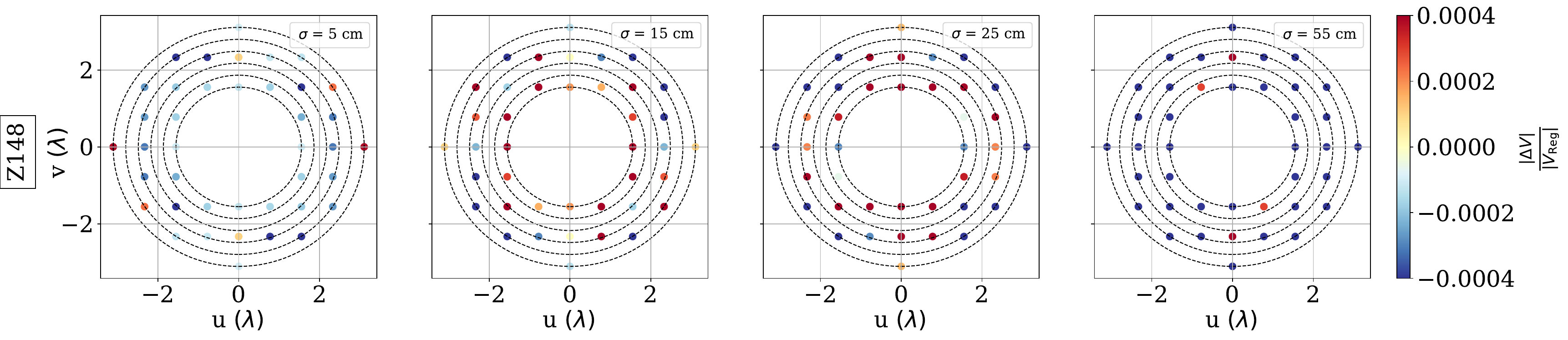}
    \vspace{0.3cm} 
    \includegraphics[width=\textwidth]{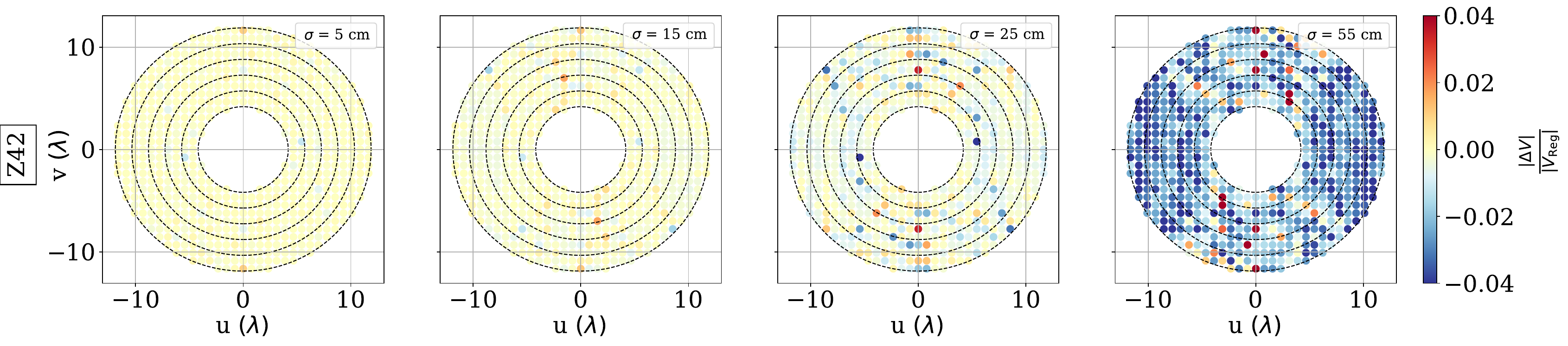}
    \caption{The fractional difference of the frequency-averaged absolute visibilities on the $u\varv$ plane, showing the effect of antenna position perturbations along the xy direction relative to the unperturbed array. Results are shown for two spectral windows (rows) and increasing perturbation (columns). Concentric circles indicating constant baseline length are drawn on the $u\varv$ plane for illustrative purposes only.}
    \label{fig:uv_xy_perturb}
\end{figure*}


Note that even when antenna position errors are along the horizontal plane (\(\Delta U = 0\)), they can induce nonzero \(\Delta w\) components due to the projection of \((\Delta E, \Delta N)\) into the line of sight component. These projected \(\Delta w\) terms contribute to the total phase error through the \(w(n - 1)\) term, which becomes increasingly large for sources away from the phase center \((l^2 + m^2 \gg 0)\). 

From Equations \ref{eq:ens_avg_vis_xy} -- \ref{eq:avg_vis_xy}, we see that in the image domain, the additional phase factor $P_{xy}$ acts as a multiplicative attenuation kernel, whose Fourier transform sets the width of the convolution kernel in $u\varv$ domain. When $P_{xy}$ is constant across the image domain, its Fourier transform is equivalent to a $\delta$-function and the kernel effectively reduces to a constant rescaling of the visibilities, i.e. negligible smoothing/smearing in $u\varv$ domain. To quantify the acceptable level of phase error required for the smoothing of $u\varv$ modes to remain negligible (i.e. equivalent to multiplication by an effectively constant kernel), we derive an analytical estimate in Appendix \ref{sec: phase_error_xy_estimate}.

Each visibility in the $u\varv$ domain is smeared by the same baseline-independent kernel that arises directly from the assumption that horizontal offsets are independent, zero-mean Gaussian perturbations applied to each antenna. Fig. \ref{fig:uv_xy_perturb} shows the fractional difference of the absolute visibilities on the $u\varv$ plane, averaged along frequency for a simulated observation as discussed in Section \ref{sec:end_to_end} for the antenna offsets along xy direction. Although the kernel is the same for all baselines for xy position offsets, we see that the $u\varv$ plane shows a slightly larger fractional difference towards longer baselines. The convolution mixes power between neighboring modes, and its effect is determined by the slope of the intrinsic sky power spectrum. Since the power spectrum of diffuse sky emission decreases rapidly with baseline length, short baselines that probe the large-scale modes are dominated by higher intrinsic power, thus, leading to leaking power into adjacent modes, contaminating slightly longer baselines. At longer baselines, which probe higher spatial frequencies (smaller angular scales) where the intrinsic sky power is lower, the convolution averages over neighboring modes of smaller amplitude, leading to a larger fractional suppression of the visibility amplitude. 



For a fixed array configuration, the projected baseline length in wavelength units scale linearly with frequency, so naturally the extent of $u\varv$ coverage of Z42  is larger than that of the Z148. This also means that Z148 has fewer independent Fourier modes but a high filling factor compared to Z42, which may affect sensitivity to large-scale cosmological modes but improve the robustness of redundant calibration at low frequencies. 
Comparing the two spectral windows, the fractional differences are observed to increase with frequency (or equivalently, shorter wavelengths), in agreement with the analytical prediction. This is because positional offsets lead to larger phase deviations at higher frequencies. For a fixed spectral window, increasing the perturbation amplitude $\sigma_{xy}$, increases the visibility suppression across the $u\varv$ plane. Even in the case of the largest $\sigma_{xy}$ considered, the fractional difference remains well below 5 per cent for Z42, and for the same $\sigma_{xy}$, the variation in fractional difference for Z148 is less than 2 per cent.

The time-dependence of the factor through $H_{0}$ arises from rotational synthesis, which changes the projection of xy positional offsets in the $u\varv w$ coordinates, changing the DD phase errors with LST. Although our simulation is for a fixed LST, this implies that full LST coverage could modulate the effective contamination in wide-FoV observations, motivating time-dependent models for calibration. 


\begin{figure*}
    \centering
    \includegraphics[width=\textwidth]{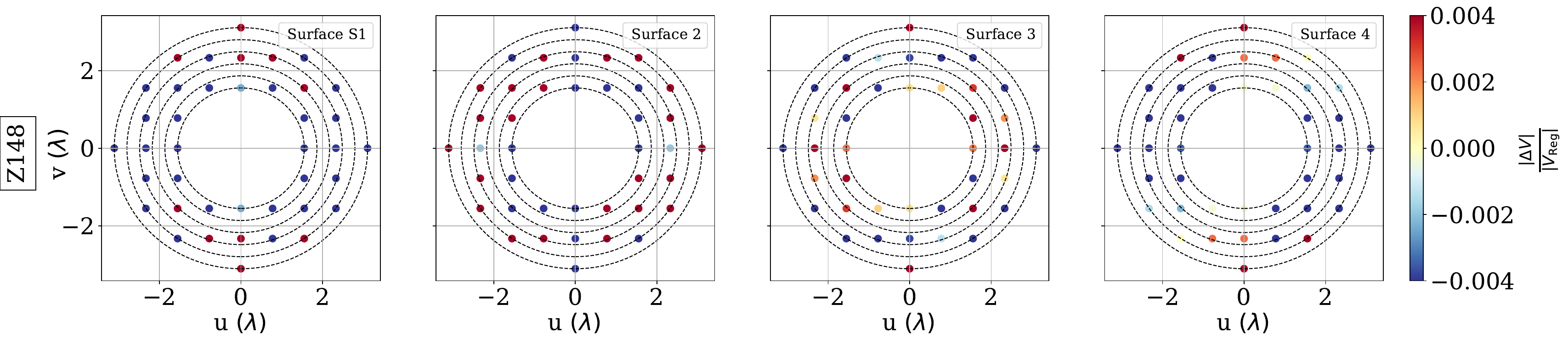}
    \vspace{0.3cm} 
    \includegraphics[width=\textwidth]{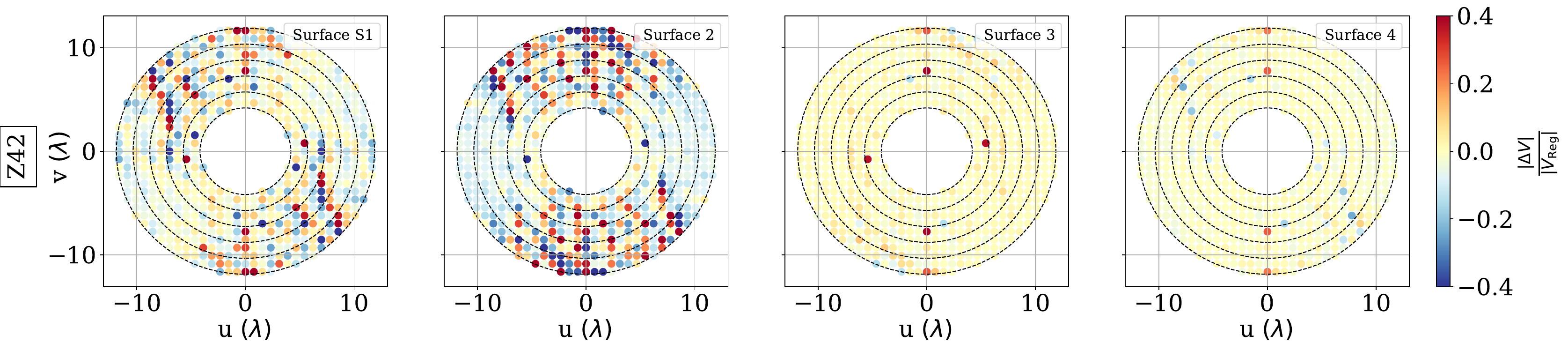}
    \caption{The fractional difference of the frequency-averaged absolute visibilities on the $u\varv$ plane, showing the effect of antenna position offsets for the four different surfaces (columns). Results are shown for two spectral windows (rows). }
    \label{fig:uv_z_perturb}
\end{figure*}

\subsubsection{Case 2: Perturbation along $z$-direction}\label{sec:ana_z_perturb}
We study the effect of array’s non-coplanarity while accounting for a realistic 3D lunar topography. To achieve this, we use the LROC NACs stereo observations yielding DTMs at scales as fine as 2 m, as discussed in Section \ref{sec:lunar_topo}. Unlike horizontal position errors, which are assumed uncorrelated between antennas, the offsets along z direction arise from the underlying lunar surface and are highly correlated between antennas. These height variations are directly sampled from the DTM and reflect the statistical properties of lunar topography, including its local roughness and spatial correlations.

For perturbations along the z direction, the variance of 
$\Delta U$ follows from Equation \ref{eq:hurst_scaling},
\begin{equation}
\mathrm{Var}(\Delta U) = \sigma_s^2 \left( \frac{r}{r_s} \right)^{2H},
\label{eq:var_height_diff}
\end{equation}

while we keep $\Delta E = \Delta N = 0$. Equation \ref{eq:var_height_diff} gives the variance of the height difference between two antennas separated by a distance $r$, so it already accounts for the pair of antennas. Hence, here no additional factor of 2 is introduced. Therefore, for vertical height offsets of the antennas, Equation \ref{eq:variance_phi} becomes

\begin{equation}
\mathrm{Var}(\Delta \varphi_{\rm z}) = \mathrm{x}^\top \mathbf{\Sigma}^{z}_{\Delta} \mathrm{x}= \frac{\sigma_s^2}{\lambda^2} \left( \frac{r}{r_s} \right)^{2H} ||A_{\mathrm{z}} ^\top \mathrm{x}{||}^2,\\
A_{\mathrm{z}}  = 
\begin{pmatrix}
a_U\\
b_U\\
c_U
\end{pmatrix}
\label{eq:var_z}.
\end{equation}

The corresponding ensemble averaged perturbed visibility can be written as





\begin{equation}
\label{eq:ens_avg_vis_z}
\left\langle V'(\nu,\mathbf{b}) \right\rangle = 
\iint K (\mathbf{s},\nu)\, 
\exp \left[ -2\pi i \varphi(\nu,\mathbf{b,s}) \right] \,
P_{z}\, \frac{dl \, dm}{n},
\end{equation}

with
\begin{align}
\label{eq:pz_kernel}
K(\mathbf{s}, \nu) &\equiv A(\mathbf{s}, \nu) \, I(\mathbf{s}, \nu),\\
P_{z}(\mathbf{s},\nu, \mathbf{b}, H_{0},\delta_{0}, \phi) &\equiv \exp\left[ -2\pi^2 \operatorname{Var}\big(\Delta \varphi_{\rm z}\big) \right]\in(0,1].
\end{align}

Equations \ref{eq:ens_avg_vis_xy} and \ref{eq:ens_avg_vis_z} have a similar functional form and dependencies when ensemble-averaged over perturbed realizations for a single baseline. Also, Equation \ref{eq:avg_vis_xy} retains the same functional form for vertical height offsets, with the additional phase factor now being replaced by $P_{z}$. The key difference is that, unlike the uncorrelated lateral offsets, which induce a baseline-independent factor (uniform mode mixing across all baselines), offsets along z direction arising from correlated surface topography introduce baseline dependence through the $r^{2H}$ scaling in Equation \ref{eq:var_height_diff}. This difference follows directly from the statistical model assumed for the different types of positional perturbations. The suppression strength of the phase factor increases with increasing angular distance from the phase center, frequency, and baseline length modulated by the Hurst exponent of the surface, $H$. For the four surfaces considered in this study, Surface 1 shows the largest height fluctuations, while Surface 4 remains the smoothest, with the lowest height difference across all scales (see Table \ref{tab:hurst_exponent}). Our analytical expression therefore predicts that Surface 4 will produce the least visibility decorrelation due to height-induced phase errors, whereas Surface 1 will produce the strongest suppression of small-scale power due to baseline-dependent phase variance.

Fig. \ref{fig:uv_z_perturb} is analogous to Fig. \ref{fig:uv_xy_perturb}, but corresponds to antenna perturbations along the z direction. The fractional difference in visibility amplitudes across the $u\varv$ plane shows a frequency scaling similar to that seen for lateral offsets. The effective width of the smearing due to the convolution in the $u\varv$ domain for height offsets is controlled by the values of $H$ and $\sigma_{s}$. This implies that smooth surfaces will show slowly varying phase decoherence, due to a compact kernel. In contrast, rougher surfaces, characterized by smaller $H$ and/or larger $\sigma_{s}$, introduce rapid phase decoherence, even on short baselines, due to a broader convolution kernel in the $u\varv$ domain. 

The anisotropy observed in the $u\varv$ plane can be directly attributed to anisotropy in the underlying surface height statistics, as evident in the detrended elevation maps in Fig. \ref{fig:surface_analysis}, where the preferred axes of maximum roughness are apparent. This leads to greater phase decoherence for baselines aligned with that axis compared to those aligned perpendicular to it. This can be seen from the different azimuthal structure of the convolution kernel for the different surfaces (columns) in the $u\varv$ map. A secondary contribution to the anisotropy may come from the projection effect, where the projected baseline length varies depending on its orientation relative to the source direction.

The positive residuals (red) in the fractional difference plots correspond to $u\varv$ cells where the perturbed visibility amplitude is higher than that of the unperturbed visibility. For offsets in the xy plane, we saw that the effect can be understood as a baseline-independent convolution in the $u\varv$ domain. This convolution smooths the visibilities by averaging over neighboring $u\varv$ cells. Positive residuals may occur near steep gradients (e.g., the regions where the amplitude of visibility is very high from bright sources or primary beam edges). For perturbations along the z direction, the resulting kernel becomes both baseline-dependent and direction-dependent, driven by fractal nature of lunar topography. This causes anisotropic smearing, with positive residuals amplified along directions where the surface height difference is largest.

Some of the positive residuals are also likely to be realization-specific and strongly driven by sample variance, owing to the limited number of independent baselines in the array. Since visibilities are formed by adding many complex phasors across the sky, small changes in phase from one realization or one LST to another can change the sign of the residual in a given $u\varv$ cell. However, the overall trend due to antenna position offsets remains statistically robust when averaged in annuli.

We note that the impact of surface roughness can be viewed, in a statistical sense, as analogous to a thin ionospheric phase screen. In both cases, if the phase fluctuations are assumed to be zero mean, Gaussian, and stationary, the ensemble-averaged visibilities take the form $\langle V \rangle = \exp[-D_\phi/2] \, V_{\mathrm{true}}$, where $D_{\phi}$ is the phase structure function and $V_{\mathrm{true}}$ is the uncorrupted visibility. 

For a fractal surface with Hurst exponent $H$, the topographic phase structure function is $D_{\phi,\mathrm{surf}} \propto \sigma_s^2 (r/r_s)^{2H} \nu^2$ scales with baseline $r$ and frequency $\nu$, while the ionospheric phase structure function for a Kolmogorov screen in compact arrays  $D_{\phi,\mathrm{iono}} \propto r^{5/3} \nu^{-2}$ \citep{vedantham2016scintillation}. With $2H \approx 5/3$, the effect of phase decoherence will be similar, though the absolute frequency scalings differ (surface roughness $\propto$ $\nu^2$; ionosphere $\propto$ $\nu^{-2}$).

A key distinction is that the phase errors due to surface roughness are anisotropic and quasi-static, set by the local topography and evolving slowly with LST as the baseline projection changes. However, ionospheric phase errors are assumed to be isotropic and time-variable, often on minute scales. This leads to different diffractive scales, $r_{\mathrm{diff, surf}} \propto \nu^{-1/H}$ and $r_{\mathrm{diff, iono}} \propto \nu^{-6/5}$. Therefore, the analogy is helpful for understanding coherence loss but highlights the distinct calibration requirements for lunar arrays (or non-coplanar arrays).

\section{End-to-End Simulation Pipeline}
\label{sec:end_to_end}
\begin{figure*}
    \centering
    \includegraphics[width=0.85\linewidth]{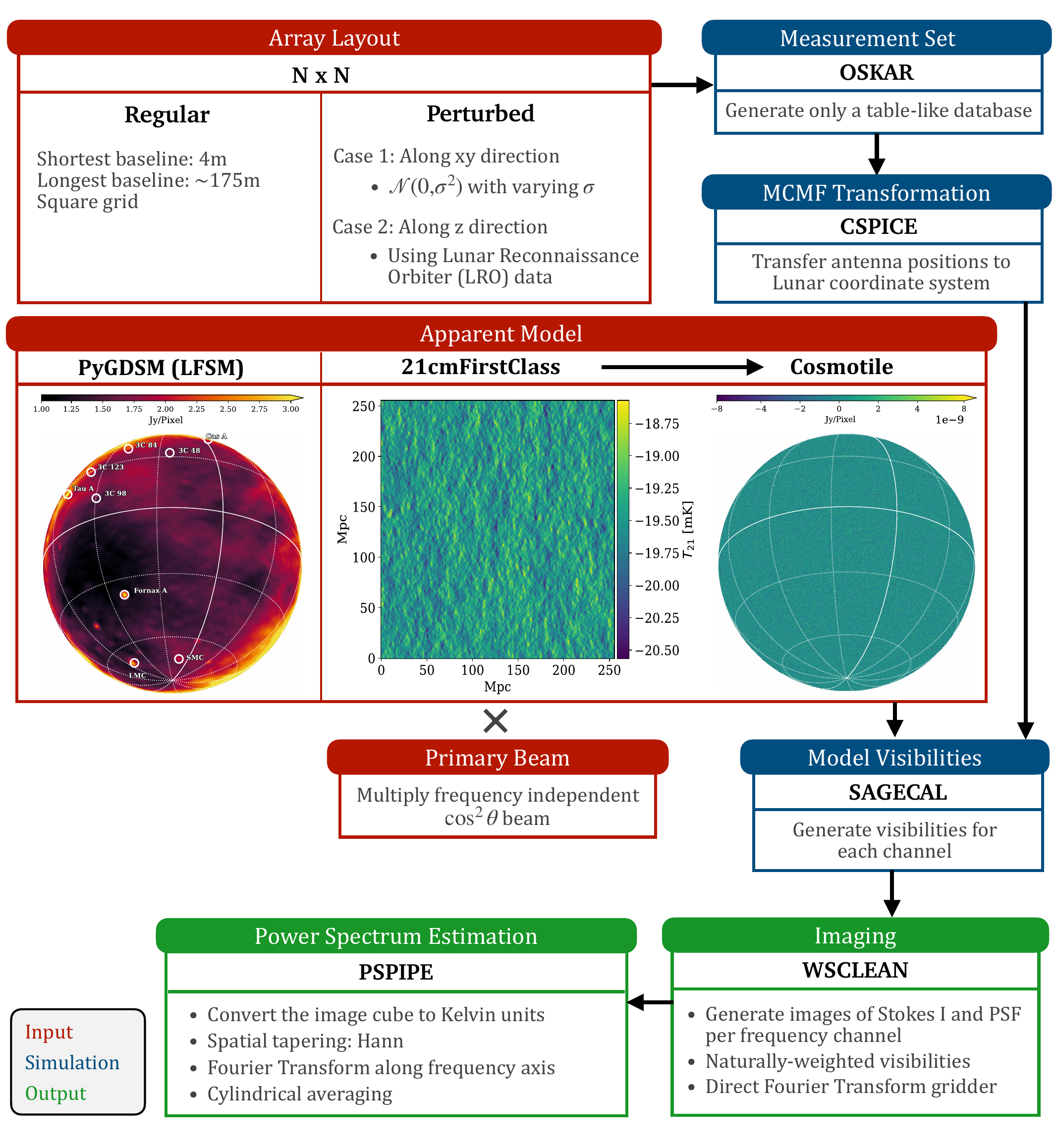}
    \caption{A flowchart of SPADE-21cm.}
    \label{fig:ALO_Forward_Pipeline}
\end{figure*}



To quantify the effect of the uncorrected antenna position deviations in the optimal performance of DEX, we develop a forward simulation pipeline, called SPADE-21cm (Simulation Pipeline for Analyzing Dark agEs using 21-cm). The full-sky simulations will provide insights into first-order tolerance levels of both lateral and vertical antenna position offsets. This pipeline includes the components necessary to generate a realistic scenario that closely resembles actual observations. Fig. \ref{fig:ALO_Forward_Pipeline} presents a schematic of the end-to-end simulation pipeline.

\subsection{Simulation Pipeline}\label{pipeline}


We simulate a zenith-pointed snapshot observation of 5 minutes centered at RA = 23.052060$^\circ$, DEC = -25.696129$^\circ$ from the Moon. The observations, Z148 and Z42, each consist of 100 channels with a resolution of 50 kHz. Only a single snapshot observation is considered in this study for two primary reasons. First, due to the near-unity filling factor of the array, the $u\varv$ coverage does not evolve significantly over the course of a full lunar synthesis. As a result, conclusions drawn from a single snapshot are expected to remain valid to good precision over longer integrations. Second, simulating a full synthesis for a 1024 antenna element array, across 100 frequency channels, incorporating both a full foreground model and a 21-cm signal model, is computationally intensive, and therefore a full synthesis is currently not feasible. For our simulations, we select the observation time to be at 16:22:30 UTC on July 30, 2040, when the center of the Galaxy is located along the horizon, far from the phase-center, but not entirely absent.  
Below we provide a detailed description of each stage of the pipeline.


\begin{figure}
    \centering
    \includegraphics[width=\columnwidth]{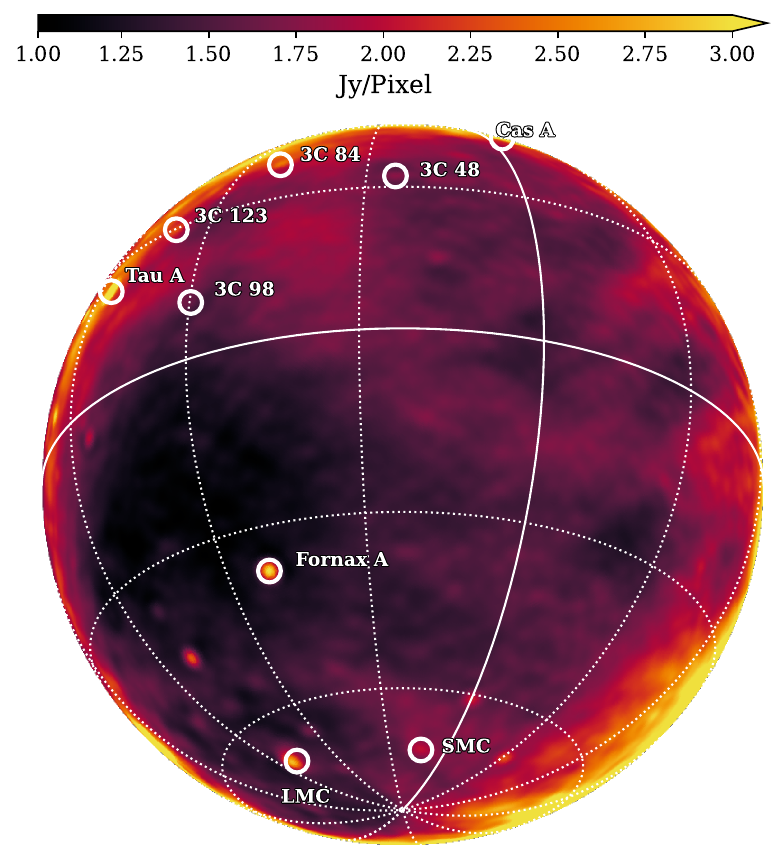}
    

    
    \caption{Orthographic projection of the diffuse sky model from \texttt{LFSM} at a reference frequency of 45 MHz, corresponding to 16:22:30 UTC on July 30, 2040, as seen from Mare Ingenii when the Galactic center lies along the horizon. To obtain the apparent sky brightness distribution, this model is multiplied by a frequency-independent $\cos^2 \theta$ beam.} 
    
    \label{fig:foreground}
\end{figure}

\subsubsection{Foreground model}\label{FG}
We adopt the Low Frequency Sky Model (LFSM) \citep{dowell2017lwa1} from the PyGDSM \citep{2016ascl.soft03013P} package as our diffuse sky model. At the angular resolution of DEX, the diffuse Galactic emission dominates the sky brightness, with point sources implicitly incorporated within the model, albeit at lower angular resolution, where they are blended with the large scale emission and contribute to the overall confusion noise. The LFSM is based on a principle component analysis of data, a similar approach to that of the Global Sky Model (GSM) \citep{de2008model,zheng2017improved}. However, a key distinction is that LFSM incorporates additional observations at 40, 50, 60, 70, and 80 MHz from the LWA1 Low Frequency Sky Survey, extending beyond the datasets used in GSM. This allows for improved interpolation of the sky towards lower frequencies. In this work, we neglect free-free absorption, which can become increasingly important at frequencies below $\sim$7 MHz, where it significantly alters the spectral structure of the diffuse Galactic foreground and similarly suppresses the 21-cm signal. To account for this effect, sky models such as the Ultralong-wavelength Sky Model with Absorption (ULSA) \citep{cong2021ultralong} can be used. In this study, we are primarily interested in isolating the relative effects of perturbations in antenna positions on the power spectrum estimation, therefore, the LFSM skymodel is a sufficient approximation of the true radio sky.

We generate our reference diffuse foreground map at 45 MHz, in the form of a HEALPIX \citep{gorski2005healpix, zonca2019healpy} map as shown in Fig. \ref{fig:foreground}.
This allows us to convert the foreground brightness distribution into pixelized cells with units of flux density in units of Jansky. Insufficient resolution of the HEALPIX map can result in undersampling of the $(l,m)$ space, leading to resampling errors when simulating visibilities, especially for long baselines. To mitigate such effects, it is recommended that the diffuse sky model be sampled well beyond the Nyquist rate (see Section 5.3 in \citealt{kittiwisit2025matvis}). We have used $N_{\rm side}$ = 512 that gives a HEALPIX pixel resolution of $\sim$ 0.11$^\circ$. This is $\sim$ 17 times finer than that the FWHM of the synthesized beam corresponding to the longest baseline of our array, which is $\sim$1.9$^\circ$. The pixels in the HEALPIX map are treated as unresolved point sources. So for $N_{\rm side}$ = 512, the number of pixels that still remain after removing the pixels below the horizon can be as large as $\sim$ 10$^{6}$. This remains computationally demanding, particularly when simulating observations for 523 776 baselines. The final maps derived from the LFSM are corrected for their missing ``zero-spacing'' data using the total power data of LEDA (see Section 2.4 in \citealt{dowell2017lwa1}). We note that in an FFT-based instrument such as DEX, autocorrelations are inherently included, resulting in a sky map with a non-zero mean that also includes contributions from receiver noise and other systematics. In practice, we can subtract the mean intensity, effectively removing the zero-spacing component. This is an important distinction compared to FX-correlated instruments, which in general do not include total power. 

To simulate the sky as observed from the lunar surface, we use lunarsky\footnote{\url{https://github.com/aelanman/lunarsky}}, an extension of astropy that provides selenocentric and topocentric reference frames for the Moon. The sky map is phased to the zenith of the array and multiplied with a frequency-independent $\cos^2 \theta$ beam model to create an apparent (instantaneous) foreground sky model. 


\subsubsection{21-cm signal model}\label{21cm}
To simulate the 21-cm signal from DA, we use 21cmFirstCLASS\footnote{\url{https://github.com/jordanflitter/21cmFirstCLASS}} \citep{flitter202321cmfirstclass,flitter202321cmfirstclass2} which is a merger of two widely used popular codes, 21cmFAST\footnote{\url{https://github.com/21cmfast/21cmFAST/tree/master}} \citep{mesinger201121cmfast, munoz2022impact} and CLASS\footnote{\url{https://github.com/lesgourg/class_public}} \citep{blas2011cosmic}. 21cmFirstCLASS begins its calculation from initial conditions at recombination (via CLASS) and evolves the 21-cm signal. During the DA, the density fluctuations of baryons ($\delta_{b}$) and cold dark matter ($\delta_{c}$) evolved differently \citep{flitter2024does}. In contrast, at lower redshifts, these differences become negligible as sufficient time has elapsed for $\delta_{b}$ and $\delta_{c}$ to evolve similarly, with their initial conditions no longer significantly influencing their dynamics. This assumption underlies in the 21cmFAST making it less suitable for modeling 21-cm signal from the DA.

\noindent
We simulate a box of [500 cMpc]$^3$ on a 256$^3$ grid giving a spatial resolution of $\sim$1.953 Mpc per voxel. 21cmFirstCLASS generates a lightcone, from which the coeval boxes can be extracted. While running, 21cmFirstCLASS uses the closest redshift given as input to generate the co-eval boxes. As a result, the redshifts of the output coeval boxes may not always exactly match those specified in the input. We use a quadratic scheme to perform pixel-wise interpolation of the boxes to the desired redshifts. The boxes are generated for the two spectral windows mentioned in Section \ref{bandwidth} with a step-size of 0.05 MHz. 


Given that DEX is sensitive to spatial modes exceeding the size of a single co-eval box, we generate a larger volume in comoving space by tiling the co-eval boxes using Cosmotile\footnote{\url{https://github.com/steven-murray/cosmotile}} \citep{kittiwisit2018sensitivity}. In addition to the tiling of the coeval boxes, Cosmotile is used to project the resulting tiled volume into angular sky coordinates. Specifically, the make\_healpix\_lightcone\_slice function in Cosmotile takes the tiled 3D Cartesian comoving cubes, selects the appropriate comoving radius corresponding to the desired redshift (i.e., the lightcone shell), and projects the data onto full-sky HEALPIX maps. These angular-frequency 21-cm maps then serve as inputs to the subsequent step for visibility simulation.

However, we note that while Cosmotile enables the construction of larger comoving volumes, it does not produce intensity modes on scales exceeding the size of the original co-eval box. As a result, spatial $k$-modes from the 21-cm signal with $k_\perp \lesssim 2\pi / (500\,\mathrm{cMpc}) \approx 0.01\,\mathrm{cMpc}^{-1}$ are absent from our simulation. For baselines corresponding to $k < 0.01\,\mathrm{cMpc}^{-1}$, only $k_\parallel$ modes are sampled, which are averaged over multiple stacked simulation boxes along the LOS. However, since the minimum accessible $k_\parallel$ in our setup is set by the finite $5$\,MHz bandwidth and exceeds $0.01\,\mathrm{cMpc}^{-1}$, the absence of transverse 21-cm signal modes on scales $k < 0.01\,\mathrm{cMpc}^{-1}$ does not impact our results.

To mitigate the edge effect at the boundaries that might arise when periodic boxes with different absolute mean levels are stitched together, the mean brightness-temperature value of each co-eval box is subtracted prior to tiling. This step effectively removes the k = 0 mode which basically mimics the absence of the zero-spacing (autocorrelation) measurement in real interferometric observations. A FFT-based correlator will retain the zero-spacing mode but including it in our simulations would introduce sharp discontinuities at the boundaries, leading to spurious artefacts in the power spectrum. Finally, the 21-cm HEALPIX maps are multiplied with the same beam model to ensure consistency with the apparent foreground sky model.

\subsubsection{Measurement Set generation}
Measurement sets (MS) are created using OSKAR\footnote{\url{https://github.com/OxfordSKA/OSKAR}} \citep{dulwich2020oskar}, a GPU-accelerated visibility simulator specifically designed for the SKA. For the purpose of this paper, we use OSKAR only to generate the template that contains the properties of the telescope array, such as antenna positions, $u\varv$-coverage, pointing directions, and frequency channels. However, we do not use the visibilities simulated by OSKAR (see Section \ref{sub: Visibility}).

Once the MSs are generated for each of the frequency channels, we update the antenna positions in the MS with MCMF coordinates using uvutils.ECEF\_from\_ENU(frame=`MCMF'). The MCMF coordinate system is often defined using two closely related coordinate systems, the Mean Earth (ME)/Polar Frame coordinate system, and the other is the Principal Axis (PA) coordinate system \citep{LRO2008WhitePaper}. In this work, we choose the ME frame, with the origin located at the barycenter of the Moon.  (see Fig. \ref{fig:mcmf}). 

Finally, the $u\varv w$ positions of the array are also updated with respect to the MCMF frame. This is done by using a new functionality in SAGECal-CO\footnote{\url{https://github.com/nlesc-dirac/sagecal}}\citep{yatawatta2015distributed}, the standard visibility prediction and calibration software of LOFAR-EoR data processing pipeline, that incorporates CSPICE\footnote{\url{https://naif.jpl.nasa.gov/naif/toolkit.html}}. CSPICE is the C component of the SPICE Toolkit, a library developed by the Navigation and Ancillary Information Facility of the Jet Propulsion Laboratory (JPL) to provide access to planetary and spacecraft ephemerides and other functions for engineering computations. With this extension, the simulated observations can use the precise information on the motion of the DEX site on the lunar surface at any epoch of time.

\begin{table}
\caption{Simulated Properties of DEX\label{tab:simulated_parameters}}
\begin{center}
\begin{tabular}{l l}
\hline
\textbf{Parameter}&\textbf{Value}\\
\hline

Phase Center&RA = 23.052060$^\circ$, DEC = -25.696129$^\circ$\\
Bandwidth [in consideration]&7 - 12 MHz [Z148] and 30 - 35 MHz [Z42]\\
Total Frequency Range&7 - 50 MHz\\
Frequency Resolution&50 kHz\\
Number of channels&100\\
Total observation time&5 minutes [snapshot]\\
Number of elements&1024\\
Antenna Configuration&32$\times 32$ regular square grid\\
Antenna Length&3 m [tip-to-tip]\\
Longest Baseline&175 m\\
Shortest Baseline&4 m\\
Location&Mare Ingenii [33$^\circ$ S, 163.5$^\circ$ E]\\
\hline
\end{tabular}
\end{center}
\end{table}

\subsubsection{Visibility prediction} \label{sub: Visibility}
The visibilities are predicted using SAGECal-CO following the RIME framework. The model visibility prediction can be computationally demanding for 1024 antenna elements. Upon comparison, a difference was observed in the wall time required for visibility prediction, with SAGECal-CO demonstrating quicker prediction in our case compared to OSKAR. However, we note that either of these can be utilized to predict model visibilities based on the availability of computing resources. With the integration of CSPICE into SAGECal-CO, the software gains the capability to forward predict visibilities in the lunar frame, enabling more accurate simulations for lunar-based interferometry. We simulate the following two data sets:
\begin{itemize}
\item \textit{Regular visibilities ($V_{\rm reg}$)}: The model visibilities of the sky and 21-cm signal for the `ideal case' when the antenna positions are perfectly regular, hence unperturbed.

   
\item \textit{Perturbed visibilities ($V_{\rm pert}$)}: This set of visibilities is generated by introducing antenna offsets along the xy direction or the z direction independently. 
\end{itemize}

\begin{figure}
    \centering
    \includegraphics[width=\columnwidth]{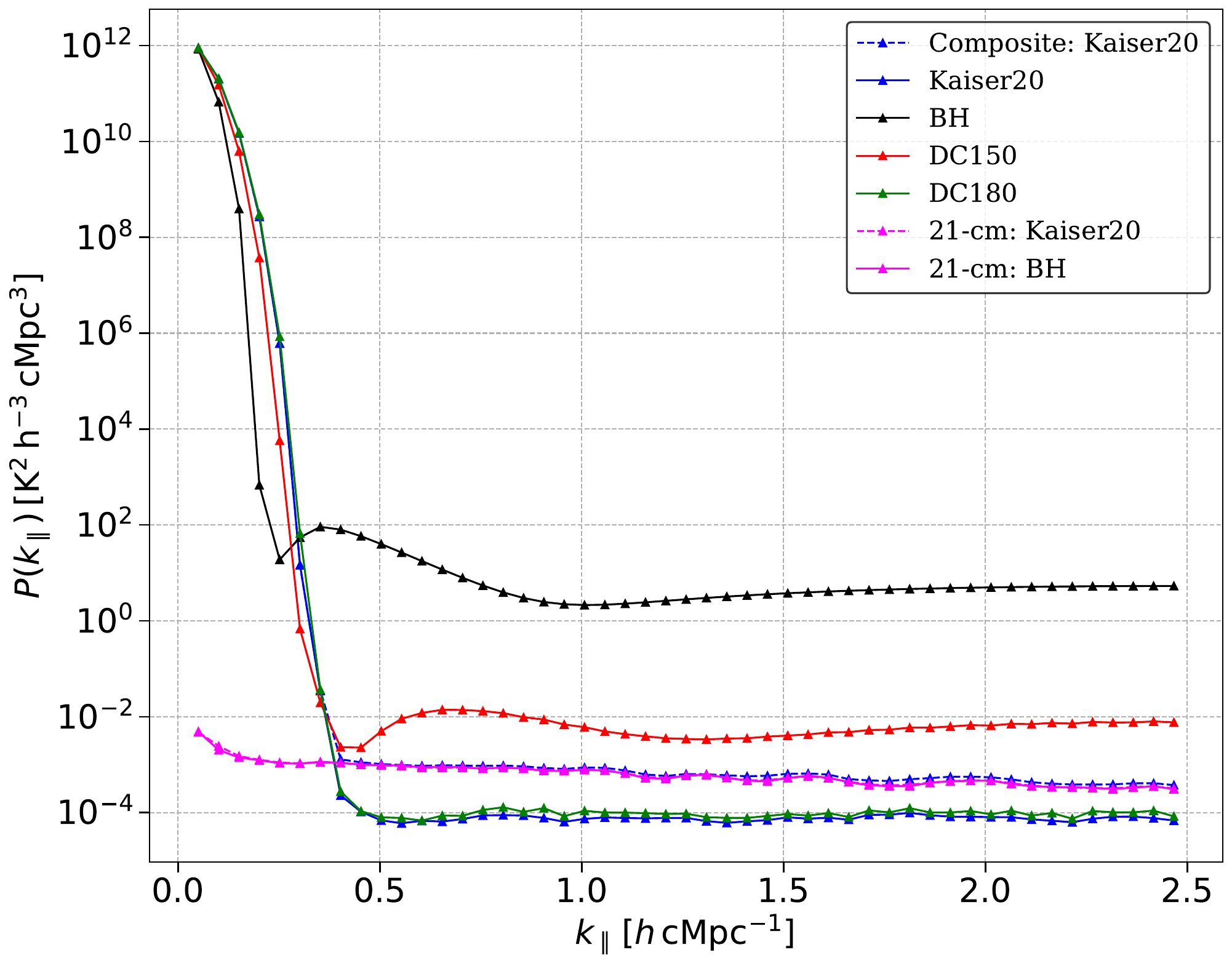}
    \caption{The cylindrically-averaged power spectra as a function of $k_{\parallel}$ (averaged over $k_{\perp}$) are shown for different spectral window functions: Blackman-Harris (BH), Dolph-Chebyshev with attenuation levels set to 150 dB (DC150) and 180 dB (DC180), and Kaiser with $\beta=20$ (Kaiser20). The coloured curves correspond to the foregrounds only simulations. For comparison, the magenta curves represent the 21-cm signal only at z $\sim$ 42.5. The dashed blue line shows the composite sky when applying the Kaiser20 window. Results are shown here for Z42.}
    \label{fig:compare_window}
\end{figure}

\subsubsection{Imaging and Power Spectrum estimation}
The visibilities are imaged with WSCLEAN\footnote{\url{https://wsclean.readthedocs.io/en/latest/}} \citep{offringa2014wsclean}. A `natural' weighting scheme is used to generate full-sky `dirty' Stokes I images (in Jy PSF$^{-1}$, where PSF denotes the point spread function), and PSF images for each of the frequency channels. No deconvolution is performed on the images. 
We used the direct-ft algorithm in WSCLEAN which evaluates the visibilities at their native baseline coordinate $(u,v,w)$. Consequently, this helps to avoid gridding related artefacts and provides more accurate imaging, albeit at a significantly higher computational cost. In addition, the direct-ft can account for the $w$-term naturally \citep{perley1999imaging}. This is important for wide-field or non-coplanar array imaging. Once the `dirty' image per channel is generated, we apply a multiplicative factor of $\sqrt{1 - l^2 - m^2}$ to each pixel to obtain accurate temperature units across the FoV. \footnote{This factor enters the conversion from Jy PSF$^{-1}$ to Kelvin, as it accounts for the projection from direction cosines to solid angle. The PSF in non-deconvolved images is direction-dependent, and this correction ensures to properly account for the spatial variation of the PSF when converting to brightness temperature.}

Prior to power-spectrum estimation, these image cubes for each channel are converted to units of Kelvin. We subsequently construct the cylindrically-averaged power spectrum using pspipe\footnote{\url{https://gitlab.com/flomertens/pspipe}} following the standard definitions of cosmological power spectra in \citealt{morales2004toward, mcquinn2006cosmological}. Following the notation as in \citealt{mertens2020improved}, the 21-cm brightness temperature field, $T(\mathbf{r})$, is defined over spatial coordinates $\mathbf{r}$. The corresponding power spectrum $P(\mathbf{k})$, as a function of the wavevector $\mathbf{k}$, is given by:
\begin{equation}
P(\mathbf{k}) = V_c |\tilde{T}(\mathbf{k})|^2\,,
\end{equation}
where, $\tilde{T}(\mathbf{k})$ is the discrete Fourier transform of the temperature field, computed as:
\begin{equation}
\tilde{T}(\mathbf{k}) = \frac{1}{N_l N_m N_\nu} \sum_{\mathbf{r}} T(\mathbf{r}) \, e^{-2i \pi kr}\,.
\end{equation}

Here, $N_{\rm{l}}$, $N_{\rm{m}}$, and $N_{\rm{\nu}}$ are number of pixels in $l$, $m$, and $\nu$ direction. The observed comoving volume $V_{\rm {c}}$ depends on the primary beam of the instrument $A_{\rm{pb}} (l, m)$, the spatial window function $A_{\rm {w}}(l, m)$ and the spectral window function $B_{\rm {w}}(\nu)$ applied to the image cube before the Fourier Transform. It is estimated as follows:
\begin{equation}
V_c = \frac{(N_l N_m N_\nu \, dl \, dm \, d\nu) D_M(z)^2 \Delta D}{A_{\text{eff}}  B_{\text{eff}}}\,,
\end{equation}
where $dl$ and $dm$ are angular pixel sizes, $d\nu$ is the channel width, $D_M(z)$ is the transverse comoving distance at redshift $z$ and $\Delta D$ represents the comoving distance corresponding to the observed frequency range. \( A_{\text{eff}} \) and \( B_{\text{eff}} \) are the effective area on the sky and effective bandwidth, respectively defined as:
\begin{align}
A_{\text{eff}} &= \langle A_{\text{pb}}(l,m)^2 A_w(l,m)^2 \rangle,
\label{eq:Aeff_window}\\
B_{\text{eff}} &= \langle B_w(\nu)^2 \rangle\,.
\label{eq:Beff_window}
\end{align}

In Equation \ref{eq:Aeff_window}, $\langle \rangle$ represent an average over spatial coordinates (the image plane), with weighting from the primary beam and spatial tapering window. On the other hand, in Equation \ref{eq:Beff_window} they represent an average over frequency channels, with weighting from the spectral tapering window. A Hann-filter with a width of 80$^\circ$ is applied as a spatial taper and the choice of spectral window function is discussed in Appendix \ref{sec: window}. 


Assuming the cosmological 21-cm signal is isotropic, the power spectrum $P({\mathbf{k}})$ can be cylindrically averaged, which preserves the separation between transverse and LOS components in Fourier space. It is defined as:
\begin{equation}
P(k_\perp, k_\parallel) = \left\langle P(\mathbf{k}) \right\rangle_{k_\perp, k_\parallel},
\end{equation}

\noindent
and is widely used as a diagnostic tool.

\section{Results}
\label{sec:results}

\begin{figure*}
    \centering
    \includegraphics[width=\textwidth]{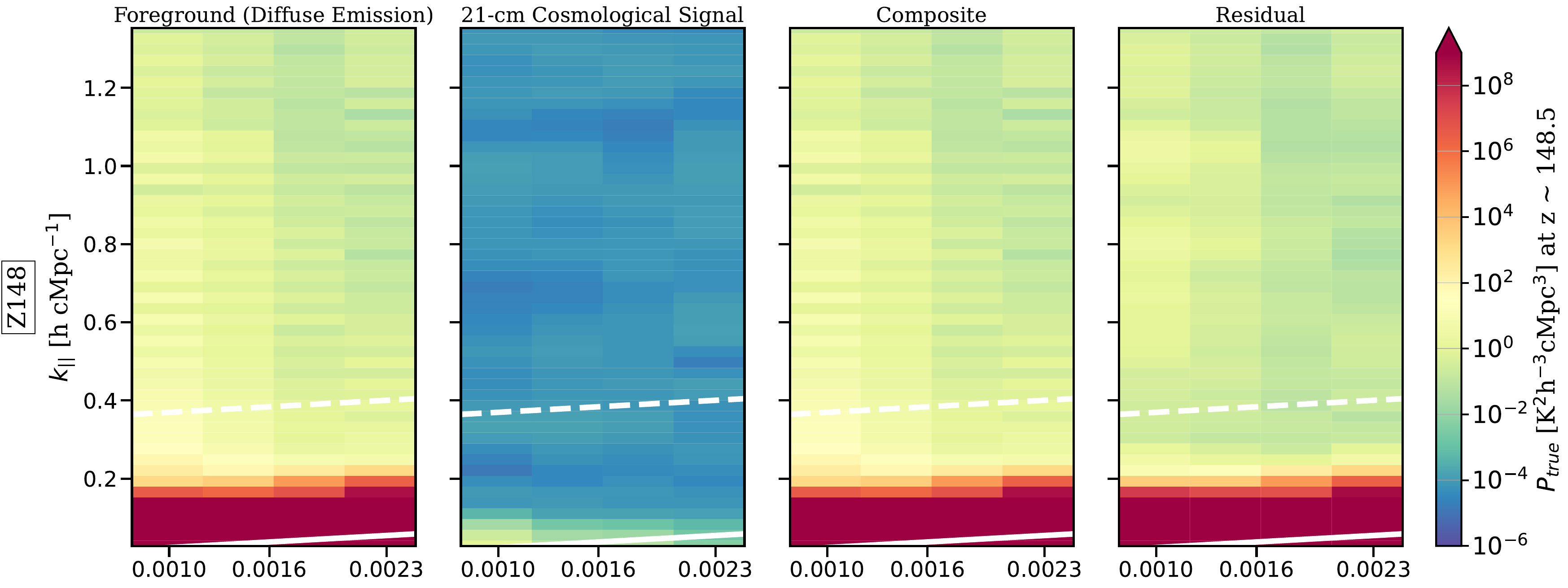}
    \vspace{0.3cm}
    \includegraphics[width=\textwidth]{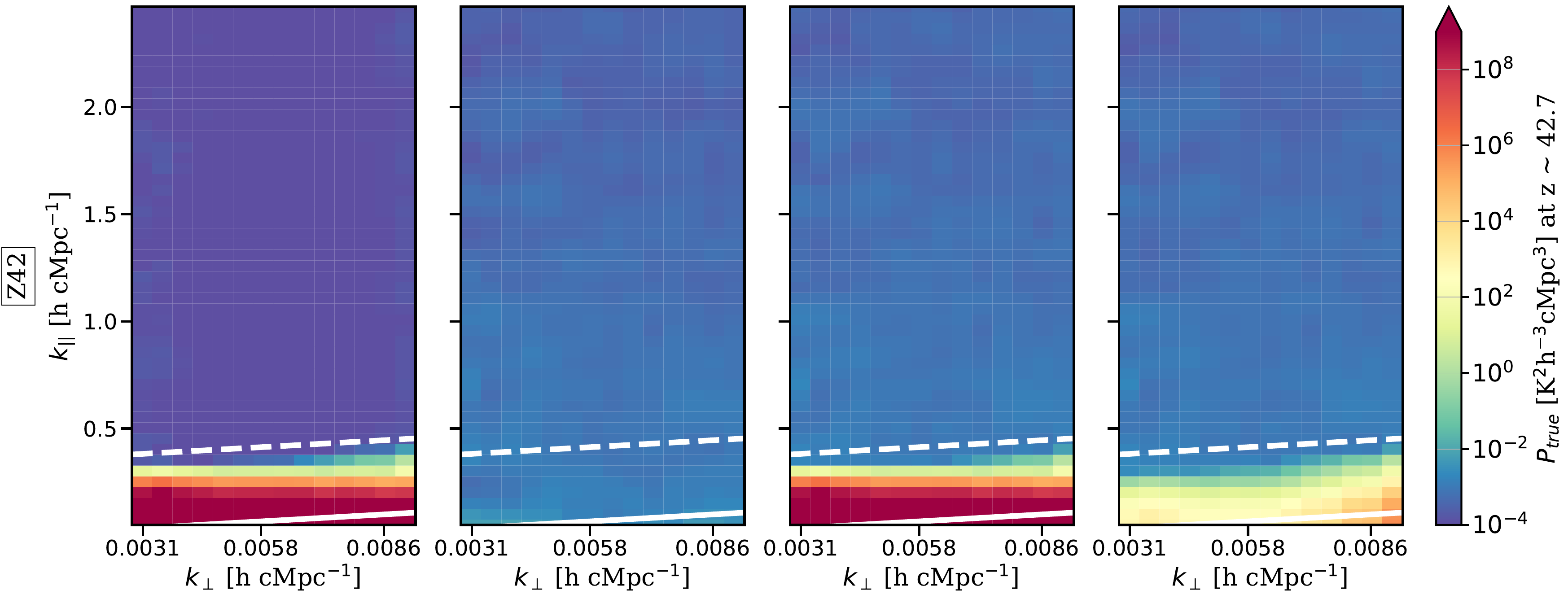}
    \caption{(From Left to Right Panel): The 2D cylindrically averaged power spectra of simulated data for DEX from a diffuse Galactic emission model, 21-cm cosmological signal, a composite model prior to polynomial subtraction, and the residuals after applying a third-order polynomial fit, as described in Section \ref{sec:end_to_end} for Z148 (Top row), and Z42 (Bottom row). 
    The solid white line represents the horizon limit, and the white dashed line represents the horizon buffer limit (for Kaiser $\beta$ = 20) equals to 346 nano seconds. The lines shown here are computed based on the improved delay formalism by \citet{munshi2025beyond}, which accounts for sky curvature. These power spectrums represent the case when all the antenna elements are perfectly on a regular grid.}
    \label{fig:2D_PS_reg}
\end{figure*}


\begin{figure*}
    \centering
    \includegraphics[width=0.99\textwidth]{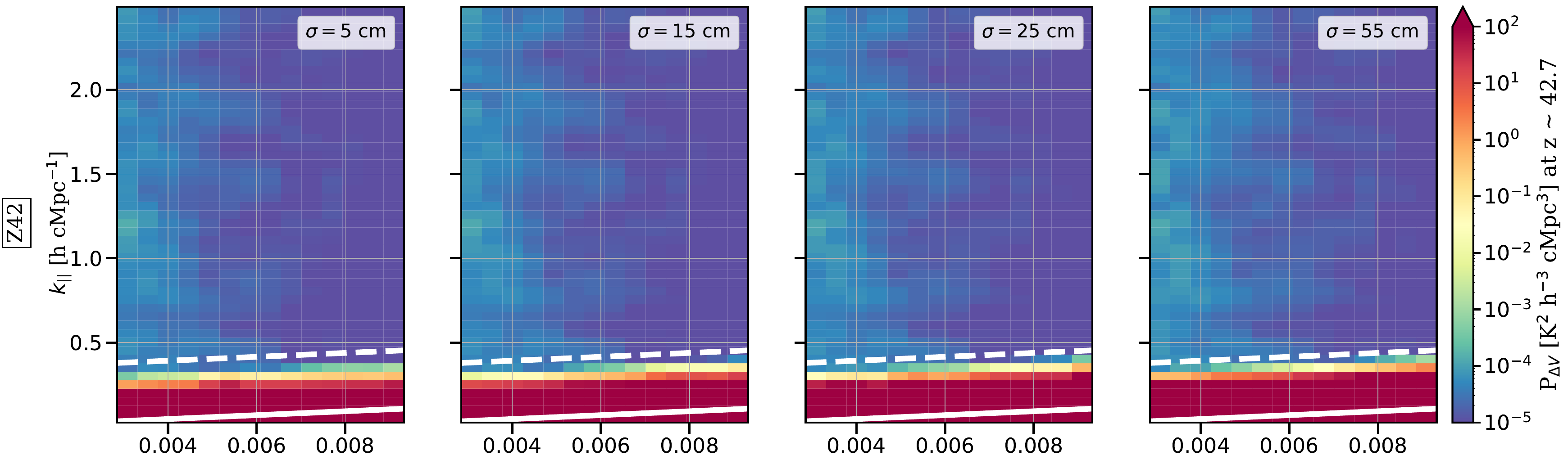}
    \vspace{0.3cm} 
    \includegraphics[width=0.99\textwidth]{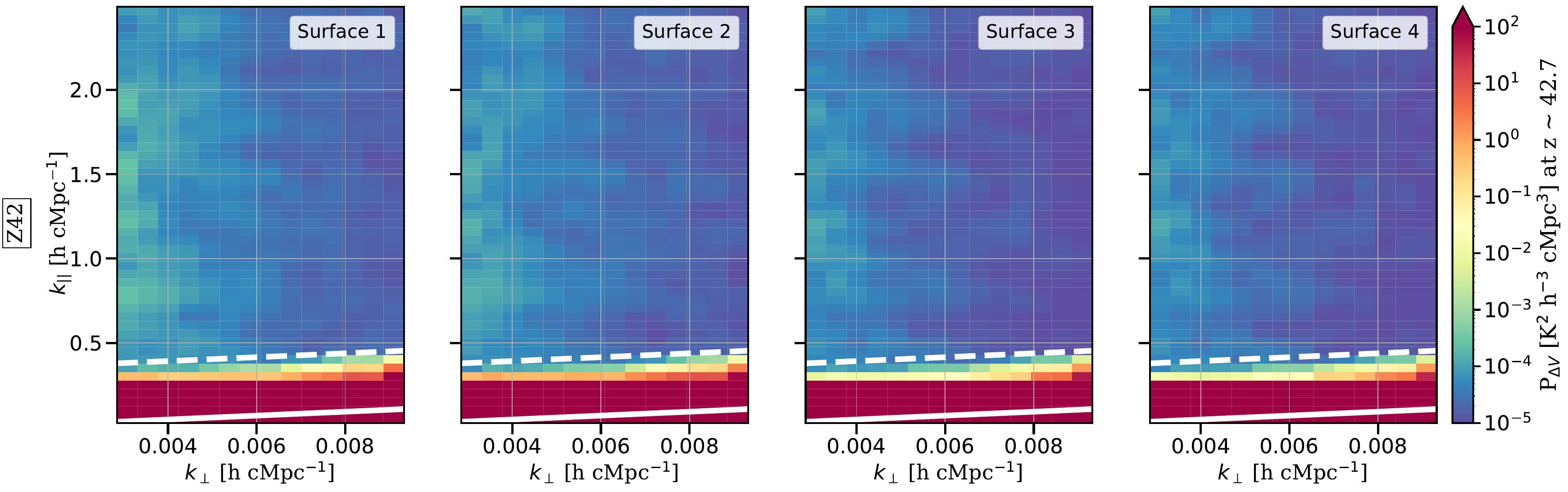}
    \caption{The cylindrically averaged power spectra of the difference in visibilities between the composite model with unperturbed antenna elements and those with perturbations on the xy plane (Top row) and those induced due to vertical height differences (Bottom row). The offsets along xy directions are drawn from a two-dimensional normal distribution $\mathcal{N}(0, \sigma^2)$ with a fixed random seed of varying standard deviation (increasing towards right). The different columns in the bottom row represents the four surfaces, each characterized by a different intrinsic surface height variance. Results are shown for Z42. The solid white line represents the horizon limit, and the white dashed line represents the horizon buffer limit.}
    \label{fig:2D_PS_xy_perturb_Z42}
\end{figure*}

In this section, we analyze the effects of antenna position offsets on the simulated data described in Section \ref{sec:end_to_end}. We first present the fiducial cylindrically averaged power spectrum for the different signal components in the case of an unperturbed array in Section \ref{sec: power_spectum_reg}. We then look into how errors arising from antenna position offsets appear in the cylindrically averaged power spectrum for offsets along the xy direction in Section \ref{sec:xy_offset}, and along the z direction in Section \ref{sec:z_offset} independently. Finally, in Section \ref{sec:detectability}, we quantify the impact of the positional offsets on the 21-cm DA power spectrum.

\subsection{Power Spectrum Characteristics (Unperturbed case)}\label{sec: power_spectum_reg}
The cylindrically averaged power spectrum (or the 2D power spectrum) in \((k_{\perp}, k_{\parallel})\) space is one of the most widely used statistical metrics for analyzing the impact of foreground contamination and systematic biases. 
While the 21-cm signal is expected to be statistically isotropic to first order, with power distributed in nearly all $k$-modes, the foregrounds are highly anisotropic and, owing to their spectral smoothness, are largely confined to low $k_{\parallel}$ modes. The inherent chromaticity of interferometers results in a leakage of foregrounds to higher $k_{\parallel}$ with increasing $k_{\perp}$, also called `mode-mixing' \citep{morales2012four}, leading to a region often referred to as the `foreground wedge'. The spectrally-smooth foregrounds are typically confined within this region. The largest possible extent of the wedge region is determined by the maximum geometric delay between two antennas in a baseline, which is achieved when the source is positioned at the horizon, for a phase center at zenith. 




Generally, in any 2D power spectrum for 21-cm experiments, a window function is applied along the LOS axis to taper the additional spectral components originating as a result of finite bandwidth. This reduces the effective bandwidth, broadens the main lobe of the window function in delay space, and increases spillover beyond the horizon. For DA experiments, the dynamic range required to suppress foreground contamination in the 21-cm signal window is at least 11 to 15 orders of magnitude in power, implying a trade-off between the level of foreground sidelobe level (SLL) suppression and the extent of low $k_{\parallel}$ modes lost due to the leakage. We find that applying a Kaiser window function (see Appendix \ref{sec: window}) with $\beta$ = 20 (hereafter denoted as Kaiser20) achieves $\sim$ 164 dB SLL suppression, compared to the commonly used 4-term Blackman-Harris which gives a SLL suppression of $\sim$ 92 dB when comparing foreground-only power outside the wedge with the peak foreground power inside the wedge. In the Kaiser window function, $\beta$ is a shape parameter that controls the trade-off between the width of the main lobe and the suppression of the SLL. Although this improvement comes at the expense of $k_{\parallel}$ resolution, any leakage into LOS modes is now governed mostly by the inherent frequency dependence of the foregrounds, rather than by windowing artefacts. A comparison of the cylindrically averaged power spectra (averaged over all baselines) with different window functions is illustrated in Fig. \ref{fig:compare_window}. Note that although the foreground-only cylindrically-averaged power spectrum (solid blue) in Fig. \ref{fig:compare_window} may lie below the 21-cm signal only power spectrum (dashed magenta) after applying the Kaiser20 window function, the composite power spectrum (dashed blue) cannot fall below the 21-cm signal. This is because it includes the intrinsic power of the 21-cm signal, and therefore sets a nonzero lower bound (or signal floor) on the composite power spectrum.

For this study, we first apply a third-order polynomial fit to the gridded visibility data cube to remove most of the spectrally-smooth foreground power, followed by Kaiser20 to approximate the level of spectral suppression typically implemented in real data analysis pipelines. Prior to power spectrum estimation, we remove the longer baselines with poor $u\varv$ coverage to avoid sampling artefacts. We show the 2D cylindrically-averaged power spectra of the foregrounds, 21-cm signal, a composite model prior to polynomial subtraction, and the residuals after applying a third-order polynomial fit in the ideal scenario for both Z148 and Z42 (rows), in Fig. \ref{fig:2D_PS_reg}. Since the primary beam considered in this work is frequency-independent, the observed wedge in the foreground emission panels arises due to the interaction of the frequency-dependent PSF with the otherwise spectrally smooth foreground emission. As expected, the amplitude of the foreground power decreases with increasing frequency due to the steep synchrotron spectral index \(T_b \propto \nu^{\beta}, \, \beta \sim -2.5\).

Spectral tapering can minimize leakage from finite-bandwidth effects, but the extent of recoverable 21-cm window is also strongly redshift-dependent. At Z148, the foreground is almost an order of magnitude higher than at Z42, while the strength of the 21-cm signal is intrinsically weaker in most theoretical models at these redshifts. Also, since the slope of the wedge is redshift-dependent, at lower frequencies (or higher redshifts), it becomes steeper, occupying a large portion of the 2D $k$-space, thereby narrowing the 21-cm DA window \citep{pober2025impact}. Therefore, the simple foreground treatment adopted here is insufficient to suppress the bright, spectrally smooth foregrounds below the expected 21-cm signal level for Z148 inside the 21-cm window. For this reason, in the remainder of this paper, we focus our analysis on Z42, where a sufficiently clean 21-cm window is achievable. To isolate the impact of position offsets on the 21-cm power spectrum, the same spectral and spatial window functions, along with polynomial fitting, are applied to both unperturbed and perturbed power spectra. As the positional perturbations considered here primarily introduce smooth spectral features (see Section \ref{sec:result_offset}), we expect that foreground mitigation techniques using Gaussian Process Regression (GPR; \citealt{mertens2018statistical, mertens2024retrieving}) would perform well. The development and testing of this technique to determine whether it enables less foreground leakage in the 21-cm window for Z148 is left for future work.

\subsection{Effect of antenna position errors}\label{sec:result_offset}
In this section, we demonstrate the impact of deviation in the antenna position in both the lateral direction and the vertical direction independently on the cylindrically-averaged power spectrum. We take a difference of the \textit{Regular visibilities} and \textit{Perturbed visibilities} where $\Delta V \equiv V_{\rm pert} - V_{\rm reg}$, and create power spectra of this difference denoted by $P_{\Delta V}(k)$.





\begin{figure*}
    \centering

    \begin{minipage}{0.49\textwidth}
        \centering
        \includegraphics[width=0.88\linewidth]{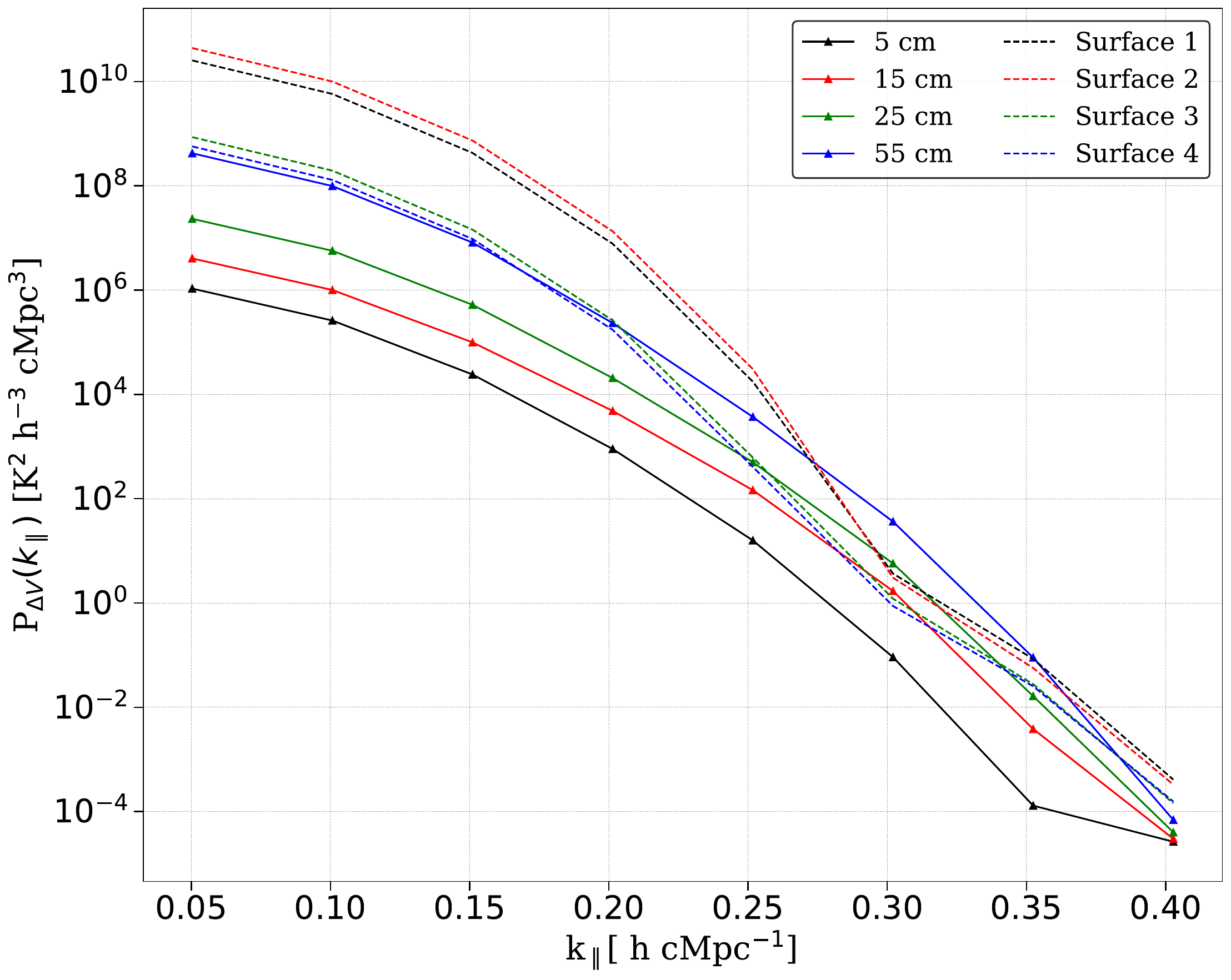}
    \end{minipage}\hfill
    \begin{minipage}{0.49\textwidth}
        \centering
        \includegraphics[width=0.88\linewidth]{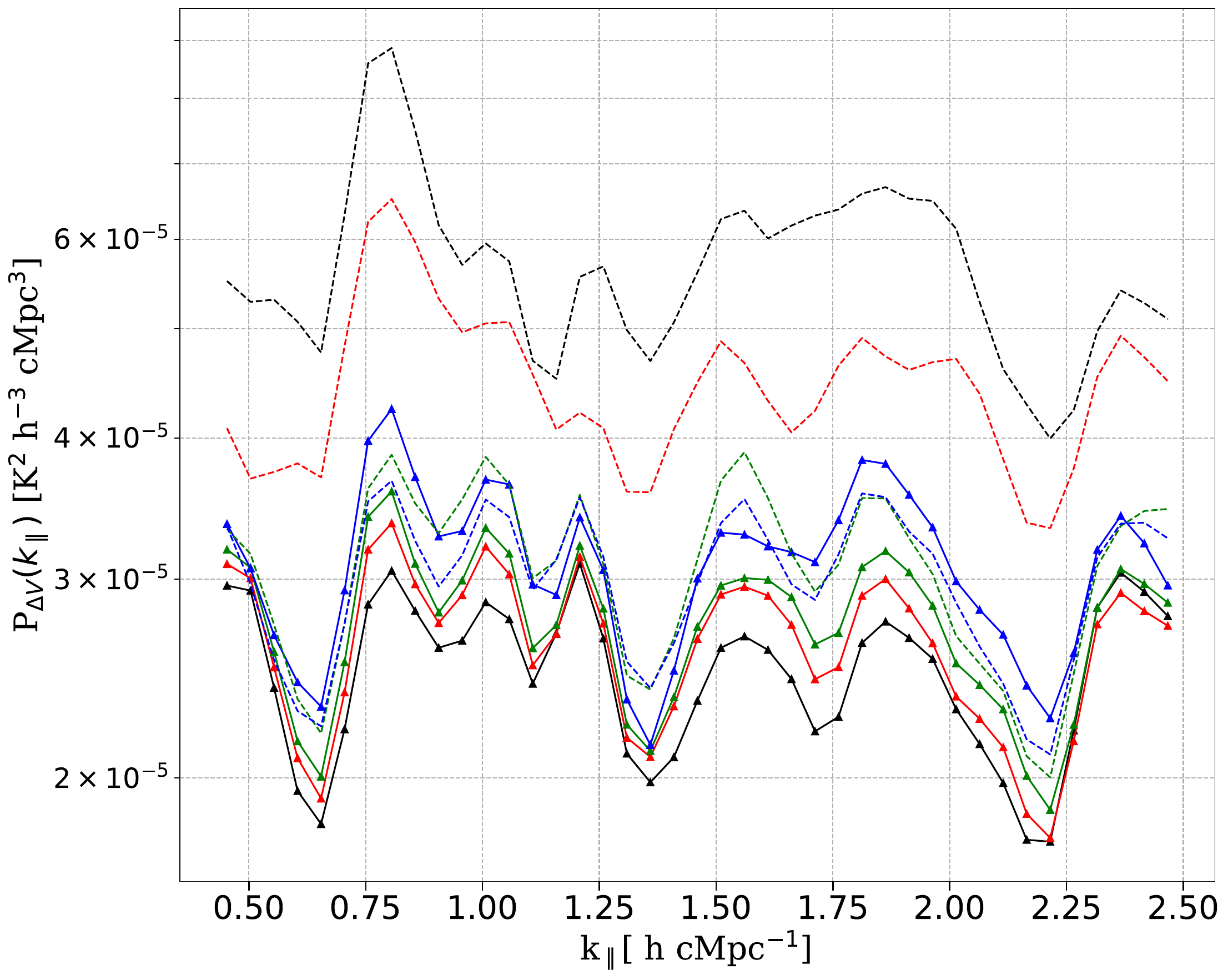}
    \end{minipage}

    \vspace{0.4cm} 

    \begin{minipage}{0.49\textwidth}
        \centering
        \includegraphics[width=0.88\linewidth]{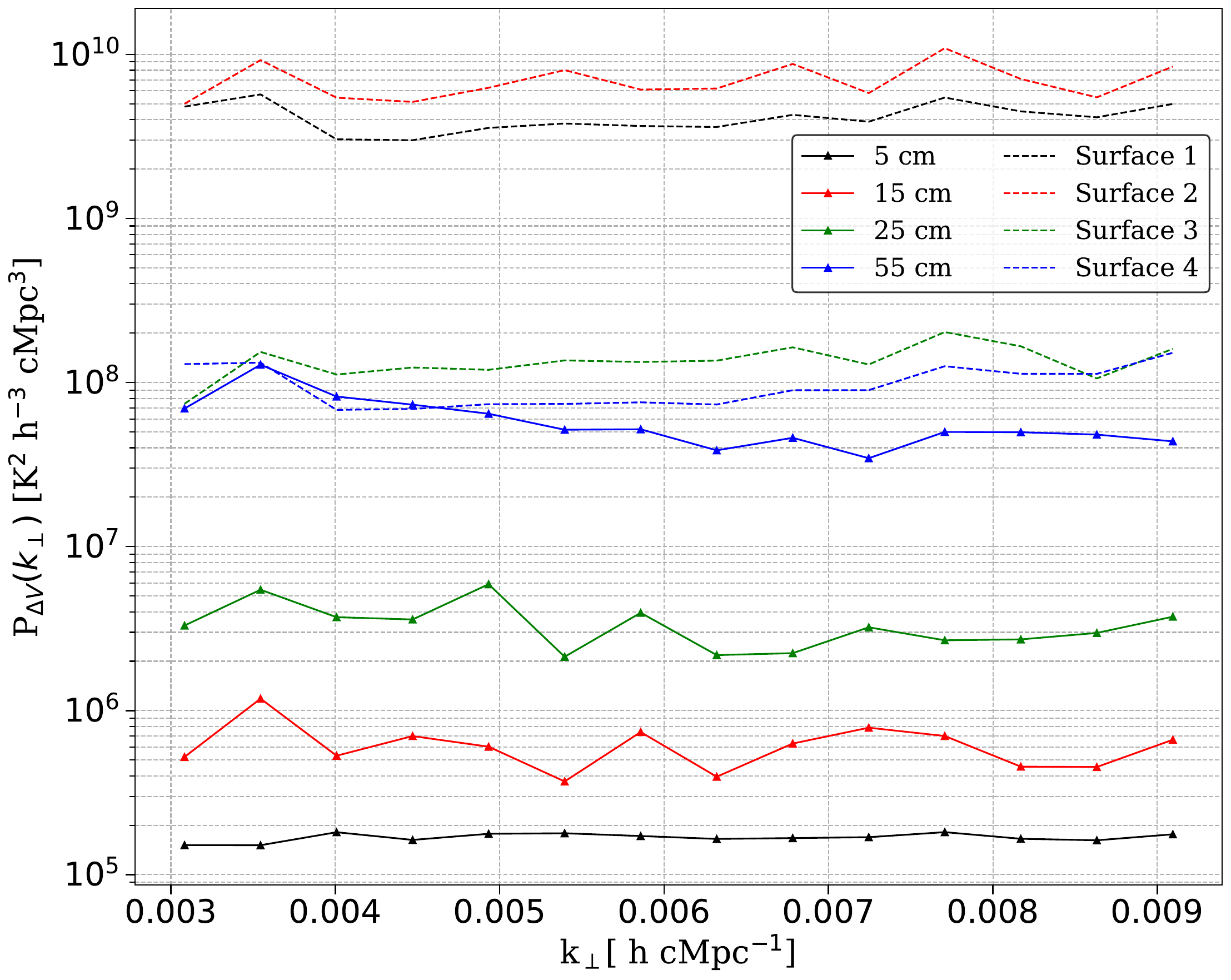}
    \end{minipage}\hfill
    \begin{minipage}{0.49\textwidth}
        \centering
        \includegraphics[width=0.88\linewidth]{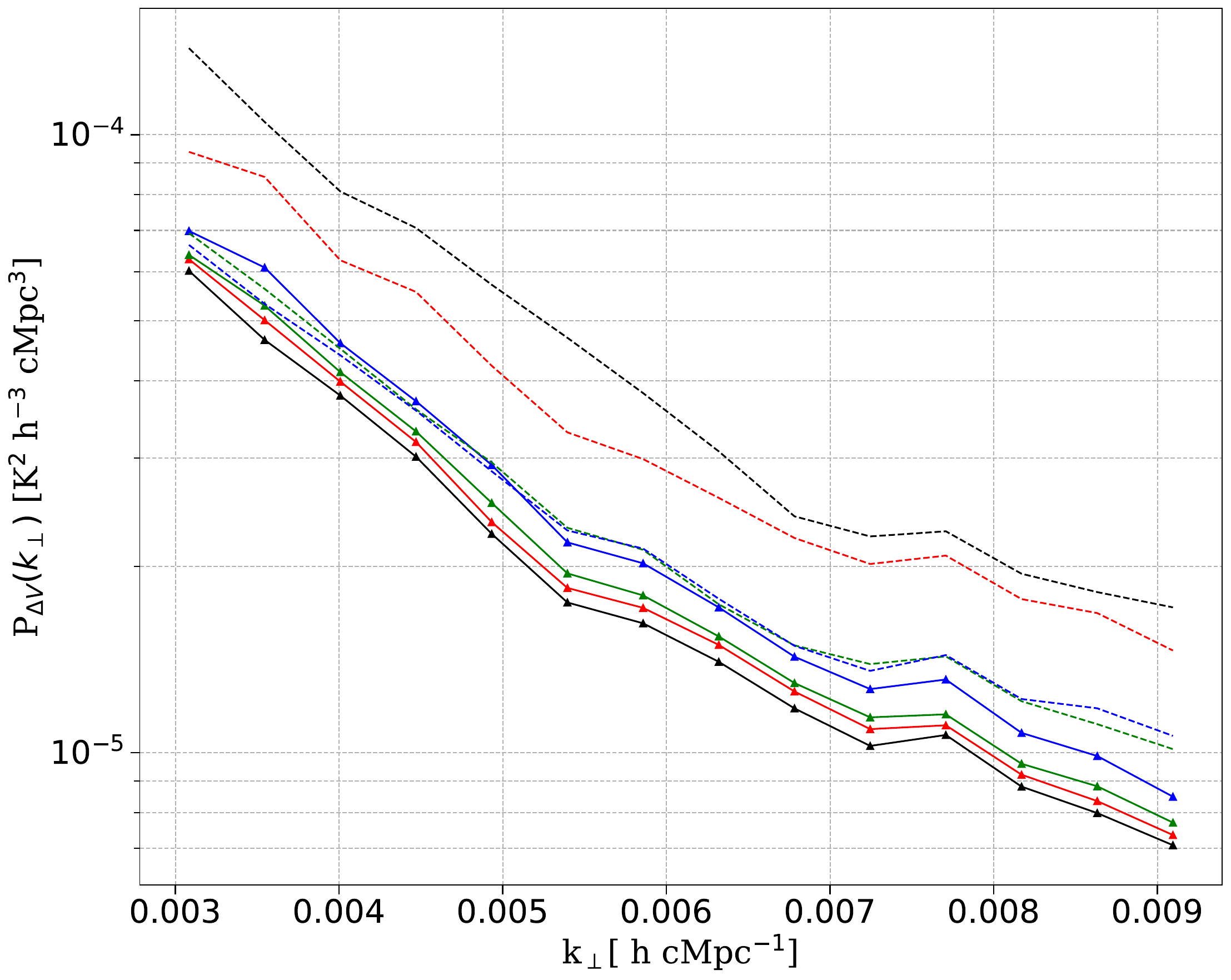}
    \end{minipage}

    \caption{The top row shows the cylindrically averaged power spectra of the difference in visibilities as a function of $k_{\parallel}$ (averaged over $k_{\perp}$), while the bottom row shows the same as a function of $k_{\perp}$ (averaged over $k_{\parallel}$). Two regions of ($k_{\perp}$,$k_{\parallel}$) space are analyzed: the cosmological window for $k_\parallel > 0.5$ h cMpc$^{-1}$ in the right column and the region inside the foreground wedge for $k_\parallel < 0.5$ h cMpc$^{-1}$ in the left column. Different colours represent the perturbation amplitude along xy direction (solid lines) and the four different surfaces (dashed lines) considered in this study. The antenna position offsets themselves change the visibilities only at the few per-cent level in amplitude for the perturbation levels considered (Section 4.1), so the difference visibilities $\Delta V = V_{\mathrm{pert}} - V_{\mathrm{reg}}$ are correspondingly small.  The oscillatory features in the cosmological window therefore represent very small absolute changes to an already foreground-suppressed region, governed by the Fourier response of the Kaiser20 window function and modulated by the small phase errors from the positional offsets.}
    \label{fig:zoom_perturb_Z42}
\end{figure*}

\subsubsection{Along xy direction}\label{sec:xy_offset} The top row of Fig. \ref{fig:2D_PS_xy_perturb_Z42} shows the cylindrically-averaged power spectra of the visibility difference for antenna position offsets along xy direction with amplitudes \( \sigma_{\text{xy}} \) = [0.05, 0.15, 0.25, 0.55] metres (from left to right). 
The results for Z42 are shown here, while the results for the Z148 are shown in the top row of Fig. \ref{fig:2D_PS_xy_perturb_Z148} in the Appendix \ref{sec:appendix2}.
The change between these columns quantifies the spatial-spectral variance introduced by the antenna position errors on the xy plane. We observe a systematic increase in the variance with the perturbation amplitude \( \sigma_{\text{xy}} \), both within the foreground wedge and in the 21-cm window. To illustrate this more clearly, in Fig. \ref{fig:zoom_perturb_Z42}, we show the cylindrically averaged power spectra of the difference in visibilities as a function of $k_{\parallel}$ (top row) and $k_{\perp}$ (bottom row), for regions within the foreground wedge ($k_\parallel < 0.5$ h cMpc$^{-1}$) and the cosmological window ($k_\parallel > 0.5$ h cMpc$^{-1}$) for xy offsets (solid). Inside the wedge, we are dominated by the foreground power at such low frequencies (see top left panel in Fig. \ref{fig:zoom_perturb_Z42}). We see that the power spectra decrease with $k_{\parallel}$ because the smooth foregrounds concentrate near low $k_{\parallel}$, and the Kaiser20 spectral window further suppresses the higher $k_{\parallel}$ modes. Any spectral feature introduced by the perturbation remains subdominant within the wedge. This is because the spectral shape there is set by the smooth foregrounds convolved with the main lobe of the chosen spectral window. On the other hand, we see that in the cosmological window, the power spectra show a weak but coherent oscillatory pattern as a function of $k_{\parallel}$ that preserves its shape across all perturbation amplitudes, with the absolute amplitude scaling with \( \sigma_{\text{xy}} \) (see top right panel in Fig. \ref{fig:zoom_perturb_Z42}). We attribute these features to the Fourier response of the Kaiser20 window, modulated by the additional phase factor introduced by the lateral offsets of the antenna position.

Now we see in the bottom left panel of Fig. \ref{fig:zoom_perturb_Z42} that inside the wedge, the perturbations act, to first order, as multiplicative scaling on the bright, spectrally smooth foreground. Any structure from phase error remains small and largely averaged out, leaving a nearly flat response in $k_{\perp}$. However, in the cosmological window, the power spectra decrease with $k_{\perp}$ because the residuals are mainly dominated by the leakage of the foregrounds through the sidelobes of the spectral window, with not much modulation induced by the positional offsets. Since diffuse foregrounds are strongest on small baselines and decrease with $k_{\perp}$, the magnitude of the leakage therefore inherits this $k_{\perp}$ trend.

\subsubsection{Along z direction}\label{sec:z_offset}
The bottom row of Fig. \ref{fig:2D_PS_xy_perturb_Z42} shows the cylindrically-averaged power spectra of the visibility differences obtained when antenna position offsets are introduced along the z direction, corresponding to the variation in the height of Surfaces 1, 2, 3 and 4 (see Section \ref{roughness result}), from left to right. Here we present the results for Z42, and the results for the Z148 are shown in the bottom row of Fig. \ref{fig:2D_PS_xy_perturb_Z148} in the Appendix \ref{sec:appendix2}. Although all four surfaces show similar spatial correlation, the RMS difference between vertical heights was seen to vary across the array scale (see Table \ref{tab:hurst_exponent}). Specifically, Surface 1 and 2 show a higher RMS height deviation than Surface 3 and 4. 

For correlated height errors arising from surface roughness, the additional phase factor scales as $\frac{\sigma_{s}^{2}}{\lambda^{2}} \left( \frac{r}{r_{s}} \right)^{2H}$ (see  Equations \ref{eq:var_height_diff} and \ref{eq:ens_avg_vis_z}). This implies that small $k_{\perp}$ modes are only weakly affected for all surfaces, while long baselines experience the strongest suppression in the image domain. Surface 1 and 2, with higher $\sigma_{s}$, lead to broader convolution kernels in the $u\varv$ domain, causing significant smearing across $u\varv$ cells. In the power spectra of the visibility difference, this appears as increase in variance towards larger $k_{\perp}$ modes and also leaks power into the 21-cm window. Surfaces 3 and 4 will have a comparatively narrower convolution kernel, even for longer baselines, thus minimizing mode mixing. As a result, the total variance introduced in both the wedge and the 21-cm window by these surfaces remains low. This is also evident in Fig. \ref{fig:zoom_perturb_Z42}, as indicated by the dashed lines. The weak oscillatory pattern in the power spectra is the same for the different surfaces as for the lateral offsets, as discussed above. Clearly, rough surfaces introduce stronger chromatic distortions compared to the largest lateral offsets considered in this study.


For non-coplanar arrays, the projection effect further amplifies phase decoherence for near horizon sources. As the sky rotates, projected baseline lengths vary with orientation, and for sources towards the horizon, the baselines experience the maximum change. Therefore, these baselines are more sensitive to small phase errors, resulting in a brighter wedge with leakage into the higher $k_{\parallel}$ modes. This geometric projection effect, in addition to the surface-induced kernel broadening in the $u\varv$ domain can be a problem for wide FoV instruments such as DEX, but could be mitigated by choosing an appropriate spatial window function to suppress the larger errors near the horizon.

\begin{figure*}
    \centering
    \begin{subfigure}[t]{0.48\textwidth}
        \centering
        \includegraphics[width=\linewidth]{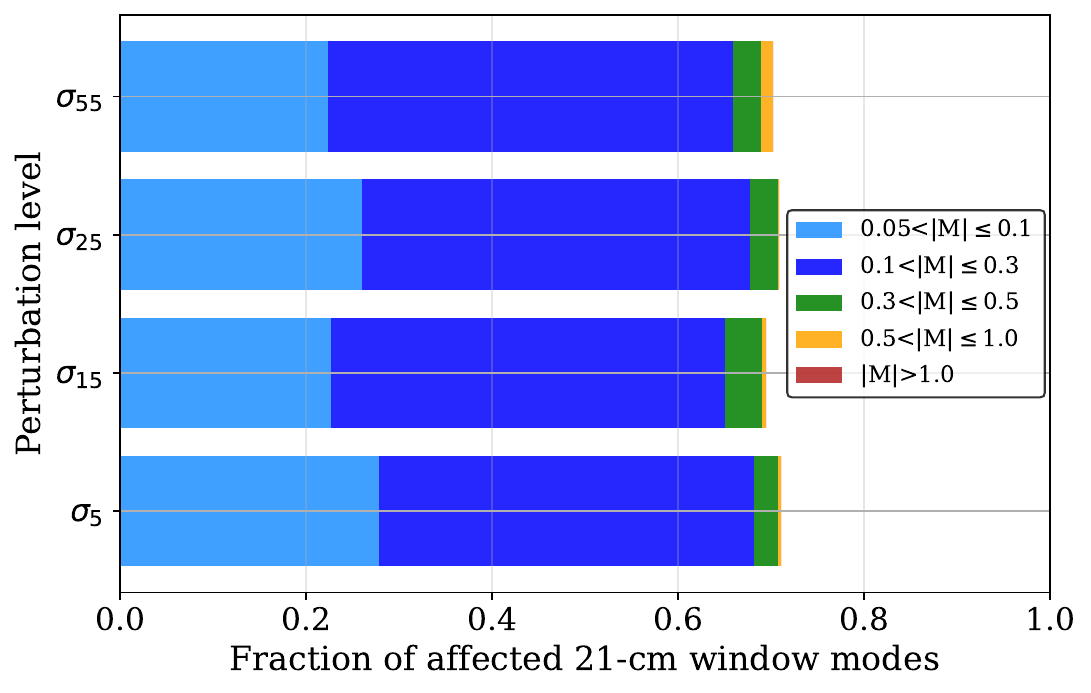}
        \label{fig:xy_contam}
    \end{subfigure}
    \hfill
    \begin{subfigure}[t]{0.48\textwidth}
        \centering
        \includegraphics[width=\linewidth]{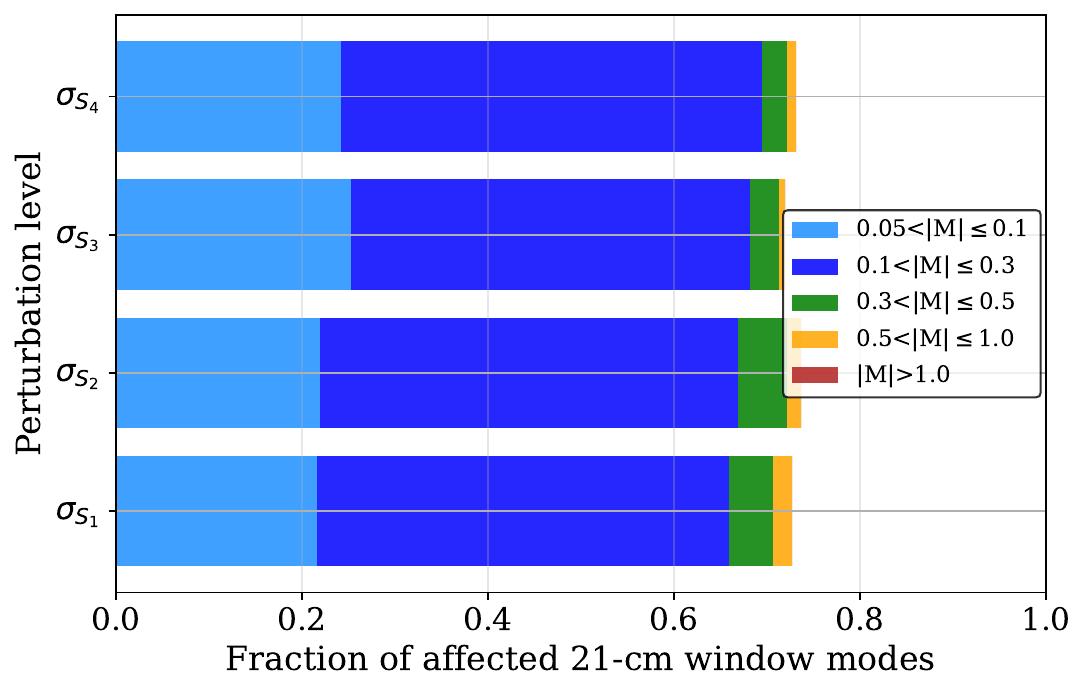}
        \label{fig:height_contam}
    \end{subfigure}

    \caption{Fraction of contaminated modes in 21-cm window for perturbations along xy direction (Left column) and along z direction (Right column). Colors distinguish increasing levels of contamination: light blue ($0.05 < |M| \leq 0.1$), dark blue ($0.1 < |M| \leq 0.3$), green ($0.3 < |M| \leq 0.5$), orange ($0.5 < |M| \leq 1.0$), and red ($|M| > 1.0$). The comparison highlights that majority of affected modes ($\sim$ 60 per cent) fall in the range ($0.1<|M|\leq0.3$) i.e., contaminated at 10 - 30 per cent level. However, antenna offsets induced by height differences of the surfaces produce a broader distribution towards higher contamination, with a slightly larger fraction of modes exceeding ($|M|>0.3$) compared to offsets along xy direction.}
    \label{fig:contamination_summary}
\end{figure*}
\noindent
\subsubsection{Impact on the Dark Ages 21-cm Power Spectrum}\label{sec:detectability}
In Sections \ref{sec:xy_offset} and \ref{sec:z_offset}, we have characterized the impact of antenna position errors using the power spectrum of visibility differences. In addition, we now define a dimensionless metric in the 21-cm window ($k_\parallel > 0.5$ h cMpc$^{-1}$),

\begin{equation}
|{\rm M}(k_\perp, k_\parallel)| \;\equiv\; 
\frac{\;|P_{\mathrm{pert}}(k_\perp,k_\parallel)| - |P_{\mathrm{reg}}(k_\perp,k_\parallel)|\;}
{|P_{21}(k_\perp,k_\parallel)|},
\label{eq:metric}
\end{equation}

\noindent
where $P_{\mathrm{reg}}$ and $P_{\mathrm{pert}}$ are the cylindrically-averaged power spectra of the composite signal for the regular and perturbed arrays, respectively, and $P_{\mathrm{21}}$ is the cylindrically-averaged 21-cm signal power spectrum. This metric provides an answer to the question: "\textit{How large is the error introduced by the antenna position offsets compared to the cosmological signal we want to detect?}". Therefore, Equation \ref{eq:metric} provides a quantitative measure of the fraction of Fourier modes that are contaminated at different levels relative to the DA 21-cm signal. As seen in the bottom panel of Fig. \ref{fig:2D_PS_reg} for Z42, within the 21-cm window, the foreground is more than a factor of ten lower than the DA 21-cm signal, so the numerator in Equation \ref{eq:metric} is dominated by the signal. A conservative threshold of |M| < 0.05 is acceptable in our case. We believe that contamination below this level is small compared to other systematic uncertainties (e.g., calibration errors, thermal noise). Modes exceeding this threshold are further subdivided into bins corresponding to increasing levels of contamination.

The left panel of Fig. \ref{fig:contamination_summary} summarizes the impact of antenna position errors along the xy direction on the 21-cm window for Z42. For all the perturbation cases considered in this study, the majority of the contaminated modes ($\approx$ 57 -- 62 per cent) depending on the perturbation level fall within the range 0.1 < |M| $\leq$ 0.3 i.e., they have a bias between 10 per cent and 30 per cent relative to the 21-cm signal. A smaller but still significant fraction of modes ($\approx$ 32 -- 39 per cent) lie in the mild contamination bin 0.05 < |M| $\leq$ 0.1. Only a few percent of modes are more strongly affected, with 0.3 < |M| $\leq$ 0.5 contributing $\approx$ 3 -- 6 per cent depending on the perturbation level. Modes with |M| > 0.5, where perturbations begin to dominate over the 21-cm signal itself, are exceedingly rare ($\leq$ 2 per cent) and no modes were found with |M| > 1 across the tested cases.

For the four representative surfaces, the impact of antenna position errors along the z direction on the 21-cm window for Z42 is shown in the right panel of Fig. \ref{fig:contamination_summary}. As before, for almost all cases, the largest affected modes ($\approx$ 60 per cent) fall within the range 0.1 < |M| $\leq$ 0.3. Between 30 -- 35 per cent of contaminated modes lie in the 0.05 < |M| $\leq$ 0.1 range, independent of the surfaces. Only 4 -- 7 per cent of modes exceed the 0.3 < |M| $\leq$ 0.5 threshold, and fewer than 3 per cent lie in the 0.5 < |M| $\leq$ 0.1 range. No modes with |M| > 1 were observed. 

We reiterate for the readers that the cylindrically averaged power spectrum of visibility differences for the composite sky (foreground + 21-cm cosmological signal) shown in Fig. \ref{fig:2D_PS_xy_perturb_Z42} and the quantity defined in Equation \ref{eq:metric} measure fundamentally different aspects of the perturbation. The power spectrum of visibility differences, $P_{\Delta V} \propto |\Delta V|^{2}$, isolates only the perturbation-induced component and therefore remains small, at the level of $\sim10^{-4}$--$10^{-5}$ in the cosmological window for $k_\parallel > 0.5\,h\,\mathrm{cMpc}^{-1}$ (as seen in Fig. \ref{fig:zoom_perturb_Z42}). In contrast, the numerator of Equation \ref{eq:metric} is the difference between the power spectra of the composite sky which gives the fractional error relative to the 21-cm signal only power spectrum. Because the 21-cm signal is intrinsically weak , even a very small absolute change in the visibilities can translate into a comparatively large fractional difference. Consequently, values within the range 0.1 < |M| $\leq$ 0.3 reflect small absolute perturbations to a foreground-dominated quantity when expressed in units of the DA 21-cm signal.

Additionally, for the very low frequencies considered here, the foreground power is extremely bright compared to the 21-cm signal. Consequently, the dominant source of leakage into the 21-cm window is the shape of the spectral window function, in our case the Kaiser20. Under these conditions, errors from antenna position offsets remain subdominant compared to the spectral correlations imposed by this window and thus do not introduce significant additional contamination. Therefore, in all cases studied, the impact of deviation from a regular, coplanar grid lies below the 21-cm power spectrum within the 21-cm window for $k_\parallel > 0.5$ h cMpc$^{-1}$, given a Kaiser20 window function applied to a 5 MHz bandwidth with 100 channels at Z42.



However, an important difference arises between the two perturbation cases. The offsets along xy direction produces fewer modes with higher contamination levels, i.e., |M| > 0.3 than height perturbations. This difference is most evident for Surfaces 1 and 2 that produce the largest fraction of modes with |M| > 0.3.

For offsets along xy direction, we can directly frame the tolerances in terms of the dimensionless ratio $\frac{\sigma}{\lambda}$. Adopting a conservative approach to ensure that contaminated modes (|M| $> 0.5$) remain below the 1 per cent level, we find from Fig. \ref{fig:contamination_summary} that this occurs when $\sigma_{xy} = 0.25 \, \mathrm{m}$. This is roughly 3 per cent of the observing wavelength ($\lambda \simeq 9.23 \, \mathrm{m}$), corresponding to $\frac{\sigma_{xy}}{\lambda}$ $\leq 0.027$. This implies that as long as deployment accuracy in the lateral direction is controlled to within 3 per cent of a wavelength, strongly contaminated modes occur only at sub-per cent levels in the 21-cm  for $k_\parallel > 0.5$ h cMpc$^{-1}$.

For vertical offsets, tolerances cannot be expressed through a single $\frac{\sigma}{\lambda}$ ratio, since the perturbations arise from the roughness of a fractal surface that depends on the reference length scale, \(r_{\rm s}\). Table \ref{tab:hurst_exponent} summarizes $H$ and $\sigma_{s}$ for the four representative lunar terrains, evaluated at two characteristic length scales: $L = 4$ m and $L = 175$ m. At the smaller scale ($L = 4$ m), $\frac{\sigma_{4\,{\rm m}}}{\lambda}$ is $0.032, 0.028, 0.016,$ and $0.015$ for Surfaces 1 -- 4, respectively, at 32.5 MHz. Clearly, the ratio for Surfaces 3 and 4 are well within 3 per cent of a wavelength. This level is comparable to acceptable offsets along xy direction and is expected to place most affected modes well below the threshold  |M| $> 0.5$. Therefore, they are unlikely to cause significant phase decoherence in the visibilities. However, at longer scales ($L = 175$ m), $\frac{\sigma_{175\,{\rm m}}}{\lambda}$ is $0.760, 0.446, 0.206,$ and $0.196$. At 32.5 MHz, this corresponds to 20 -- 76 per cent of a wavelength, which will produce phase errors sufficient for a significant fraction of modes to exceed the |M| $> 0.5$ threshold. Surfaces 3 and 4, although smoother on small scales and therefore less problematic for smaller baselines, show height differences that is a significant fraction of a wavelength at larger scales. Therefore, while such surfaces are preferred for the deployment of compact radio arrays, careful calibration will still be required to correct for the phase errors introduced by height variations at large scale. Thus, lateral offsets can be controlled through deployment tolerances ($\frac{\sigma}{\lambda} \lesssim 0.027$), but vertical offsets must be mitigated primarily through the choice of deployment site.

\section{Summary and Conclusions}
\label{sec:conclusions}
DEX is a proposed low frequency radio interferometer concept for deployment on the lunar farside, designed to probe the Dark Ages using 21-cm signal. The current minimum baseline configuration consists of a compact, regular $32 \times 32$ grid of zenith-pointing, co-located cross-dipole antennas of 3 m length with near-unity filling factor, operating in the 7 -- 50 MHz. band. A key feature of DEX is the intention to use an FFT correlator, which allows efficient data processing with reduced power consumption, an essential advantage for large arrays operating under the resource constraints of a lunar surface mission. 

However, the effectiveness of such a correlation architecture is dependent on the assumption of a perfectly regular grid. Even small deviations from this geometry can have important consequences. First, position errors will break the redundancy required for FFT correlation, potentially prohibiting its use, and thereby greatly increasing both computational and energy costs. This is an important limitation for large arrays based on the lunar surface. Second, such offsets lead to decoherence of the visibilities, leading to spectral contamination, and thereby biasing the Fourier modes used for 21-cm signal extraction. Finally, they complicate calibration, since the relative positions and orientations of antennas must be known to within a small fraction of a wavelength.

In this work, we focus on lateral and vertical position offsets independently, which are the most likely errors to arise from rover deployment and local topographic undulations on the lunar surface. To do so, we adopt a visibility-based approach. This is a practical consideration because full forward simulations at the electric field level for DEX-like arrays would be computationally intensive. 

To quantify the impact of such perturbations on the 21-cm power spectrum for DEX, we first derive and generalize the equations describing the perturbed visibilities, without taking a flat-sky approximation and thereby enabling a wide-field treatment. Additionally, these equations provide a theoretical basis for interpreting the trends observed in the simulations obtained with the SPADE-21cm (Simulation Pipeline for Analyzing Dark agEs using 21-cm). This end-to-end pipeline incorporates a cosmological signal model, a realistic sky, lunar topography data, and a complete lunar topocentric coordinate system. Mare Ingenii, an impact basin located in the lunar southern hemisphere on the farside of the Moon, is selected as the representative deployment site for our simulations. To our knowledge, this constitutes the first systematic assessment of antenna position errors for a lunar surface array, providing practical guidance for array design, deployment strategy, and site selection. Our main conclusions are summarized below:


\textbf{Criteria for reliable deployment site}: A preferred deployment site should ideally depend on RMS deviation in height differences \(\sigma_{\rm s}\), Hurst exponent $H$. These quantities are coupled and degenerate, and their interpretation also depends on the choice of a reference distance $r_s$ between two points on the surface. At all times, \(\sigma_{\rm s}\) must remain low across all baseline lengths of interest such that the phase errors due to the height differences are minimized. The impact of $H$ depends on array configuration: higher $H$ is preferable for compact arrays as it introduces least possible variance on small baselines, whereas lower $H$ is advantageous for longer baselines.

\textbf{Impact of antenna position offsets on the $u\varv$ plane}: Random antenna position errors can be described as multiplicative attenuation kernels in the image domain that act as an additional window function, equivalent to convolutional kernels in the $u\varv$ domain. For offsets along xy direction, the ensemble-averaged kernel is baseline-independent, producing relatively uniform mode mixing across all baselines. However, projection effects generate non-zero $\Delta w$ terms that increase suppression at larger angular distances from the phase center, particularly in wide-field observations with maximum impact towards the horizon.

In contrast, offsets along z direction introduce baseline-dependent suppression due to the statistical model of correlated surface topography, where the height-difference scales as $r^{2H}$. The kernel width and degree of attenuation therefore increase with baseline length, observing frequency, and angular distance from the phase center, and are further modulated by the Hurst exponent and RMS surface height deviation. Anisotropy in the $u\varv$ plane also directly reflects anisotropy in the underlying surface roughness, with stronger phase decorrelation for baselines aligned along axes of maximum roughness. Thus, while xy and z offsets can both be described within the same analytical framework, their different statistical models give rise to a baseline-independent kernel and baseline-dependent kernel, in the image and $u\varv$ domains, respectively.

\textbf{Impact of antenna position offsets on the power spectrum}:
For both the perturbation scenarios, we see the power to increase with increasing perturbation levels in both the wedge and in the 21-cm window, and dominated fully by residual leakage due to the chosen spectral window function (Kaiser20 in our case). For the offsets along xy direction, the increment remains modest up to $\sigma$ $\leq$ 0.25 m, but becomes measurable beyond this level. For positional perturbations along the z direction, the phase decoherence is governed by the distinct surface topography, characterized by $\sigma_{s}$ and $H$. In this study, all the four surfaces have similar values of $H$, indicating comparable spatial correlation. However, Surfaces 1 and 2 show larger $\sigma_{s}$ across all scales, leading to larger phase errors, which scale as $\sigma_{s}^{2}\ (r/r_{s})^{2H}$. In contrast, Surfaces 3 and 4 are comparatively smoother, with lower $\sigma_{s}$. Consequently, Surfaces 1 and 2 show increased leakage into the 21-cm window, while Surfaces 3 and 4 show leakage at a reduced level.

Our simplified treatment provides a first-order estimate of the impact of the positional offsets along xy direction and along z direction independently. More realistic scenarios with correlated offsets could be incorporated in future, with more detailed simulations and more complex spatial error patterns. But as long as the level of the perturbations remains comparable to the values explored in this study, we do not expect a larger impact on the power spectrum. However, we do note that the vertical height offsets are expected to dominate the resulting phase errors and hence distort the power spectrum.

\textbf{Position induced errors in the 21-cm Dark ages power spectrum}: 
Antenna position errors typically bias the 21-cm signal at the 10 - 30 per cent level, with about one-third of modes affected at 5 - 10 per cent and only a few percent exceeding 30 per cent. Modes where the error is larger than 0.5 times the 21-cm signal (|M| > 0.5) are rare ($\leq$ 2 per cent), and no modes were found with errors larger than the 21-cm signal itself (|M| > 1). For lateral offsets, keeping ($\frac{\sigma_{xy}}{\lambda} \lesssim 0.027$) (corresponding to $\sigma_{xy}$ = 0.25 metres at 32.5 MHz) limits the fraction of modes with |M| > 0.5 to less than 1 per cent. For vertical offsets, the impact depends on the surface roughness. The small-scale variations correspond to 1 - 3 per cent of a wavelength, comparable to xy offsets. But long-scale roughness corresponds to 20 - 76 per cent of a wavelength, producing a large fraction of modes  |M| > 0.3. This conclusion holds for the 21-cm window with $k_\parallel > 0.5$ h cMpc$^{-1}$ over the range of $k_\perp = 0.003 - 0.009$ h cMpc$^{-1}$ while applying a Kaiser spectral window with $\beta$ =  20.

Thus, errors from lateral height offsets can be reduced by maintaining placement accuracy within the derived tolerance limits, whereas errors from vertical offsets introduce more severe phase errors and must be taken care through careful site selection.

\section*{Acknowledgements}
We dedicate this work to the memory of our co-author, Albert-Jan Boonstra$^\dagger$, whose insight and guidance help shaped this work, but who sadly passed away during its preparation. SG, LVEK, JKC, SAB, and SM acknowledge the financial support from the European Research Council (ERC) under the European Union's Horizon 2020 research and innovation programme (Grant agreement No. 884760, `CoDEX'). EC would like to acknowledge support from the Centre for Data Science and Systems Complexity (DSSC), Faculty of Science and Engineering at the University of Groningen and from the Ministry of Universities and Research (MUR) through the PRIN project `Optimal inference from radio images of the epoch of reionization'. FGM acknowledges support from the I-DAWN project, funded by the DIM-ORIGINS programme.  SG is grateful to Prasun Mahanti for directing to the relevant lunar surface topography data sources used in this work.  

\noindent
This work extensively uses the \textsc{\scriptsize NUMPY} \citep{harris2020array}, \textsc{\scriptsize SCIPY} \citep{virtanen2020scipy}, \textsc{\scriptsize PYTHON-CASACORE} \citep{bean2022casa}, \textsc{\scriptsize ASTROPY} \citep{Astropy2013, Astropy2018, Astropy2022}, and \textsc{\scriptsize MATPLOTLIB} \citep{hunter2007matplotlib} Python packages.

\section*{Data Availability}

The data underlying this article will be shared on reasonable request
to the corresponding author.



\bibliographystyle{mnras}
\bibliography{references2} 




\appendix

\section{Rotation Matrices for ENU to MCMF to uvw}
\label{sec:appendix1}
\begin{figure}
    \centering
    \includegraphics[width=\columnwidth]{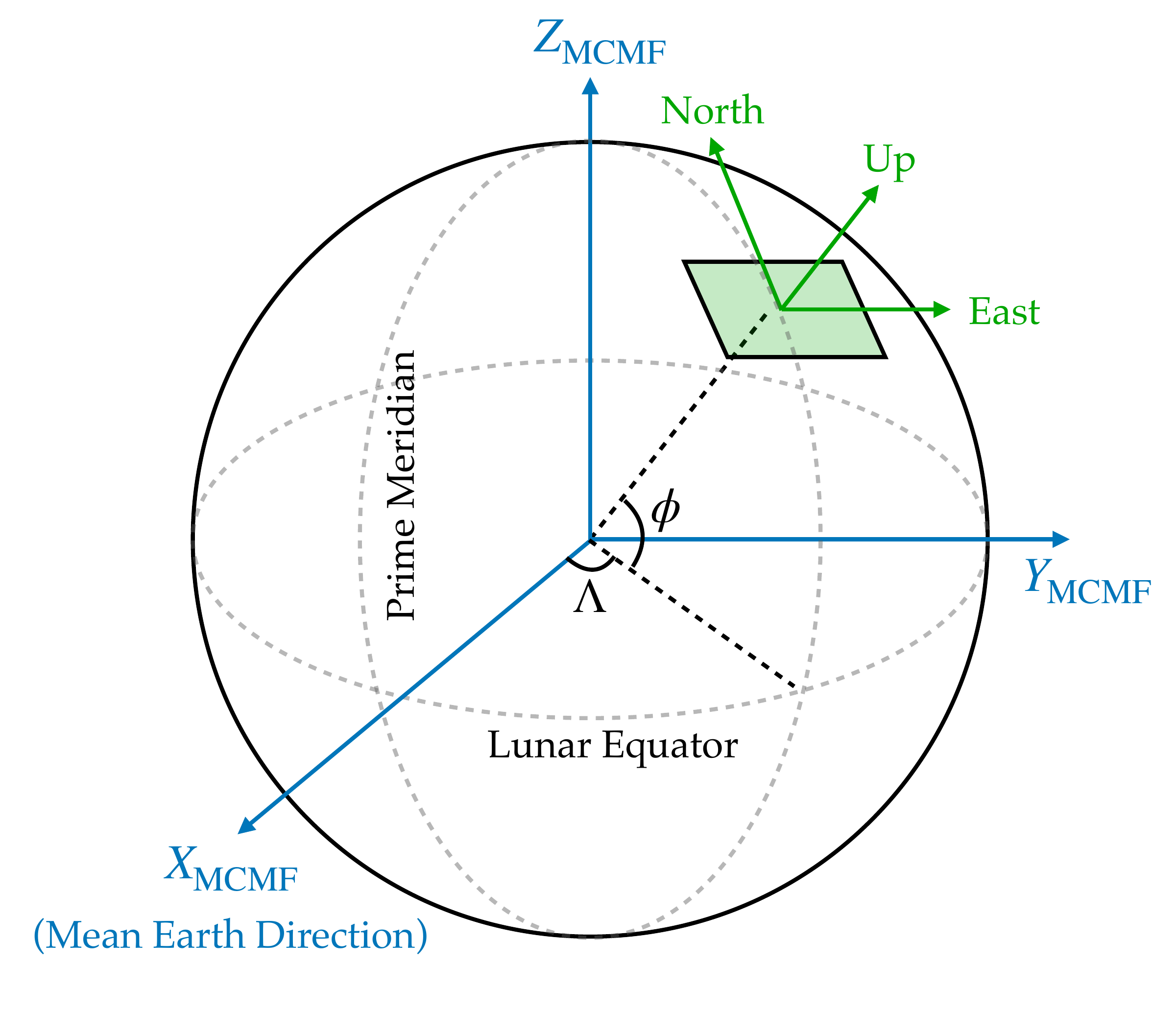}
    \caption{The MCMF and lunar topocentric frames.  }
     \label{fig:mcmf}
\end{figure}

\begin{figure}
    \centering
    \includegraphics[width=0.9\columnwidth]{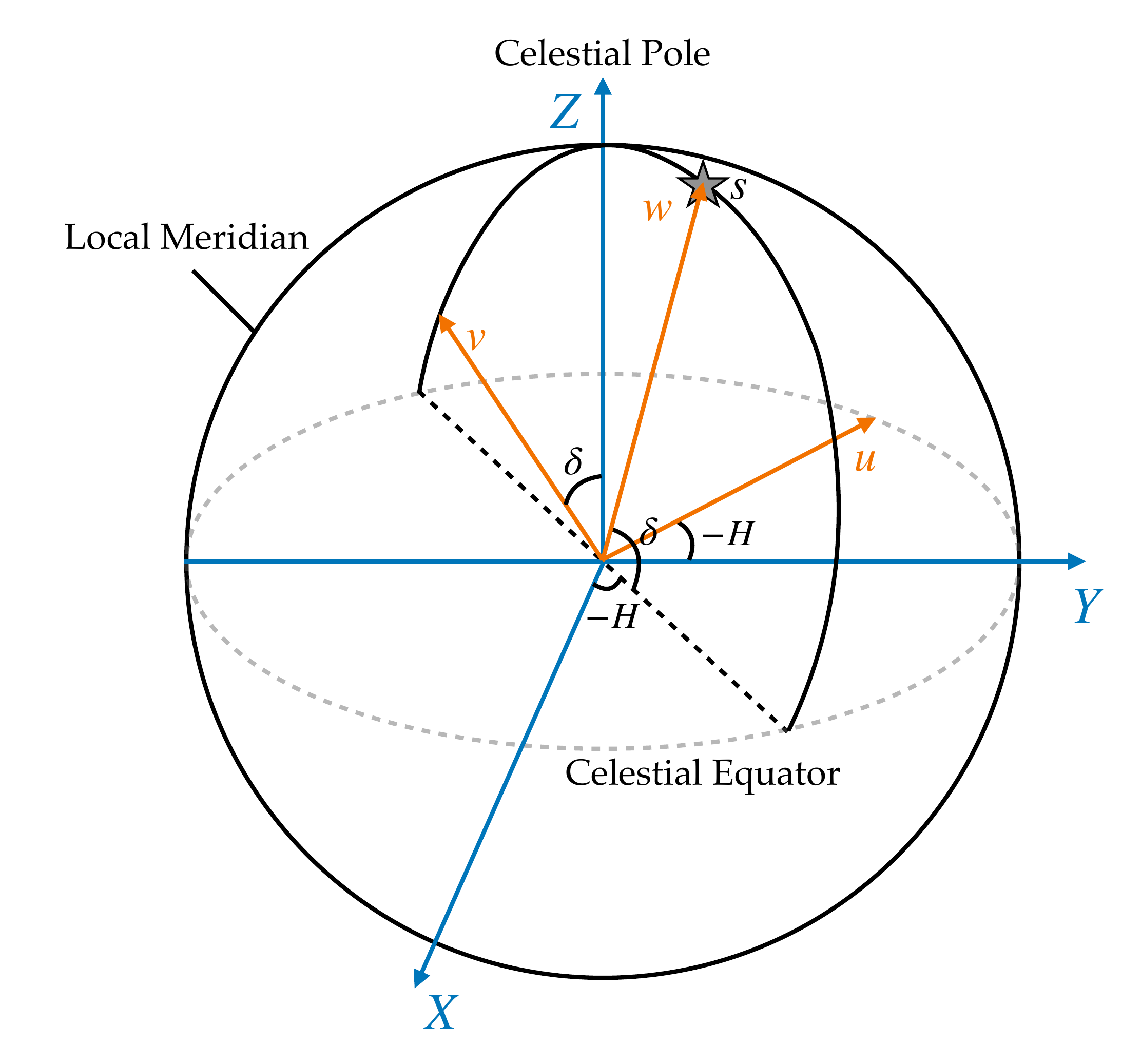}
    \caption{Geometric relation between the celestial and interferometric coordinate systems.
 }
     \label{fig:uvw}
\end{figure}

Consider a point on the lunar surface defined by selenographic latitude \(\phi\) and longitude \(\Lambda\), with \(\Lambda\) increasing westward. The unit position vector in the MCMF frame is given by
\begin{equation}
\hat{\mathbf r} =
\begin{pmatrix}
\cos\phi\,\cos \Lambda\\
-\cos \phi \sin \Lambda\\
\sin\phi
\end{pmatrix}.
\end{equation}

\noindent
The outward radial unit vector \(\hat{\mathbf u}\), is identical to the position vector. The \(\hat{\mathbf n}\) points along the meridian towards the North Pole, obtained by projecting the \(\hat{Z}_{\rm {MCMF}}\) represented by \( [0, 0, 1]^T \) onto the tangent plane perpendicular to \( \hat{\mathbf u} \). The projection is

\begin{equation}
\hat{\mathbf n} = \frac{[0, 0, 1]^T - (\hat{u} \cdot [0, 0, 1]^T) \hat{u}}{\| [0, 0, 1]^T - (\hat{u} \cdot [0, 0, 1]^T) \hat{u} \|}
\label{eq:n_cap}
\end{equation}

\noindent
where \( \hat{\mathbf u} \cdot [0, 0, 1]^T = \sin \phi \). Solving the numerator in Equation \ref{eq:n_cap}, we get

\begin{align}
&[0, 0, 1]^T - \sin \phi \hat{\mathbf u} \\
&= [0, 0, 1]^T - \sin \phi [\cos \phi \cos \Lambda, -\cos \phi \sin \Lambda, \sin \phi]^T \notag \\
&= [-\sin \phi \cos \phi \cos \Lambda, \sin \phi \cos \phi \sin \Lambda, 1 - \sin^2 \phi]^T \notag \\
&= [\sin \phi \cos \phi (-\cos \Lambda), \sin \phi \cos \phi \sin \Lambda, \cos^2 \phi]^T
\end{align}

\noindent
The denominator in Equation \ref{eq:n_cap} is simply the norm given by,

\begin{align}
&\| [-\sin \phi \cos \phi \cos \Lambda, \sin \phi \cos \phi \sin \Lambda, \cos^2 \phi]^T \|\\
&= \sqrt{(\sin \phi \cos \phi)^2 (\cos^2 \Lambda + \sin^2 \Lambda) + \cos^4 \phi} \notag \\
&= \sqrt{\sin^2 \phi \cos^2 \phi + \cos^4 \phi} = \cos \phi
\end{align}

\noindent
Finally, we get

\begin{align}
&\hat{\mathbf n} = \frac{1}{\cos \phi} [\sin \phi \cos \phi (-\cos \Lambda), \sin \phi \cos \phi \sin \Lambda, \cos^2 \phi]^T \\
&= [-\sin \phi \cos \Lambda, \sin \phi \sin \Lambda, \cos \phi]^T
\end{align}

\noindent
The \( \hat{\mathbf e} \), perpendicular to \( \hat{\mathbf u} \) and \( \hat{\mathbf n} \), lies along the parallel (constant \(\phi\)) and points eastward (toward decreasing \(\Lambda\)). Using the right-hand rule,
\begin{equation}
\hat{\mathbf e} = \hat{\mathbf n} \times \hat{\mathbf u} = \begin{vmatrix}
\hat{i} & \hat{j} & \hat{k} \\
-\sin \phi \cos \Lambda & \sin \phi \sin \Lambda & \cos \phi \\
\cos \phi \cos \Lambda & -\cos \phi \sin \Lambda & \sin \phi
\end{vmatrix}
\end{equation}

Expanding the determinant:

\begin{align}
&= \hat{i} [(\sin \phi \sin \Lambda)(\sin \phi) - (\cos \phi)(-\cos \phi \sin \Lambda)] \notag \\
&- \hat{j} [(-\sin \phi \cos \Lambda)(\sin \phi) - (\cos \phi)(\cos \phi \cos \Lambda)] \notag \\
&+ \hat{k} [(-\sin \phi \cos \Lambda)(-\cos \phi \sin \Lambda) - (\sin \phi \sin \Lambda)(\cos \phi \cos \Lambda)] \notag \\
&= \hat{i} [\sin \Lambda] + \hat{j} [\cos \Lambda] + \hat{k} [0] = [\sin \Lambda, \cos \Lambda, 0]^T
\end{align}

\noindent
We note that the unit vectors $(\hat{e}, \hat{n}, \hat{u})$ 
corresponds directly to the unit vectors in standard spherical-polar coordinate system. The vector $\hat{u}$ is identical to the radial unit vector $\hat{r}$. The eastward tangent vector $\hat{e}$, along a line of constant latitude, coincides with the azimuthal unit vector $\hat{\phi}$. The northward tangent vector $\hat{n}$ points toward increasing latitude and is therefore opposite to the usual polar unit vector $\hat{\theta}$, which increases toward the south. Thus, $\hat{u} = \hat{r}$, $\hat{e} = \hat{\phi}$, and $\hat{n} = -\hat{\theta}$.
The rotation matrix \(\mathbf{R}_{\mathrm{MCMF}}\) is finally expressed as
\begin{equation}
\mathbf{R}_{\mathrm{MCMF}} =
\begin{pmatrix}
\hat{\mathbf{e}} \\
\hat{\mathbf{n}} \\
\hat{\mathbf{u}}
\end{pmatrix}
=
\begin{pmatrix}
\sin \Lambda & -\sin \phi \cos \Lambda & \cos \phi \cos \Lambda \\
\cos \Lambda & \sin \phi \sin \Lambda & -\cos \phi \sin \Lambda \\
0 & \cos \phi & \sin \phi
\label{eq:rot_mcmf}
\end{pmatrix}
\end{equation}

We now define the direction to a source \textbf{s} in the MCMF coordinate system. A source located at hour angle \(H\) (west positive) and declination \(\delta\) has unit vector
\begin{equation}
\hat{\mathbf s}=\hat{\mathbf w}(H,\delta)=
\begin{pmatrix}
\cos\delta\,\cos H\\
-\cos\delta\,\sin H\\
\sin\delta
\end{pmatrix}.
\end{equation}

\noindent
The tangent plane at the phase center is perpendicular to \( \hat{\mathbf w} \). The \( \hat{\mathbf v} \) point towards the Celestial pole. The \( \hat{\mathbf v} \) is obtained by projecting the \(\hat{\mathbf Z}_{\rm {MCMF}}\) = \( [0, 0, 1]^T \) onto this plane and normalizing

\begin{equation}
\hat{\mathbf v}(H,\delta)=
\begin{pmatrix}
-\sin\delta\,\cos H\\
\sin\delta\,\sin H\\
\cos\delta
\end{pmatrix}.
\end{equation}

The \( \hat{\mathbf u} \) lies in the tangent plane, orthogonal to \( \hat{\mathbf w} \)  and \( \hat{\mathbf v} \). By the right hand convention, 
\begin{equation}
\hat{\mathbf u} = \hat{\mathbf v} \times \hat{\mathbf w} =
\begin{pmatrix}
\sin H\\
\cos H\\
0
\end{pmatrix}
\end{equation}

\noindent
The rotation matrix from MCMF to $u\varv w$ coordinate system is then given by
\begin{equation}
\mathbf{R}_{\mathrm{uvw}} =
\begin{pmatrix}
\hat{\mathbf{u}}^\top \\
\hat{\mathbf{v}}^\top \\
\hat{\mathbf{w}}^\top
\end{pmatrix}
=
\begin{pmatrix}
\sin H & \cos H & 0 \\
- \sin\delta \cos H &  \sin\delta \sin H & \cos\delta \\
\cos\delta \cos H & -\cos\delta \sin H & \sin\delta
\label{eq:rot_uvw}
\end{pmatrix}.
\end{equation}

\noindent
For a fixed array at longitude \(\Lambda\), observing a phase center \((\alpha_0,\delta_0)\), we absorb \(\Lambda\) by setting \(\Lambda\) =  0 in the local frame and define the local hour angle $H_0$ $\equiv$ LST (t, \(\Lambda\)) - \(\alpha_0\). The Equations \ref{eq:rot_mcmf} and \ref{eq:rot_uvw} then become

\begin{equation}
\mathbf{R}_{\mathrm{MCMF_{\Lambda=0}}} = 
\begin{pmatrix}
0 & -\sin \phi & \cos \phi \\
1 & 0 & 0 \\
0 & \cos \phi & \sin \phi
\label{eq:rot_mcmf_array}
\end{pmatrix}
\end{equation}

\begin{equation}
\mathbf{R}_{\mathrm{uvw}} =
\begin{pmatrix}
\sin H_0 & \cos H_0 & 0 \\
- \sin\delta_0 \cos H_0 &  \sin\delta_0 \sin H_0 & \cos\delta_0 \\
\cos\delta_0 \cos H_0 & -\cos\delta_0 \sin H_0 & \sin\delta_0
\label{eq:rot_uvw_array}
\end{pmatrix}.
\end{equation}


Let $\Delta\mathbf{r}_{\rm ENU} = (\Delta E, \Delta N, \Delta U)^{\top}$ denote antenna position offsets in metres.
From the main text [Eq.~(6) and Eq.~(7)] 
\begin{equation}
\begin{pmatrix}
\Delta u\\[2pt] \Delta \varv\\[2pt] \Delta w
\end{pmatrix}
= \frac{1}{\lambda}\;
\mathbf{R}_{\rm uvw}\;\mathbf{R}_{\rm MCMF}\;
\begin{pmatrix}
\Delta E\\[2pt] \Delta N\\[2pt] \Delta U
\end{pmatrix}.
\qquad\qquad\tag{A18}
\end{equation}

\section{Impact of Window Functions}
\label{sec: window}
The choice of spectral window functions in the 21-cm data analysis is closely related to the foreground mitigation strategy adopted, as it shapes the instrument's response along the $k_{\parallel}$ direction in Fourier space. 



\begin{figure*}
    \centering
        \includegraphics[width=0.99\textwidth]{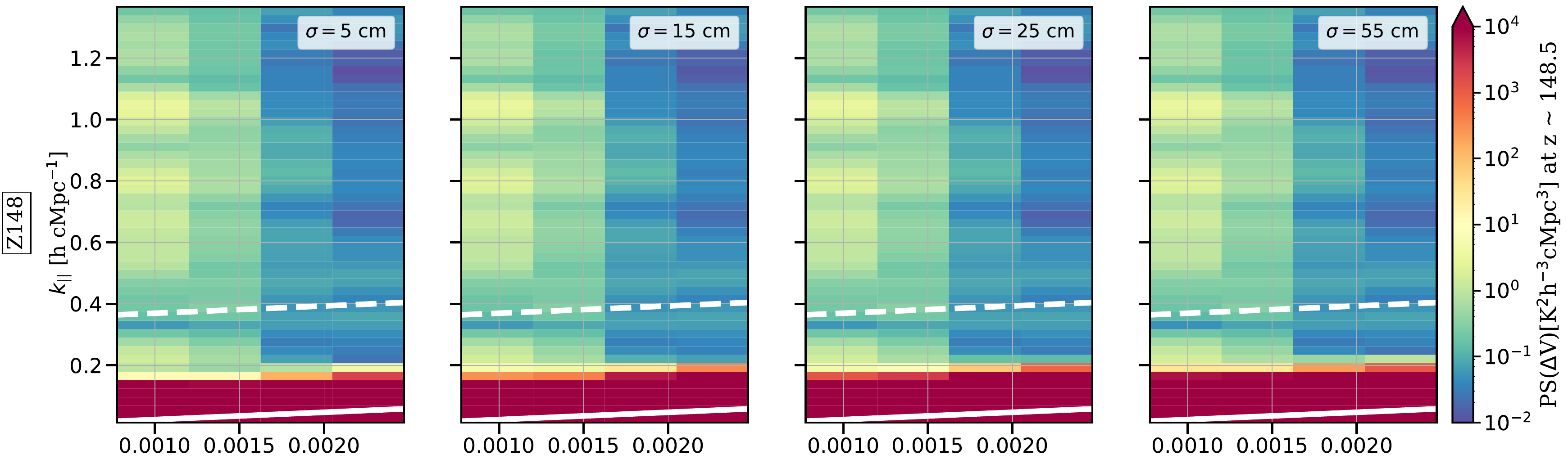}
        \vspace{0.3cm} 
        \includegraphics[width=0.99\textwidth]{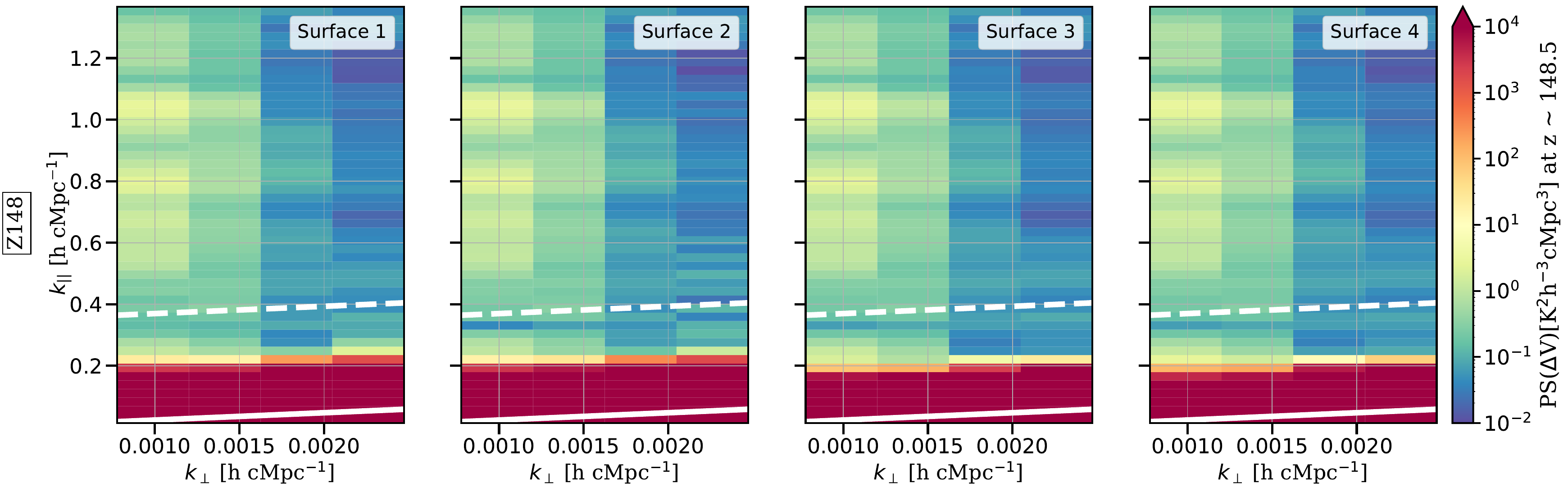}
    \caption{The cylindrically averaged power spectra of the difference in visibilities between the composite model with unperturbed antenna elements and those with perturbations on the xy plane (\textbf{Top row}) and those induced due to the surface irregularities (\textbf{Bottom row}). The offsets along xy directions are drawn from a two-dimensional normal distribution $\mathcal{N}(0,\sigma^2)$ with a fixed random seed of varying standard deviation (increasing towards right). The different columns in the bottom row represents the variance in the power spectra for the different surfaces. Results are shown for Z148. The solid white line represents the horizon limit, and the white dashed line represents the horizon buffer limit.}
    \label{fig:2D_PS_xy_perturb_Z148}
\end{figure*}

In this paper, we test four spectral window functions: the 4-term Blackman-Harris (BH), Dolph-Chebyshev with 150 dB and 180 dB sidelobe suppression (DC150, DC180), and the Kaiser window with $\beta$ = 20 (Kaiser20). Each offers distinct trade-offs between spectral leakage suppression and resolution along $k_{\parallel}$ in the noiseless scenario. The equivalent-noise bandwidth (ENBW) of a window is defined as 

\begin{equation}
\mathrm{ENBW}=N\,\frac{\sum_n w_n^{2}}{\left(\sum_n w_n\right)^{2}},
\label{eq:4}
\end{equation}

\noindent
where $N$ is the number of frequency channels and $w_{n}$ are the coefficients of the chosen spectral window function. Therefore, this ratio gives the effective width of the main lobe of the window in units of frequency bins, and provides a measure of the spectral resolution along $k_{\parallel}$. Narrower ENBW values correspond to sharper resolution in $k_{\parallel}$, but that comes with weaker suppression of SLL, whereas wider ENBW values provide stronger suppression of SLL at the cost of coarser $k_{\parallel}$ resolution.


Fig. \ref{fig:compare_window} shows the cylindrically averaged power spectra for Z42, averaged over all baselines, comparing the effect of different window functions. The BH window gives the narrowest ENBW ($\approx$ 2 bins), providing the sharpest $k_{\parallel}$ resolution. However, with a SLL of $\approx$ 92 dB, it lacks the required dynamic range to suppress foregrounds at the levels required for DA experiments. By contrast, the DC150, DC180, and Kaiser20 windows achieve a SLL of $\gtrsim$ 150 dB. The DC150 window has an ENBW of 2.38 bins and provides strong SLL suppression relative to BH. The Kaiser20 window and DC180 window have nearly identical ENBW of 2.59 bins, and a SLL suppression sufficient to reach the sensitivity beyond the expected 21-cm level at $z$ $\sim$ 42.5 in the noiseless regime. Although their ENBWs are similar, they show different spectral behavior. The DC180 window produces an equiripple sidelobe, whereas the Kaiser20 window shows a monotonic sidelobe roll-off that continues to decrease towards higher $k_{\parallel}$ modes. For foreground-dominated measurements during DA, Kaiser20 is preferable as it minimizes the risk of isolated contamination peaks and provides a smooth roll-off at higher delays. The solid magenta and dashed lines in Fig. \ref{fig:compare_window} represent the cylindrically averaged power spectra of the 21-cm signal only using BH and DC180, respectively. The two curves are in close agreement with each other, showing that the choice of window functions does not affect the intrinsic spectral features of the 21-cm signal, at least in the absence of thermal noise and foregrounds.

We note that the most suitable window function will depend on the approach taken for foreground mitigation. Since our aim in this paper is to quantify the impact of antenna position offset on the 21-cm power spectrum, the Kaiser20 window is found to be sufficient for our purposes. It is also important to note that when strongly tapered windows are used, a sufficient number of frequency channels ($\geq$ 100) are used for the window to work effectively. This is because such windows require a certain minimum width in the spectral domain to fully realize their sidelobe suppression characteristics and avoid numerical artefacts such as poor frequency response or incomplete tapering near the window edges \citep{harris2005use}.

\section{Impact of position error for Z148}
\label{sec:appendix2}
The top row of Fig. \ref{fig:2D_PS_xy_perturb_Z148} presents the cylindrically-averaged power spectra of the visibility difference for antenna position offsets along xy direction with amplitudes \( \sigma_{\text{xy}} \) = [0.05, 0.15, 0.25, 0.55] metres (from left to right). The bottom row of Fig. \ref{fig:2D_PS_xy_perturb_Z148} shows the corresponding results for vertical height offsets, where the variation in the antenna height is determined by Surfaces 1, 2, 3 and 4 (see Section \ref{roughness result}), from left to right. The results for Z148 are shown here.

\section{Phase-error threshold for lateral offset}
\label{sec: phase_error_xy_estimate}
In this section, we provide an asymptotic estimate of the phase accuracy required for the smoothing of $u\varv$ modes to remain small in effect. To quantify this, we Taylor expand the exponential for small phase variance using $\exp(-x) = 1 - x$ for $|x|\ll 1$, then Equation \ref{eq:pxy_kernel} can be written as,
\begin{equation}
P_{xy} \simeq 1 - 2\pi^{2}\,\mathrm{Var}(\Delta\phi_{xy}).
\label{eq:Pxy_approx}
\end{equation}
If there are no positional offsets, then $P_{xy}$ = 1. Therefore, the fractional difference from 1 is
\[
\frac{1 - P_{xy}}{P_{xy}} \simeq 2\pi^{2}\,\mathrm{Var}(\Delta\phi_{xy}).
\]
We define `negligible' smoothing as the regime where
$P_{xy}$ deviates from 1 by less than about 1\% across the finite field of view, i.e.\ $|1 - P_{xy}/P_{xy}|\lesssim 0.01$. This implies
\[
2\pi^{2}\,\mathrm{Var}(\Delta\phi_{xy}) \lesssim 0.01.
\]
Therefore,
\[
\mathrm{Var}(\Delta\phi_{xy}) \lesssim \frac{0.01}{2\pi^{2}}
\approx 5\times 10^{-4},
\]
corresponding to an rms phase error $\lesssim 0.02\,\mathrm{rad} \approx 1^\circ$.

Using Equation \ref{eq:var_xy}, and assuming that $\|\mathbf{A}_{xy}^{\mathsf T}\mathbf{x}\|$ is $\sim$ 1 over the finite field of view, this translates to $\frac{\sigma_{xy}}{\lambda} \lesssim 0.02.$

Thus, if the rms positional offset along xy direction is below a few per cent of the wavelength, $P_{xy}$ is flat at the 1\% level and the associated smoothing in $u\varv$ domain is negligible. For e.g. at $\nu = 32.5$\ MHz, we have $\lambda \simeq 9.2$\,m, then for our largest perturbation along xy direction, $\sigma_{xy}=0.55$\,m, $\sigma_{xy}/\lambda$ corresponds to 0.06, with an rms phase error of $\approx 4.8^\circ$ and $P_{xy}\approx 0.87$ (a $\sim 13\%$ deviation). This test case therefore lies outside the `negligible smoothing' regime by our 1\% criterion.

Now let us consider $\sigma_{xy} = 0.25$\,m. At the same observing frequency, this corresponds to $\frac{\sigma_{xy}}{\lambda} = \frac{0.25}{9.2} \approx 0.027$. Therefore,\[
\mathrm{Var}(\Delta\phi_{xy}) \simeq (0.027)^{2}
\approx 7.3\times 10^{-4}.\]Substituting in Equation \ref{eq:Pxy_approx} gives,

\[
P_{xy} \simeq 1 - 2\pi^{2}(7.3\times 10^{-4})
           \approx 1 - 0.014
           \approx 0.986.
\]
This corresponds to a deviation of $\sim 1.4\%$, only slightly above the 1\% threshold. Therefore, a positional offset of $\sigma_{xy}=0.25$\,m along xy direction produces a small level of smoothing of $u\varv$ modes and lies close to the boundary of the `negligible smoothing' regime as defined above.


\bsp	
\label{lastpage}
\end{document}